\documentclass[a4paper]{revtex4}
\usepackage{hyperref}
\usepackage{bbding}
\usepackage[T1]{fontenc}
\usepackage[utf8]{inputenc}
\usepackage{textcomp}
\usepackage{amsmath}
\usepackage{amssymb}
\usepackage[amsmath,amsthm,thmmarks]{ntheorem}
\usepackage{amsfonts}
\usepackage{amssymb,epic,eepic}
\usepackage{amsmath}
\usepackage{stmaryrd}
\usepackage{capt-of}
\usepackage{graphicx}
\usepackage[usenames,dvipsnames]{pstricks}
\usepackage{epsfig}
\usepackage{pst-grad}
 \usepackage{pst-plot}
\usepackage{here} 
\usepackage{paralist}
\usepackage{color}

\newtheorem{theorem}{Theorem}[section]
\newtheorem{lemma}[theorem]{Lemma}
\newtheorem{corollary}[theorem]{Corollary}
\newtheorem{definition}[theorem]{Definition}
\newtheorem{remark}[theorem]{Remark}

\newtheorem{example}[theorem]{Example}

\begin{document}
\title{Secret
Message Transmission over Quantum Channels under Adversarial Quantum 
Noise: Secrecy Capacity and   Super-Activation  }
\author{Holger Boche}
\email{boche@tum.de}
\affiliation{
Lehrstuhl f\"ur Theoretische
Informationstechnik,
 Technische Universit\"at M\"unchen,
Munich, Germany}
\author{Minglai Cai}
\email{minglai.cai@tum.de} 
\affiliation{
Lehrstuhl f\"ur Theoretische
Informationstechnik,
 Technische Universit\"at M\"unchen,
Munich, Germany}
\author{Christian Deppe}
\email{christian.deppe@tum.de} 
\affiliation{
Lehrstuhl f\"ur
  Nachrichtentechnik,
Technische Universit\"at M\"unchen,
Munich,  Germany}
\author{Janis N\"otzel}
\email{janis.noetzel@tu-dresden.de}
\affiliation{
Technische Universit\"at Dresden
 Dresden, Germany}

\begin{abstract}
We determine the secrecy capacities  of AVQCs. 
Both secrecy capacity with average error probability
and  with maximal error probability are
derived.      Both derivations are based on one common code construction.
   The code we construct fulfills a  stringent secrecy requirement,
which is called the strong code concept.     As an application
of our result for secret message transmission over AVQCs,     we 
determine when the secrecy
capacity is a continuous function of the system parameters and
completely characterize its discontinuity   points both for average error criterion
and for maximal error criterion. Furthermore, we prove the phenomenon ``super-activation''
for  secrecy capacities  of arbitrarily varying quantum
 channels, i.e., two quantum channels both with zero  secrecy capacity, which, if used together, allow
 secure transmission with positive capacity.      We give therewith an answer to the question
``When is
the secrecy capacity a continuous function of the system parameters?'', which has been
listed as an open problem in quantum information problem page of the ITP Hannover. We also
discuss the relations between the entanglement distillation capacity,
the entanglement generating capacity, and the strong subspace 
transmission capacity for AVQCs.  R. Ahlswede,  I. Bjelakovi\'{c}, H. Boche,  and J. N\"otzel
made in 2013 the  conjecture  that
the
 entanglement generating capacity of an AVQC is equal to
its
 entanglement generating capacity 
  under shared randomness  assisted quantum coding.
     We state the conditions     under which the secrecy
capacity is a continuous function of system parameters.
We demonstrate that   the validity of this conjecture implies that the entanglement generating capacity, the
entanglement distillation capacity, and the strong subspace transmission capacity of an AVQC are
continuous functions of the system parameters. Consequently, under the premise of this conjecture, 
the  secrecy capacities  of an AVQC
  differ significantly from the
	general quantum capacities.  
\end{abstract}
\maketitle

\section{Introduction}

In the last few years,
the developments in modern communication systems have produced 
many results in a short amount of time.   Quantum communication systems
especially,   allow us to exploit new possibilities while at the same time imposing fundamental limitations.

Quantum mechanics differs significantly from classical   mechanics,  
it has its own laws.  Quantum information theory unifies information theory with quantum mechanics,
generalizing classical information theory to the quantum world. 
The unit of quantum information is   called    the "qubit", the quantum analogue of the classical ``bit''.
Unlike a bit, which is either ``0'' or ``1'', a qubit can be in   a   ``superposition'',
i.e.,   two     states at the same   time. This property has turned into one of the fundamental resources 
of quantum information processing.

A quantum
channel is  a communication  channel which can carry quantum information, e.g., photons.
    Two of the most standard
ways    to represent a quantum  channel with linear algebraic tools are
 a sum of several
transformations, or as a single     isometry,      which
explicitly includes the unobserved environment.

A quantum  channel  can transmit both classical
and quantum information.
If the sender wants to transmit a classical message of a finite set $\mathbf{A}$ to
the receiver using a quantum channel $N$, his encoding procedure will
include a classical-to-quantum encoder 
to prepare a quantum message state $\rho$ suitable as an input for the channel. If the sender's
encoding is restricted to transmit an  indexed finite set of
 quantum states     $\{\rho_{x}: x\in \mathbf{A}\}$,      then we can consider the choice of the signal
quantum states     $\rho_{x}$      as a component of the channel.
This is equivalent  to  considering
 the classical capacity of  quantum channels.

Our goal is
  to   investigate    communication that
	takes place over a quantum channel 
which is, in addition to   the   noise from the environment, subjected 
to the action of a jammer   who   actively manipulates the states.
The messages   should   also  be kept secret from an 
eavesdropper.

Preceding work in quantum information theory  has mostly focused on either 
of the two attacks. Our goal is to deliver a more general theory considering 
both channel robustness and security in quantum information theory. By doing 
so, we build on the preceding works \cite{Bl/Ca} and \cite{Bo/No}. 
Furthermore, we are interested in the delivery of large volumes of 
messages over many channel uses, so that we study the asymptotic behavior of the system.

A channel with a jammer is called an arbitrarily varying  channel,
where the jammer may change his input in every channel use and is not restricted to use a 
repetitive probabilistic strategy. 
In  this model  we consider such a channel  which is not stationary  
and    can change  with every use. 
     Communication over this channel works   as follows:   At first  the sender and the receiver have to select their
coding scheme. After that, the jammer makes his choice of the channel state to sabotage the message  transmission.
However, due to the physical properties, we assume
that  the jammer's changes only take  place in a set which is known to       the sender and the receiver.

The arbitrarily varying channel was first introduced
 in \cite{Bl/Br/Th2}. 
 \cite{Ahl1}  showed a surprising result
which is  known as the Ahlswede Dichotomy:
Either the 
capacity for message transmission
of an arbitrarily varying channel is zero or it equals its shared randomness assisted capacity. 
After the discovery in   \cite{Ahl1} it   has remained  an open question as to when the deterministic capacity is positive
      for several years.
     In   \cite{Rei},    a sufficient condition for that has been given, and in \cite{Cs/Na}, it is
 proved that this condition is also necessary.
The  Ahlswede Dichotomy demonstrates the importance of shared randomness for communication in a very clear
 form.

A quantum channel with a jammer is called an arbitrarily varying 
quantum channel.  It is defined as a 
family of  indexed channels $\{N_{\theta}:{\theta}=1,\ldots,T\}$,
where ${\theta}$ is called a channel state of the 
channel pair. This channel state ${\theta}$, 
which varies from symbol to symbol in an arbitrary manner, governs  the 
 channel.
In \cite{Ahl/Bli}, the classical capacity of arbitrarily varying 
quantum channels has been analyzed, and      a      lower bound on the capacity has been given.
 An alternative proof  and a proof of the strong converse are both  given 
in \cite{Bj/Bo/Ja/No}. In \cite{Ahl/Bj/Bo/No}, the Ahlswede Dichotomy for the arbitrarily varying 
classical-quantum channels is established, and a sufficient and necessary condition for the zero 
deterministic capacity is given. In \cite{Bo/No}, a simplification of this condition for the 
arbitrarily varying classical-quantum channels is given. 
\cite{Ahl/Bj/Bo/No} and \cite{Bo/Ca/De}
 complete the characterization of message transmission capacity  for the arbitrarily varying channel.
In \cite{Ahl/Bj/Bo/No}, message transmission, key transmission, and
strong subspace transmission have been considered. A full description of  these  
transmission tasks has been completely established.
 In \cite{Bo/No} an example has been given showing there are indeed arbitrarily varying classical-quantum
channels which have zero deterministic capacity and positive random capacity.
In this work we rendered these results to  secrecy  message transmission over arbitrarily varying quantum
 channels, which is the full description of secure message transmission and
 key transmission through arbitrarily varying quantum
wiretap  channels.

The noise affecting the transmission over a noisy quantum channel
can be     interpreted as
 interaction with the environment.
Following the general protocol of \cite{De}, we say
this environment  is completely under the control of the eavesdropper.
Secure communication over a classical channel with an eavesdropper was first introduced   in \cite{Wyn}.
A classical-quantum  channel with an eavesdropper is called a
classical-quantum wiretap channel.
The secrecy capacity for
classical-quantum 
channel with an eavesdropper
has been determined in \cite{De} and \cite{Ca/Wi/Ye}.

This work is an   extension   of our
previous papers  \cite{Bo/Ca/De}, \cite{Bo/Ca/De2}, and \cite{Bo/Ca/De3}, where  we considered  channel robustness against
  jamming,   and   concurrently   security against  eavesdropping
for classical-quantum channels (cf. Section \ref{pdamr}),  where classical-quantum channels are quantum  channels whose
sender's inputs are classical variables. 
In our earlier works \cite{Bo/Ca/De}, \cite{Bo/Ca/De2}, and \cite{Bo/Ca/De3}, 
we  investigated  secret message transmission over a
classical-quantum  channel.  The messages were kept secret from an eavesdropper. Communication took place over a quantum channel which was, 
in addition to noise from the environment, subjected to the action of a jammer, which actively manipulated the states.
 A classical-quantum channel with both a jammer and an eavesdropper is called 
an arbitrarily varying classical-quantum wiretap channel. It is defined as a 
family of pairs of indexed channels $\{(W_{\theta},V_{\theta}):{\theta}=1,\ldots,T\}$ with a common 
input alphabet and possible different output systems, connecting a sender with two 
receivers, a legal one and a wiretapper. The legitimate receiver accesses the output of 
 the first channel $W_{\theta}$ in the pair, and the wiretapper observes the output of the 
 second channel $V_{\theta}$, respectively. ${\theta}$ governs both the 
legal receiver's channel and the wiretap channel. A code for the channel 
conveys information to the legal receiver such that      the wiretapper's knowledge of the transmitted information
can be kept arbitrarily small.      In  \cite{Bo/Ca/De}, \cite{Bo/Ca/De2}, and \cite{Bo/Ca/De3}, 
the Ahlswede Dichotomy for arbitrarily varying
classical-quantum
wiretap channels is established, i.e. either the deterministic 
capacity of an arbitrarily varying channel is zero or is equal to its shared randomness assisted capacity.
We delivered the formula for secrecy capacity of the arbitrarily varying classical-quantum
wiretap    channel. A full description of the arbitrarily varying classical-quantum
wiretap  channels is thus established     with these earlier contributions \cite{Bo/Ca/De}, \cite{Bo/Ca/De2}, and \cite{Bo/Ca/De3}
when we   combine their results.        Currently only the message transmission capacity of  arbitrarily varying quantum 
channels can be completely characterized
(cf. \cite{Ahl/Bj/Bo/No}, \cite{Bo/No}, and \cite{Bo/No2}).
The
entanglement distillation capacity, the entanglement generating capacity,
and the strong subspace transmission capacity 
of arbitrarily varying quantum 
channels can only be determined for the case when
the respective capacity is positive by now (\cite{Ahl/Bj/Bo/No}),
i.e., the characterization of these capacities when
the respective capacity is equal to zero
is still an open problem.

 In this work we  determine the secrecy capacities  of arbitrarily varying quantum
 channels. 
We also give an example showing that
there are indeed arbitrarily varying quantum
 channels which have zero deterministic secrecy capacity 
and positive      randomness assisted      secrecy capacity.       The 
deterministic secrecy capacity 
and the     randomness assisted      secrecy capacity
of an AVQC is thus in general not equal.
This behavior comes as quite a surprise
because in \cite{Ahl/Bj/Bo/No} the authors
conjectured  equality of
deterministic capacity 
and  randomness assisted capacity for
entanglement distillation, entanglement generating,
and  strong subspace transmission
for
arbitrarily varying quantum 
channels.

The capacities of classical arbitrarily varying channel under
maximal error criterion and under
the average error criterion are in general, not equal.
The capacity formula of classical arbitrarily varying channels under
maximal error criterion is still an open problem.
Interestingly,  \cite{Bo/No}
shows that 
the capacities of an arbitrarily varying quantum channel under
maximal error criterion and under the average error criterion are equal
(cf. Remark \ref{fcavccutec}, Remark \ref{icwavqcwa}, and Remark \ref{iutcoavcqcum}).      In
Section \ref{protmr}  we extend this observation: We show that 
the secrecy capacities of an arbitrarily varying quantum channel under
maximal error criterion and under the average error criterion
are equal. For the proof we have to construct two sets
of superposition codes to show      the positivity of the     secure capacity
under
maximal error criterion.

As an application of our results, we turn to the
question: Is
the secrecy capacity a continuous function of the system parameters?
The analysis of the continuity of  capacities of     quantum     channels is
motivated by the question of
whether small changes in the channel system are able to cause dramatic losses in the performance.
The continuity of the message and entanglement transmission capacity of a stationary memoryless quantum channel has been
listed as an open problem in \cite{WynSite} and was solved in \cite{Le/Sm}. 
 Considering      quantum      channels with active jamming faces an especially  new difficulty.
 The reason is that
 the  capacity in this case is, in
general, not solely specified by entropy quantities.
In \cite{Bo/No2}      the conditions under which
the message transmission capacity of an
 arbitrarily varying quantum channel is continuous have been
delivered.
The condition for continuity of message transmission capacity of a classical 
 arbitrarily varying  wiretap channel has been given 
in \cite{Wi/No/Bo}. 
We shall discuss the context
of the
entanglement distillation capacity, the entanglement generating capacity,
and the strong subspace transmission capacity 
in Section \ref{mare} and in Section \ref{ccoscoqc}.

The  continuity of
the secrecy capacity of a classical arbitrarily varying channel 
under  randomness assisted quantum coding has been shown
 in  \cite{Bo/Sch/Po}. This proof is still 	capable of     improvements, 
     since  in general, the legal channel users do not have control over
the   the eavesdropper's channel.
However, this proof requires that the output alphabet  of the
 eavesdropper's channel is of finite cardinality.
In this work we show that the  continuity of the secrecy capacity of
 an arbitrarily varying 
quantum channel under  randomness assisted quantum coding
only depends on the legal channel.      Moreover, we 
improve the result of  \cite{Bo/Sch/Po} when we give a  generalized
 control function. This  control function only depends on the legal channel.

Furthermore, we show     as
   a  consequence of our results that      there  is     a  phenomenon  called
``super-activation'' for the secrecy capacity of
 arbitrarily varying 
quantum channels, i.e., two arbitrarily varying quantum
 channels,     each
useless
for secure message
transmission, can be super-activated 
to acquire positive    secrecy  capacity when used together.

\section{Preliminaries}
\label{bdacs2}
\subsection{Basic properties, Communication Scenarios, and Notations}\label{commscen}

For  a finite set $\mathbf{A}$ we denote  the
set of probability distributions on $\mathbf{A}$ by $P(\mathbf{A})$.
Let $\rho_1$ and  $\rho_2$ be  Hermitian   operators on a  finite-dimensional
complex Hilbert  space $G$.
We say $\rho_1\geq\rho_2$ and $\rho_2\leq\rho_1$ if $\rho_1-\rho_2$
is positive-semidefinite.
 For a finite-dimensional
complex Hilbert space  $G$, we denote
the (convex) space 
of  density operators on $G$ by
\[\mathcal{S}(G):= \{\rho \in \mathcal{L}(G) :\rho  \text{ is Hermitian, } \rho \geq 0_{G} \text{ , }  \mathrm{tr}(\rho) = 1 \}\text{ ,}\]
where $\mathcal{L}(G)$ is the set  of linear  operators on $G$, and $0_{G}$ is the null
matrix on $G$. Note that any operator in $\mathcal{S}(G)$ is bounded.\vspace{0.2cm}

For any finite set $\mathbf{A}$, any finite-dimensional
complex Hilbert space $H$, and  $n\in\mathbb{N}$, we define ${\mathbf{A}}^n:= \Bigl\{(a_1,\cdots,a_n): a_i \in \mathbf{A}
\text{ } \forall i \in \{1,\cdots,n\}\Bigr\}$, and $H^{\otimes n}:=
span\Bigl\{v_1\otimes \cdots \otimes v_n: v_i \in H
\text{ } \forall i \in \{1,\cdots,n\}\Bigr\}$.    We  write    $a^n$ 
for the elements of
${\mathbf{A}}^n$.

We denote the identity operator on a space $H$ by $\mathrm{id}_H$
and the symmetric group on $\{1,\cdots,n\}$ by $\mathsf{S}_n$.
   For a set $\mathbf{A}$ on a  Euclidean space $G$  we define the  convex hull of $\mathbf{A}$
by $Conv(\mathbf{A})$. 
\vspace{0.2cm}

For a discrete random variable $X$  on a finite set $\mathbf{A}$ and a discrete
random variable  $Y$  on  a finite set   $\mathbf{B}$,   we denote the Shannon entropy
of $X$ by
$H(X)=-\sum_{x \in \mathbf{A}}p(x)\log p(x)$ and the mutual information between $X$
and $Y$ by  
$I(X;Y) = \sum_{x \in \mathbf{A}}\sum_{y \in \mathbf{B}}  p(x,y) \log{ \left(\frac{p(x,y)}{p(x)p(y)} \right) }$.
Here $p(x,y)$ is the joint probability distribution function of $X$ and $Y$, and 
$p(x)$ and $p(y)$ are the marginal probability distribution functions of $X$ and $Y$ respectively,
and ``$\log$''  means logarithm to base $2$.\vspace{0.2cm}

For a quantum state $\rho\in \mathcal{S}(H)$ we denote the von Neumann
entropy of $\rho$ by \[S(\rho)=- \mathrm{tr}(\rho\log\rho)\text{
.}\]\vspace{0.2cm}

Let $\mathfrak{P}$ and $\mathfrak{Q}$ be quantum systems. We 
denote the Hilbert space of $\mathfrak{P}$ and $\mathfrak{Q}$ by 
$G^\mathfrak{P}$ and $G^\mathfrak{Q}$, respectively. Let $\phi^\mathfrak{PQ}$ be a bipartite
quantum state in $\mathcal{S}(G^\mathfrak{PQ})$. 
We denote the partial
trace over $G^\mathfrak{P}$ by
\[\mathrm{tr}_{\mathfrak{P}}(\phi^\mathfrak{PQ}):= 
\sum_{l} \langle l|_{\mathfrak{P}} \phi^\mathfrak{PQ} |  l \rangle_{\mathfrak{P}}\text{ ,}\]
where $\{ |  l \rangle_{\mathfrak{P}}: l\}$ is an orthonormal basis
of $G^\mathfrak{P}$.
We denote the conditional entropy by
\[S(\mathfrak{P}\mid\mathfrak{Q})_{\rho}:=
S(\phi^\mathfrak{PQ})-S(\phi^\mathfrak{Q})\text{
,}\]
where $\phi^\mathfrak{Q}=\mathrm{tr}_{\mathfrak{P}}(\phi^\mathfrak{PQ})$.\vspace{0.2cm}

Let $\Phi := \{\rho_x : x\in \mathbf{A}\}$  be a set of quantum  states
labeled by elements of $\mathbf{A}$. For a probability distribution  $Q$
on $\mathbf{A}$, the     Holevo quantity $\chi$      is defined as
\[\chi(Q;\Phi):= S\left(\sum_{x\in \mathbf{A}} Q(x)\rho_x\right)-
\sum_{x\in \mathbf{A}} Q(x)S\left(\rho_x\right)\text{ .}\]
Note that we can always associate a state 
$\rho^{XY}=\sum_{x}Q(x)|x\rangle\langle x|\otimes \rho_x$ to
$(Q;\Phi)$ such that $\chi(Q;\Phi)=I(X;Y)$ holds for the quantum
mutual information.

For a probability distribution $P$ on a finite set $\mathbf{A}$  and a positive constant $\delta$,
we denote the set of typical sequences by 
\[\mathcal{T}^n_{P,\delta} :=\left\{ a^n \in \mathbf{A}^n: \left\vert \frac{1}{n} N(a'\mid a^n)
- P(a') \right\vert \leq \frac{\delta}{|\mathbf{A}|}\forall a'\in \mathbf{A}\right\}\text{ ,}\]
where $N(a'\mid a^n)$ is the number of occurrences of the symbol $a'$ in the sequence $a^n$. 
\vspace{0.2cm}

 For  finite-dimensional
complex Hilbert spaces  $G$ and  $G'$,  a quantum channel $N$:
$\mathcal{S}(G) \rightarrow \mathcal{S}(G')$, $\mathcal{S}(G)  \ni
\rho \rightarrow N(\rho) \in \mathcal{S}(G')$ is represented by a
completely positive trace-preserving map $\mathcal{L}(G) \rightarrow \mathcal{L}(G')$
 which accepts input quantum states in $\mathcal{S}(G)$ and produces output quantum
states in  $\mathcal{S}(G')$. \vspace{0.2cm}

\begin{definition}
Let $\mathfrak{P}$ and $\mathfrak{Q}$ be
quantum systems. We denote the Hilbert space of $\mathfrak{P}$ and
$\mathfrak{Q}$ by $H^\mathfrak{P}$ and  $H^\mathfrak{Q}$,
 respectively, and let
  $\Theta$ $:=$ $\{1,\cdots,T\}$ be a finite set. \vspace{0.2cm}

For every ${\theta} \in \Theta$,		
	let ${N}_{\theta}$    be a quantum channel
$\mathcal{S}(H^\mathfrak{P}) \rightarrow \mathcal{S}(H^\mathfrak{Q})$.
We call the set of the  quantum
channels  $\{{N}_{\theta}: {\theta} \in \Theta\}$ an \bf arbitrarily varying
quantum  channel \it  when the state $\theta$ varies from
symbol to symbol in an  arbitrary manner.
\end{definition}\vspace{0.2cm}

The following 	Definition
\ref{symmetL} plays a
very important role for
the characterization 
of the capacity of
arbitrarily varying quantum channels.       
			The intuitive meaning
of this is that the jammer can choose the state of the
channel such that any two send sequences of quantum states of any length, may be
confused by the receiver.

\begin{definition}\label{symmetL} Let $L\in\mathbb{N}$.
We say that the arbitrarily varying quantum channel
$\{N_{\theta} : {\theta} \in \Theta\}$ is $L$-symmetrizable if
for every 
$\rho^L$, ${\rho'}^L\in$ $\mathcal{S}(H^{\mathfrak{P}^L})$
 there exists a map $\tau$ which maps from $\{\rho^L,{\rho'}^L\}$ 
to the set of distributions  on $\Theta^L$ such that 
\begin{equation}\sum_{\theta^L\in\Theta^L}\tau(\rho^L)(\theta^L) {N}_{\theta^L}({\rho'}^L)
=\sum_{\theta^L\in\Theta^L}\tau({\rho'}^L)(\theta^L) {N}_{\theta^L}(\rho^L)\text{ .}
\label{stlitltrl}\end{equation}

We say that the arbitrarily varying quantum channel
$\{N_{\theta} : {\theta} \in \Theta\}$ is symmetrizable if
$\{N_{\theta} : {\theta} \in \Theta\}$ is $L$-symmetrizable 
for all $L\in \mathbb{N}$.

\end{definition}\vspace{0.2cm}

\begin{remark} In \cite{Bo/No2} the $L$-symmetrizability was defined as follows:
We say that 
$\{N_{\theta} : {\theta} \in \Theta\}$ is $L$-symmetrizable if
for every finite set $\{\rho_1^L, \cdots, \rho_K^L  \}$
 $\subset$  $\mathcal{S}(H^{\mathfrak{P}^L})$
 there exists a $\tau$ such that for all 
$\rho^L$, ${\rho'}^L$  $\in\{\rho_1^L, \cdots, \rho_K^L  \}$, (\ref{stlitltrl}) holds.
The authors then showed that we might limit the cardinality $K$ to $K=2$,        which is 
the condition for $L$-symmetrizability defined in (\ref{stlitltrl}).
Thus the two conditions for $L$-symmetrizability are equivalent.      
\label{icbntlswdaf} \end{remark}\vspace{0.2cm}



\subsection{Problem Definition and Basic  Definitions} 
\label{pdamr}

Two of the common ways to represent a quantum  channel,    i. e. a
completely positive trace-preserving map $\mathcal{L}(H^\mathfrak{P}) \rightarrow \mathcal{L}(H^\mathfrak{Q})$,
with linear algebraic tools, are:\\[0.2cm]
\it 1. Operator sum decomposition  (Kraus representation)\rm
\begin{equation}N(\rho)= \sum_{i=1}^{K} A_i\rho{A_i}^*\text{ ,}\label{krausrep}\end{equation}
where  $A_1,\dots,A_K$ (Kraus operators) are linear operators  $H^{\mathfrak{P}}$
$\rightarrow$  $H^{\mathfrak{Q}}$ (cf.\cite{Kr}, \cite{Ba/Ni/Sch}, and \cite{Ni/Ch}). They
satisfy the       trace-preserving       relation
$ \sum_{i=1}^{K} {A_i}^*A_i=\mathrm{id}_{H^\mathfrak{P}}$.
The representation of a quantum channel $N$ according to (\ref{krausrep})
is not unique.
Let
 $A_1,\dots,A_K$  and  $B_1,\dots,B_{K'}$  be two sets
 of Kraus operators (by appending zero operators to the shorter list
of operation elements we may ensure that $K'=K$).   Suppose $A_1,\dots,A_K$    represents $N$,
then  $B_1,\dots,B_{K}$  also represents $N$ if and only if
there exists a $K\times K$ unitary matrix $\left(u_{i,j}\right)_{i,j=1,\dots,K}$ such that for all $i$ we
have $A_i = \sum_{j=1}^K u_{i,j}B_{j}$ (cf. \cite{Ni/Ch}).
\\[0.15cm]
\it 2. Isometric extension (Stinespring dilation)\rm
\begin{equation}N(\rho)= \mathrm{tr}_{\mathfrak{E}}\left(U_{N} \rho U_{N}^*\right)\text{ ,}\label{stinespringdi}\end{equation}
 where $U_{N}$ is a  linear operator
$H^{\mathfrak{P}}$ $\rightarrow$
$H^{\mathfrak{QE}}$ such that $U_{N}^*U_{N}=\mathrm{id}_{H^\mathfrak{P}}$, and $\mathfrak{E}$ is the quantum
system of  the environment (cf. \cite{Sho}, \cite{Ba/Ni/Sch}, and
 also cf. \cite{St} for a more general  Stinespring dilation     theorem). $H^{\mathfrak{E}}$ can be chosen such that $\dim H^{\mathfrak{E}} \leq (\dim H^{\mathfrak{P}} )^2$.
 The isometric extension  of a quantum channel $N$ according to (\ref{stinespringdi})
is not  unique either. Let $U$ and $U'$ be two linear operators
$H^{\mathfrak{P}}$ $\rightarrow$
$H^{\mathfrak{QE}}$.  Suppose $U$   represents $N$,
then  $U'$  also represents $N$ if and only if
 $U$ and $U'$ are unitarily equivalent.
\vspace{0.15cm}

It is well known that we can deduce each of these two representations of 
the quantum channel  from the other one. Let  $A_1,\dots,A_K$ be a
set of Kraus operators which  represents $N$. Let
$\{|j\rangle^\mathfrak{E}:j=1, \dots, K\}$ be an orthonormal system
on $H^\mathfrak{E}$. Then $U_{N}= \sum_{j=1}^{K} {A_j}\otimes
|j\rangle^\mathfrak{E}$ is an isometric extension which  represents
$N$, since $\left(\sum_{j=1}^{K} {A_j}\otimes
|j\rangle^\mathfrak{E}\right)$ $\rho$  $ \left(\sum_{k=1}^{K}
{A_k}\otimes |k\rangle^\mathfrak{E}\right)^*$ $=$ $\sum_{j=1}^{K}
A_j\rho{A_j}^*$ and $ \left(\sum_{j=1}^{K} {A_j}\otimes
|j\rangle^\mathfrak{E}\right)^*$ $\left(\sum_{k=1}^{K} {A_k}\otimes
|k\rangle^\mathfrak{E}\right)$ $=$ $\sum_{j=1}^{K} {A_j}^*A_j$. For
the other way around, every isometric extension $U_{N}$ that
represents $N$ can be written  in the form $U_{N}= \sum_{j=1}^{K}
{A_j}\otimes |j\rangle^\mathfrak{E}$, i.e. if the sender sends
$\rho$, and if the  environment's measurement gives
$|i\rangle^\mathfrak{E}$, the receiver's outcome will be
$\frac{1}{|A_i\rho{A_i}^*|_1}A_i\rho{A_i}^*$. Here $A_1,\dots,A_K$ is a set of Kraus operators
which  represents $N$, and
 $\{|j\rangle^\mathfrak{E}:j=1, \dots, K\}$ is an orthonormal system on $H^\mathfrak{E}$.\vspace{0.2cm}

\begin{center}\begin{figure}[H]
\includegraphics[width=0.9\linewidth]{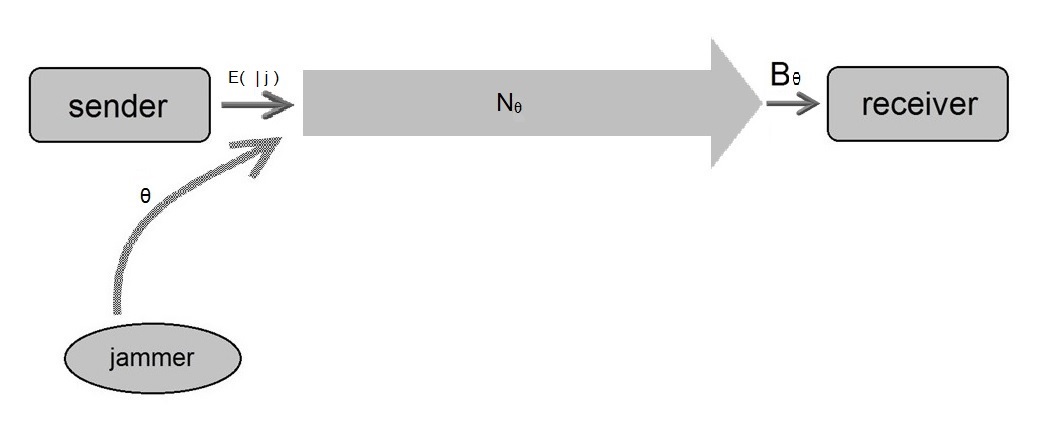}
\caption{Arbitrarily varying quantum channel}
\label{qqpic1}\end{figure}\end{center}

We aim to analyze secret message transmission
over a quantum channel. This channel connects the sender and the legal receiver,
conveying information from the former to the latter.
It is subject to active  attacks: A jammer may change his
input in every channel use and is not restricted to use a repetitive probabilistic
strategy. If we do not consider security,
we have a situation as shown in Figure \ref{qqpic1},
where public messages are transmitted. 
This scenario is described by a tripartite system
with the sender, the legal receiver, and the jammer.

Following \cite{De} we  now
define the secret message transmission protocol
when we
assume that the environment is completely under the control of the eavesdropper.

\begin{definition}
Let $\mathfrak{P}$ and $\mathfrak{Q}$ be
quantum systems and
  $\Theta$ $:=$ $\{1,\cdots,T\}$ be a finite set.
Let $\mathfrak{I}$
$=\{N_{\theta}: {\theta}\in\Theta\}$
be an arbitrarily varying
quantum  channel. Following \cite{De} we  assume that the environment $\mathfrak{E}$ is completely under the control of the eavesdropper
in the following sense:

Let $\rho^{\mathfrak{P}} \rightarrow U_{N_{\theta}}\rho^{\mathfrak{P}}U_{N_{\theta}}^*$ be an isometric transformation
which represents $N_{\theta}$, where $U_{N_{\theta}}$ is a  linear operator $\mathcal{S}(H^{\mathfrak{P}})$
$\rightarrow$  $\mathcal{S}(H^{\mathfrak{QE}})$, and $\mathfrak{E}$ is the quantum system of  the environment.
 $H^{\mathfrak{E}}$ can be chosen such that $\dim H^{\mathfrak{E}} \leq (\dim H^{\mathfrak{P}} )^2$.
Fix a $ \rho^{\mathfrak{P}}$ with  eigen-decomposition
$\sum_{x \in \mathcal{X}} p(x)|\phi_x\rangle^{\mathfrak{P}} \langle \phi_x|^{\mathfrak{P}}$.
If the channel state is $\theta$,  the local
output density matrix seen by the receiver is \[
\mathrm{tr}_{\mathfrak{E}} \left(\sum_{x } p(x)U_{N_{\theta}} |\phi_{x}\rangle\langle
\phi_{x}|^{\mathfrak{P}}U_{N_{\theta}}^*\right) \text{ ,}\] and
the local output density matrix seen by the environment (which we
interpret as the wiretapper)
 is \[ \mathrm{tr}_{\mathfrak{Q}}
  \left( \sum_{x } p(x)U_{N_{\theta}}|\phi_{x}\rangle\langle \phi_{x}|^{\mathfrak{P}}U_{N_{\theta}}^*\right)
\text{ .}\]
Therefore $\{{V}_{\theta}': {\theta}\in\Theta\}$ defines an arbitrarily varying wiretap quantum channel
\[\{(N_{\theta},{V}_{\theta}') : {\theta}\in \Theta\}\text{ ,}\]  where
$N_{\theta}: H^{\mathfrak{P}} \rightarrow H^{\mathfrak{Q}}$,
$\sum_{x \in \mathcal{X}} p(x)|\phi_x\rangle \langle \phi_x|^{\mathfrak{P}}$
$\rightarrow$ $\mathrm{tr}_{\mathfrak{E}} \left(\sum_{x } p(x)U_{N_{\theta}} |\phi_{x}\rangle\langle
\phi_{x}|^{\mathfrak{P}}U_{N_{\theta}}^*\right)$, and
${V}_{\theta}': H^{\mathfrak{P}} \rightarrow H^{\mathfrak{E}}$,
$\sum_{x \in \mathcal{X}} p(x)|\phi_x\rangle \langle \phi_x|^{\mathfrak{P}}$
$\rightarrow$ $\mathrm{tr}_{\mathfrak{Q}} \left(\sum_{x } p(x)U_{N_{\theta}} |\phi_{x}\rangle\langle
\phi_{x}|^{\mathfrak{P}}U_{N_{\theta}}^*\right)$.

\end{definition} 
\vspace{0.2cm}

\begin{center}\begin{figure}[H]
\includegraphics[width=0.9\linewidth]{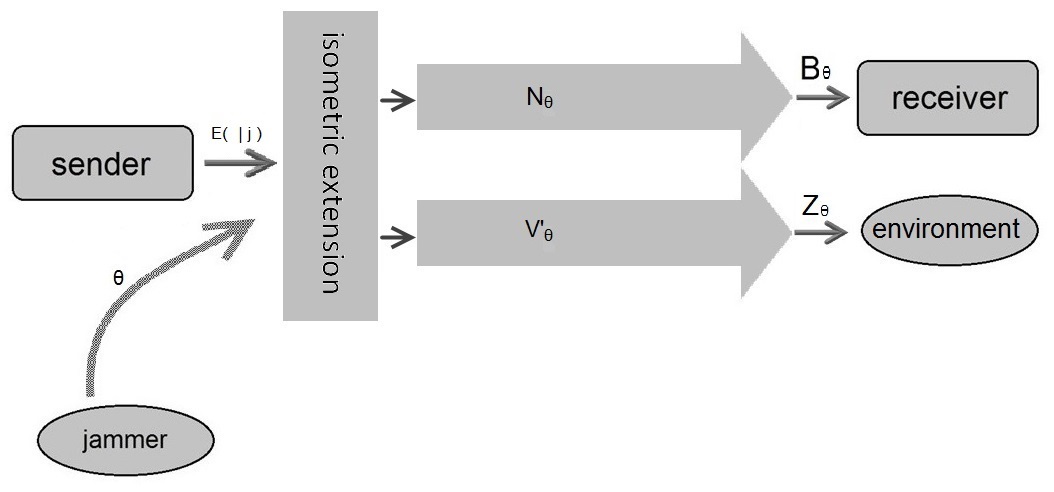}
\caption{Arbitrarily varying wiretap quantum channel
defined by an arbitrarily varying quantum channel}
\label{qqpic2}\end{figure}\end{center}

The jammer and the eavesdropper
here act as two additional channel users.
Instead of the tripartite system of 
 Figure \ref{qqpic1}, we now have
a quadripartite system in 
 Figure \ref{qqpic2}. 
The quantum channel used by the
legal parties (sender and receiver) is subject to two attacks at
the same time: one passive (eavesdropping), 
and one active (jamming).
The task for the legal transmitters is to transmit
private information (secret messages or secret keys) despite the jamming attacks
and
at the same time to keep it secret against eavesdropping.

 We consider the secrecy capacity of
quantum channels   carrying private information.
We assume that
the jammer is an additional
hostile channel user. It is effectively given by an interference channel where the legal sender and the
jammer are allowed to make inputs to the system, and the legal receiver as well
as the eavesdropper, receive the corresponding outputs.
Furthermore, we assume that
the eavesdropper controls the 
complete environment. 
The jammer and the eavesdropper aim
to prevent the
legal parties from privately communicating,
each with their own attacking strategies, respectively.
\vspace{0.2cm}

Quantum  channels  can transmit both classical
and quantum information.
For the transmission of classical information via a quantum channel,
we  first have to convert a classical
message into a quantum state. 
      We assume that
the states to
be produced in the input system are given by the set $\{\rho_{x}: x\in \mathbf{A}\}\subset
\mathcal{S}(H^{\mathfrak{P}})$, where $\mathbf{A}$ is a finite set of letters.
By Section \ref{commscen}, we can define
the map $F:$ $\mathbf{A}$ $\rightarrow \mathcal{S}(H^{\mathfrak{P}})$, which is defined by
$\mathbf{A} \ni x \rightarrow F(x)$ $= \rho_{x}$ $\in \mathcal{S}(H^\mathfrak{P})$,
meaning that
each classical input of $x\in \mathbf{A}$ leads to a distinct quantum output
$\rho_{x} \in \mathcal{S}(H^\mathfrak{P})$. $F$
defines a linear map (a classical-quantum channel)  
$\mathbf{N} \circ F :=$
 $W:$ $P(\mathbf{A})\rightarrow\mathcal{S}(H^\mathfrak{Q})$, 
\[P(\mathbf{A})  \ni P
\rightarrow W(P) =  \sum_x P(x)N(\rho_{x})  \in \mathcal{S}(H^\mathfrak{Q})\text{ .}\]
In view of this, we have the following definition.
\vspace{0.2cm}

\begin{definition}
 An $(n, J_n)$   (deterministic)    \bf     code   \it    $\mathcal{C}$ for an
arbitrarily
varying quantum channel $\mathfrak{I}$ $=\{N_{\theta} : {\theta}\in \Theta\}$
with  classical input $F:$ $\mathbf{A}$ $\rightarrow \mathcal{S}(H^{\mathfrak{P}})$
consists of a stochastic encoder $E$ : $\{
1,\cdots ,J_n\} \rightarrow P({\mathbf{A}}^n)$, 
$j\rightarrow E(\cdot|j)$,
 specified by
a matrix of conditional probabilities $E(\cdot|\cdot)$, and
 a collection of positive-semidefinite operators $\left\{D_j: j\in \{ 1,\cdots ,J_n\}\right\}$
on ${H}^{\otimes n}$,
which is a partition of the identity, i.e. $\sum_{j=1}^{J_n} D_j =
\mathrm{id}_{{H}^{\otimes n}}$. We call these  operators the decoder operators.

The  average probability of the decoding error of a
deterministic code $\mathcal{C}$ is defined as 
\[P_e(\mathcal{C}, {\theta}^n) := 1- \frac{1}{J_n} \sum_{j=1}^{J_n}
\mathrm{tr}(N_{{\theta}^n}\circ F^n(E(~|j))D_j)\text{ .}\]

The maximal probability of the decoding error of a
deterministic code $\mathcal{C}$ is defined as
  \[P_e^{(max)}(\mathcal{C}, \theta^n) := 1- \min_{j\in\{1\cdots J_n\}}
\mathrm{tr}(N_{{\theta}^n}\circ F^n(E(~|j))D_j)\text{ .}\]
\end{definition} \vspace{0.2cm}

A code is created by the sender and the legal receiver before the
message transmission starts. The sender uses the encoder to encode the
message that he wants to send, while the legal receiver uses the
decoder operators on the channel output to decode the message. \vspace{0.2cm}

In \cite{De} it has been shown that codes for   secure
message transmission over a classical-quantum wiretap channel can be
used to build  codes for   entanglement transmission over a
quantum  channel. Please see \cite{Bo/Ca/Ca/De} for a discussion
of entanglement generation over compound quantum channels,
which are channels 
when the channel states do not change with every use
as in case of AVQCs, but are stationary over the time. \vspace{0.2cm}

\begin{remark}\label{fcavccutec}
For classical arbitrarily varying channels,
capacities under
the average error criterion and under
maximal error criterion are  distinguished from
each other. The capacity formula for the
latter one is still unknown (cf. \cite{Ahl1}).
This statement is not true for the capacities of  arbitrarily varying classical-quantum channel under
maximal error criterion and under the average error criterion (cf.  
Remark \ref{icwavqcwa}, Remark \ref{iutcoavcqcum}, and  \cite{Bj/Bo/Ja/No}).  
\end{remark}
\vspace{0.2cm}

We now consider, in addition to a
faithful message transmission, a security requirement, i.e., 
a code for the channel
conveying private information to the legal receiver such that
the wiretapper's knowledge of the transmitted information
can be kept arbitrarily small.

\begin{definition}
Let $\mathfrak{I}$
$=\{N_{\theta}: {\theta}\in\Theta\}$
be an arbitrarily varying
quantum  channel.
A non-negative number $R$ is an achievable \bf (deterministic) secrecy
rate \it  for the arbitrarily varying quantum channel
$\mathfrak{I}$
\bf  under the average error criterion   \it
 if for every $\epsilon>0$, $\delta>0$,
$\zeta>0$ and sufficiently large $n$ there exists 
 a finite set $\mathbf{A}$, a map $F:$ $\mathbf{A}$ $\rightarrow \mathcal{S}(H^{\mathfrak{P}})$,
and
an  $(n, J_n)$
code $\mathcal{C} = \bigl(E, \{D_j^n : j = 1,\cdots J_n\}\bigr)$  
such that
 $\frac{\log
J_n}{n}> R-\delta$, and
\begin{equation} \max_{{\theta}^n \in \Theta^n} P_e(\mathcal{C}, {\theta}^n) < \epsilon\text{ ,}\label{annian1qq}\end{equation}
\begin{equation}\max_{{\theta}^n\in\Theta^n}
\chi\left(R_{uni};Z_{{\theta}^n}\right) < \zeta\text{
,}\label{b40qq}\end{equation}     where $R_{uni}$ is a random variable  uniformly
distributed on $\{1,\dots ,J_n\}$ and  
\[Z_{{\theta}^n}=\Bigl\{{V'}_{{\theta}^n}\circ F^n(E(~|i)):
i\in\{1,\cdots,J_n\}\Bigr\}\text{ ,}\] where 
 ${V}_{\theta}'$ is a channel to the environment
defined by  $N_{\theta}$.

The supremum on achievable  (deterministic) secrecy rates of
$\mathfrak{I}$  under the average error criterion
is called the (deterministic)  secrecy
capacity of $\mathfrak{I}$ under the average error criterion, denoted by
$C_{s}(\mathfrak{I})$.
\label{defofrateqq}
\end{definition}\vspace{0.2cm}

\begin{remark} A weaker and widely used
 security criterion is
obtained if we replace  (\ref{b40qq})  with
$\max_{{\theta}^n\in\Theta^n}\frac{1}{n}
 \chi\left(R_{uni};Z_{{\theta}^n}\right) < \zeta$. In this paper  we will follow
  \cite{Bj/Bo/So2} and use  (\ref{b40qq}).
\label{remsc1}\end{remark}\vspace{0.2cm}

\begin{remark} 
 Let ${V}_{\theta}':$ $H^{\mathfrak{P}} \rightarrow H^{\mathfrak{E}}$
be a channel to the environment
defined by the quantum channel  $N_{\theta}:$ $H^{\mathfrak{P}} \rightarrow H^{\mathfrak{Q}}$.
We may choose another Stinespring dilation for  $N_{\theta}$
and obtain another channel ${V}_{\theta}:$ $H^{\mathfrak{P}} \rightarrow H^{\mathfrak{E}}$
 to the environment.
${V}_{\theta}'$ and ${V}_{\theta}$
 are
equivalent in the sense that there is a partial isometry
$W$ such that for all $\rho \in  \mathcal{S}(H^{\mathfrak{P}})$
we have ${V}_{\theta}'(\rho)= W^{*} {V}_{\theta}(\rho) W$
(ca. \cite{Pa} and \cite{Ho2}). Thus the security criteria in 
Definition \ref{defofrateqq} does not depend on which
Stinespring dilation we choose to define
the channel to the environment.
\end{remark}\vspace{0.2cm}

\begin{remark}
It is understood that the sender and the receiver
have to select their coding scheme first.
We assume that this coding scheme is  known by
the jammer and the eavesdropper.
The jammer can make use of this knowledge 
 to advance his attacking  strategy.
Our Definition  \ref{annian1qq} for the
criterion of decoding error probability
and  Definition  \ref{b40qq} for the
security  criterion require that a
private information transmission can be
guaranteed even in the worst case, i.e.,
under the assumption that the jammer and the eavesdropper
know the complete coding scheme and will
choose attacking  strategies which are most
advantageous for themselves.
\label{qotciiuttsatr}\end{remark}
\vspace{0.2cm}

\begin{definition}
Let $\mathfrak{I}$
$=\{N_{\theta}: {\theta}\in\Theta\}$
be an arbitrarily varying
quantum  channel.
A non-negative number $R$ is an achievable \bf (deterministic) secrecy
rate \it  for the arbitrarily varying quantum channel
$\mathfrak{I}$  \bf   under the 
maximal error criterion \it   if for every $\epsilon>0$, $\delta>0$,
$\zeta>0$ and sufficiently large $n$ there exists 
 a finite set $\mathbf{A}$, a map $F:$ $\mathbf{A}$ $\rightarrow \mathcal{S}(H^{\mathfrak{P}})$,
and
an  $(n, J_n)$
code $\mathcal{C} = \bigl(E, \{D_j^n : j = 1,\cdots J_n\}\bigr)$  
such that
 $\frac{\log
J_n}{n}> R-\delta$, and
\begin{equation} \max_{{\theta}^n \in \Theta^n} P_e^{(max)}(\mathcal{C}, {\theta}^n) < \epsilon\text{ ,}\label{annian1qqmax}\end{equation}
\begin{equation}\max_{{\theta}^n\in\Theta^n}
\chi\left(R_{uni};Z_{{\theta}^n}\right) < \zeta\text{
,}\label{b40qqmax}\end{equation}    
where $R_{uni}$ and $Z_{{\theta}^n}$ are defined as in
Definition \ref{defofrateqq}.

The supremum on achievable  (deterministic) secrecy rates of
$\mathfrak{I}$  under the maximal error criterion
is called the (deterministic)   secrecy
capacity of $\mathfrak{I}$ under the maximal error criterion, denoted by
$C_{s, max}(\mathfrak{I})$.
\label{defofrateqqmax}
\end{definition}\vspace{0.2cm}

The maximal error criterion
is the strongest error criterion
in the sense that when
we consider secret message transmission
under the maximal error criterion,
we assume that the jammer not only
knows the  coding scheme of
the     legal channel users
      (cf. Remark \ref{qotciiuttsatr}), but
also the actual message.
Definition \ref{defofrateqqmax}
requires that a
private information transmission can be
guaranteed even 
under the assumption that the jammer, 
knowing both the complete coding scheme and 
the message,
will
choose the most
advantaged jamming attacking  strategy
depending on this knowledge.

With this interpretation in mind it is intuitively clear that, if sender 
and receiver have the ability to use one of many codes and the jammer is 
not informed about this choice anymore, the maximum- and average error 
criterion become equivalent again. This intuition is sharpened in
\cite{Bo/Ca/Ca}.\vspace{0.2cm}

\begin{definition}
Let $\mathfrak{I}$ $=\{N_{\theta} : {\theta}\in \Theta\}$
be an arbitrarily varying  quantum channel.
We denote  the set of $(n, J_n)$ deterministic  codes for  $\mathfrak{I}$ by $\Lambda$.

A non-negative number $R$ is an achievable  \bf  secrecy rate  \it  for the
arbitrarily varying quantum  channel
$\mathfrak{I}$ \bf under randomness  assisted  coding   \it   if  for
every
 $\delta>0$, $\zeta>0$, and  $\epsilon>0$, if $n$ is sufficiently large,
there is 
 a finite set $\mathbf{A}$, a map $F:$ $\mathbf{A}$ $\rightarrow \mathcal{S}(H^{\mathfrak{P}})$,
and
 a distribution $G$ on
$\left(\Lambda,\sigma\right)$ such that
$\frac{\log J_n}{n} > R-\delta$, and
\[ \max_{{\theta}^n\in\Theta^n}\int_{\Lambda}P_{e}(\mathcal{C}^{\gamma},{\theta}^n)d
G(\gamma) < \epsilon\text{ ,}\]
\[\max_{{\theta}^n\in\Theta^n} \int_{\Lambda}
\chi\left(R_{uni};Z_{\mathcal{C}^{\gamma},{\theta}^n}\right)dG(\gamma) < \zeta\text{
.}\] 
Here  $\sigma$ is a   sigma-algebra,   so
chosen such that the functions $\gamma \rightarrow P_e(\mathcal{C}^{\gamma},
{\theta}^n)$ and  $\gamma \rightarrow \chi
\left(R_{uni};Z_{\mathcal{C}^{\gamma},{\theta}^n}\right)$ are both $G$-measurable
with respect to $\sigma$ for every   ${\theta}^n\in\Theta^n$.  
       $R_{uni}$ and $Z_{{\theta}^n}$ are defined as in
Definition \ref{defofrateqq}.

 The supremum on achievable secrecy   rates    for 
$\mathfrak{I}$ under randomness  assisted  coding
is called 
the randomness  assisted secrecy  capacity of    $\mathfrak{I}$,  
denoted by  $C_{s}(\mathfrak{I},r)$.
\label{wdtsonjdccl} 
\end{definition}

\begin{remark}
 The randomness  assisted code technique is not to be confused with the random encoding  technique.
 For the  random encoding  technique, only the sender, but not the receiver, 
 randomly chooses a code word in ${\mathbf{A}}^n$ to encode a message $j$ according to a probability distribution.
 The receiver should be able to  decode $j$ even when he only knows the  probability distribution,
 but not which code word is actually chosen by the sender.
 For the randomness  assisted code technique,
 the sender randomly chooses a stochastic encoder $E^{\gamma}$, and the receiver
  chooses a set of the decoder operators $\{D_j^{\gamma'}:j=1,\cdots,J_n\}$.
 The receiver can decode the message if and only  if  he knows the actual random 
outcome of the sender's random variable.      The use of randomized encoding diffuses the 
knowledge of a jammer that has access to the code-words that are transmitted has 
regarding the actual message, and makes it more difficult for him to prevent reliable 
communication between sender and receiver.
\end{remark}\vspace{0.2cm}

\begin{definition}
Let $\mathfrak{I}$ $=\{N_{\theta} : {\theta}\in \Theta\}$
be an arbitrarily varying quantum channel.
We denote  the set of $(n, J_n)$ deterministic  codes for  $\mathfrak{I}$ by $\Lambda$.

A non-negative number $R$ is an achievable  \bf  secrecy rate  \it  for the
arbitrarily varying quantum channel
$\mathfrak{I}$ \bf under randomness  assisted  classical coding   and the maximal error criterion \it   if  for
every
 $\delta>0$, $\zeta>0$, and  $\epsilon>0$, if $n$ is sufficiently large,
there is 
 a finite set $\mathbf{A}$, a map $F:$ $\mathbf{A}$ $\rightarrow \mathcal{S}(H^{\mathfrak{P}})$,
and
a distribution $G$ on
$\left(\Lambda,\sigma\right)$ such that
$\frac{\log J_n}{n} > R-\delta$, and
\[ \max_{{\theta}^n\in\Theta^n}\int_{\Lambda}P_e^{(max)}(\mathcal{C}, {\theta}^n)d
G(\gamma) < \epsilon\text{ ,}\]
\[\max_{{\theta}^n\in\Theta^n} \int_{\Lambda}
\chi\left(R_{uni};Z_{\mathcal{C}^{\gamma},{\theta}^n}\right)dG(\gamma) < \zeta\text{
.}\] 
Here  $\sigma$ is a   sigma-algebra,   so
chosen such that the functions $\gamma \rightarrow P_e(\mathcal{C}^{\gamma},
{\theta}^n)$ and  $\gamma \rightarrow \chi
\left(R_{uni};Z_{\mathcal{C}^{\gamma},{\theta}^n}\right)$ are both $G$-measurable
with respect to $\sigma$ for every   ${\theta}^n\in\Theta^n$.  

 The supremum on achievable secrecy   rates    for 
$\mathfrak{I}$ under randomness  assisted  classical coding  and the maximal error criterion
is called 
the randomness  assisted secrecy classical capacity  under the maximal error criterion of    $\mathfrak{I}$,  
denoted by  $C_{s, max}(\mathfrak{I},r)$.
\label{wdtsonjdcclmax} 
\end{definition}

\vspace{0.2cm}

\begin{definition}

The code concept for  arbitrarily varying
wiretap  quantum channels
is similar to the code concept for  arbitrarily varying classical-quantum wiretap channels in
\cite{Bo/Ca/De} and \cite{Bo/Ca/De3}.
We build a two-part code word, the first part is used to
create the  randomness for the sender and the legal receiver, 
the second is used to transmit the message to the legal
receiver. We call it a  \bf  weak code concept  \rm  when
the first part to synchronize the second part  is public,
 and  a \bf  strong code concept \rm   when the first part is secure.

\label{stccwecc}
\end{definition} \vspace{0.2cm}

\begin{definition}
Let 
$\{N_{\theta}: {\theta}\in\Theta\}$ and
$\{\hat{N}_{\theta'}: {\theta'}\in\Theta'\}$ 
be two arbitrarily varying
quantum  channels.


We say  super-activation occurs to the  secrecy
capacity for 
two  arbitrarily varying classical-quantum wiretap  channels 
$\{N_{\theta}: {\theta}\in\Theta\}$ and    
 $\{\hat{N}_{\theta'}: {\theta'}\in\Theta'\}$ when
the following hold:
\[C_{s}(\{N_{\theta}: {\theta}\in \Theta\})=0\text{ ,}\]
\[C_{s}(\{\hat{N}_{\theta'}): {\theta'}\in \Theta'\})=0\text{ ,}\]
and
\[C_{s}(\{N_{\theta}\otimes \hat{N}_{\theta'}: (\theta, \theta')  \in \Theta\times \Theta'\})>0\text{ .}\]
 
\end{definition} \vspace{0.2cm}

\begin{definition}

Let $\mathfrak{I}$ $=\{N_{\theta}: {\theta}\in\Theta\}$ be  an arbitrarily varying
quantum  channel.
We define
\begin{equation} 
\mathfrak{C}(\mathfrak{I}) =
  \limsup_{n\rightarrow \infty} \frac{1}{n} \sup_{\mathbf{A}, F, P}\left(
\inf_{q \in P({\Theta})}  \chi(P,B_q^n) -\max_{{\theta}^{n}\in \Theta^{n}}\chi (P,Z_{{\theta}^{n}})\right)\text{ .}
\end{equation}

The supremum is taken over all 
chosen  finite sets $\mathbf{A}$, 
maps $F:$ $\mathbf{A}$ $\rightarrow \mathcal{S}(H^{\mathfrak{P}})$, and
probability distributions
$P$ on the input quantum  states $\{F(x):x\in \mathbf{A}\}$.
Here $B_{\theta}$ and  $Z_{\theta}$ are the resulting  quantum states at the output of $N_{\theta}$
and ${V}_{\theta}'$, respectively,  where 
 ${V}_{\theta}'$ is the channel to the environment
defined by  $N_{\theta}$ and $B_q = \sum_{\theta} q(\theta) B_{\theta}$.\vspace{0.2cm}

	Let $\mathfrak{L}$ $=(W_{\theta},{V}_{\theta})_{ {\theta} \in \Theta}$ 
	be an arbitrarily varying classical-quantum wiretap channel.
We define
\begin{align}&\mathfrak{C}(\mathfrak{L}): =
\lim_{n\rightarrow \infty} \frac{1}{n} \max_{U\rightarrow A \rightarrow \{B_q,Z_{\theta}:q,\theta\}}
	\Bigl(\inf_{B_q \in Conv((B_{\theta})_{{\theta}\in \Theta})}\chi(p_U;B_q^{\otimes n})-\max_{{\theta}^n\in \Theta^n}\chi(p_U;Z_{{\theta}^n})\Bigr)
	\text{ .}\end{align}

	Here $B_{\theta}$ are the resulting  quantum states at the output of the
legitimate receiver's channels. $Z_{{\theta}^n}$ are the resulting  quantum states  at
the output of wiretap channels.
 The maximum is taken over all random
variables  that satisfy the Markov chain relationships:
$U\rightarrow A \rightarrow B_q Z_{\theta}$ for every $B_q \in Conv((B_{\theta})_{{\theta}\in \Theta})$ and
$ {\theta}\in \Theta$.   Here $A$   is  a random
variable taking values on $\mathbf{A}$,   and  $U$ a random
variable taking values on some finite set $\mathbf{U}$
with probability  distribution $p_U$.
	
 \end{definition} \vspace{0.2cm}

\begin{definition}

For  finite sets $\mathbf{A}$  and $\mathbf{B}$,
we define a (discrete) \bf classical channel \it   $\mathsf{V}$:  $P(\mathbf{A}) \rightarrow P(\mathbf{B})$,
$ P(\mathbf{A}) \ni p \rightarrow \mathsf{V}(p) \in P(\mathbf{B})$
 to be a system characterized by a probability transition matrix $\mathsf{V}(\cdot|\cdot)$.
For $p \in P(\mathbf{A})$ and $y \in \mathbf{B}$,
$\mathsf{V}(y|p)$  expresses the probability of  the output symbol $y$
when we send the symbol $x$ through the channel. The channel is said to be memoryless
if the probability distribution of the output depends only on the input at
that time and is conditionally independent of previous channel inputs and
outputs. Further, we can extend this definition when we define a 
 classical channel to a map  $\mathsf{V}$:  $P(\mathbf{A}) \rightarrow P(\mathbf{B})$
by denoting $\mathsf{V}(y|p)$ $:=$ $\sum_{x\in\mathbf{A}} p(x)\mathsf{V}(y|x)$.
\end{definition} \vspace{0.2cm}

\begin{definition}
Let $\mathbf{A}$, $\mathbf{B}$, and $\mathbf{C}$ be finite sets.
Let $\Theta$ := $\{1,\dots,T\}$ be a finite set. For every $\theta \in \Theta$, let
$\mathsf{W}_{\theta}$ be a classical  channel $P(\mathbf{A}) \rightarrow P(\mathbf{B})$ and
 $\mathsf{V}_{\theta}$ be a  classical  channel $P(\mathbf{A}) \rightarrow P(\mathbf{C})$.

We call the set of the classical   channel pairs  $\{(\mathsf{W}_\theta,\mathsf{V}_\theta):\theta \in \Theta\}$
 a \bf
(classical) arbitrarily varying    wiretap channel  \it    when the channel state $\theta$ varies from
symbol to symbol in an  arbitrary manner.
When
the sender inputs $p  \in P(\mathbf{A})$ into the channel,  the receiver receives the output $y \in B$ with probability
$\sum_x p(x)\mathsf{W}_\theta(y|x)$, while the wiretapper  receives the output $z \in Z$ with probability
$\sum_x p(x)\mathsf{V}_\theta(z|x)$.
\end{definition} \vspace{0.2cm}

\begin{definition} Let $\mathfrak{K}$ $=\{(\mathsf{W}_\theta,\mathsf{V}_\theta):\theta \in \Theta\}$ be
a classical arbitrarily varying    wiretap channel.
An $(n, J_n)$ code   for $\mathfrak{K}$ consists of a stochastic encoder
$E$ : $\{ 1,\dots ,J_n\} \rightarrow P(\mathbf{A}^n)$,  specified by
a matrix of conditional probabilities $E(\cdot|\cdot)$,  and
 a collection of mutually disjoint sets $\left\{D_j
\subset \mathbf{B}^n: j\in \{ 1,\dots ,J_n\}\right\}$ (decoding sets).

The  average probability of the decoding error of a
deterministic code $\mathcal{C}$ is defined as 
\[P_e(\mathcal{C}, {\theta}^n) :=  \frac{1}{J_n} \sum_{j=1}^{J_n}
E(x^n|j)\mathsf{W}_{\theta}^{n}(D_j^c|x^n)\text{ .}\] 

\end{definition} \vspace{0.2cm}




\begin{definition}\label{symmetc}
We say that the arbitrarily varying channel
$\{\mathsf{W}_{\theta} : {\theta} \in \Theta\}$ is symmetrizable if
 there exists a
parametrized set of distributions $\{\tau(\cdot\mid a):
 a\in \mathbf{A}\}$ on $\Theta$ such that for all $a$, ${a'}\in \mathbf{A}$, and $b \in \mathbf{B}$
\[\sum_{{\theta}\in\Theta}\tau({\theta}\mid a)\mathsf{W}_{{\theta}}(b\mid {a'})=\sum_{{\theta}\in\Theta}\tau({\theta}\mid {a'})\mathsf{W}_{{\theta}}(b\mid a)\text{ .}\]

We say that the arbitrarily varying classical-quantum  channel
$\{W_{\theta} : {\theta} \in \Theta\}$ is symmetrizable if
 there exists a
parametrized set of distributions $\{\tau(\cdot\mid a):
 a\in \mathbf{A}\}$ on $\Theta$ such that for all $a$, ${a'}\in \mathbf{A}$,
\[\sum_{{\theta}\in\Theta}\tau({\theta}\mid a)W_{\theta}({a'})=\sum_{{\theta}\in\Theta}\tau({\theta}\mid {a'})W_{\theta}(a)\text{ .}\]
\end{definition}

\begin{definition}
Let $\mathbf{A}$ and  $\mathbf{B}$ be finite sets, and  $H$ be a finite-dimensional
complex Hilbert   space.   Let
  $\Theta$ $:=$ $\{1,2, \cdots\}$ be an index set. For every   $\theta \in \Theta$,   let $\mathsf{W}_{\theta}$   be a classical channel
$P(\mathbf{A}) \rightarrow P(\mathbf{B})$ and ${V}_{\theta}$ be a classical-quantum channel $P(\mathbf{A})
\rightarrow \mathcal{S}(H)$. We call the set of the  classical channel/classical-quantum 
channel pairs  $\{(\mathsf{W}_{\theta},{V}_{\theta}): \theta \in \Theta\}$   a  \bf  classical   arbitrarily varying quantum wiretap   channel  \it    when the
state $t$ varies from symbol to symbol in an  arbitrary
manner, while  the legitimate
receiver accesses the output of  $W_{\theta}$, and the wiretapper observes the output of 
  ${V}_{\theta}$, respectively.\end{definition} \vspace{0.2cm}

\begin{definition} Let $\mathfrak{J}$ $=\{(\mathsf{W}_{\theta},{V}_{\theta}): \theta \in \Theta\}$ 
be a  classical arbitrarily varying quantum wiretap channel.
 An $(n, J_n)$   \bf    code   \it   $\mathcal{C}$ for the
 classical arbitrarily varying quantum wiretap channel $\mathfrak{J}$
consists of a stochastic encoder $E$ : $\{
1,\cdots ,J_n\} \rightarrow P({\mathbf{A}}^n)$, 
$j\rightarrow E(\cdot|j)$,
 specified by
a matrix of conditional probabilities $E(\cdot|\cdot)$, and
  a collection of mutually disjoint sets $\left\{D_j
\subset \mathbf{B}^n: j\in \{ 1,\dots ,J_n\}\right\}$ (decoding sets).
\end{definition} \vspace{0.2cm}

\begin{definition} 
A non-negative number $R$ is an achievable   \bf    secrecy
rate    \it   for the  classical arbitrarily varying quantum wiretap channel $\mathfrak{J}$
$=\{(\mathsf{W}_{\theta},{V}_{\theta}): \theta \in \Theta\}$  \bf  under the 
maximal error criterion  \it  if for every $\epsilon>0$, $\delta>0$,
$\zeta>0$ and sufficiently large $n$ there exist an  $(n, J_n)$
code $\mathcal{C} = \bigl(E, \{D_j : j = 1,\cdots J_n\}\bigr)$  such that $\frac{\log
J_n}{n}
> R-\delta$, and
\begin{equation} \label{b3tm} \max_{j\in\{1,\cdots,J_n\}}\max_{\theta \in
\Theta} 
\mathsf{W}_{\theta}^{n}(D_j^c|E(\cdot |j))\leq  \varepsilon\text{ ,}\end{equation} and
\begin{equation}\max_{{\theta}^n\in\Theta^n}
\chi\left(R_{uni};Z_{{\theta}^n}\right) < \zeta\text{ .}\label{b40tm}\end{equation} 

 The supremum on achievable secrecy   rates    for 
$\mathfrak{J}$ is called 
the secrecy capacity of   $\mathfrak{J}$
under the maximal error criterion,  
denoted by  $C_{s,max}(\mathfrak{J})$.
\label{defofrate2tm}
\end{definition}\vspace{0.2cm}

\begin{definition} Let $\mathfrak{P}$ and $\mathfrak{Q}$ be
quantum systems. We denote the Hilbert space of $\mathfrak{P}$ and
$\mathfrak{Q}$ by $H^\mathfrak{P}$ and $H^\mathfrak{Q}$,
respectively. Let $\Theta$ := $\{1,\dots,T\}$ be a finite set. For every ${\theta} \in \Theta$,		
	let ${N}_{\theta}$    be a quantum channel
$\mathcal{S}(H^\mathfrak{P}) \rightarrow \mathcal{S}(H^\mathfrak{Q})$.

We call the set of the  quantum   channel   $\{{N}_{\theta}: {\theta} \in \Theta\}$
a
\bf quantum compound
 channel\it. When
the channel state is $t$   and the sender inputs a  quantum state $\rho^{\mathfrak{P}} \in \mathcal{S}({H^\mathfrak{P}})$ into the channel,
 the receiver receives an output  quantum state ${N}_{\theta}(\rho^{\mathfrak{P}}) \in \mathcal{S}({H^\mathfrak{Q}})$.
\end{definition}\vspace{0.2cm}

\begin{definition}
 An $(n, J_n)$   code $\mathcal{C}$ for the
classical-quantum compound
wiretap channel $\{N_{\theta}: \theta \in \theta\}$
consists of a stochastic encoder $E$ : $\{
1,\cdots ,J_n\} \rightarrow P({\mathbf{A}}^n)$, 
$j\rightarrow E(\cdot|j)$ and
 a collection of positive-semidefinite operators $\Bigl\{D_j$ $: j$ $\in \{ 1,$ $\cdots ,J_n\}\Bigr\}$
on ${H}^{\otimes n}$,
which is a partition of the identity.
\end{definition}\vspace{0.2cm}

We  deal  with two communication scenarios. In the first
one, only the sender is informed about the index $t$, or in other words,  has
CSI, where CSI is an abbreviation for
 ``channel state information''. In the second scenario,
neither  sender nor  receiver has any information about that index at all.\vspace{0.2cm}

\begin{definition}

A non-negative number $R$ is an achievable  \bf secrecy
rate \it  with  CSI  at the encoder for the compound
wiretap quantum channel
$\{N_{\theta}: \theta \in \Theta\}$ if for every $\epsilon>0$, $\delta>0$,
$\zeta>0$ and sufficiently large $n$ there exist
 a finite set $\mathbf{A}$, a map $F:$ $\mathbf{A}$ $\rightarrow \mathcal{S}(H^{\mathfrak{P}})$,
and
 an  $(n, J_n)$
code $\mathcal{C}_{\theta} = \bigl(E_{\theta}, \{D_{j}^{({\theta})} : j = 1,\cdots J_n\}\bigr)$ 
for each $\theta \in \Theta$  such that $\frac{\log
J_n}{n}
> R-\delta$, and
\begin{equation} \max_{\theta \in \Theta}  1- \frac{1}{J_n} \sum_{j=1}^{J_n}
\mathrm{tr}(N_{\theta}^{\otimes n}\circ F^n(E_{\theta}(~|j))D_j^{({\theta})}) < \epsilon\text{ ,}\label{annian1comp1}\end{equation}
\begin{equation}\max_{\theta\in\Theta}
\chi\left(R_{uni};Z_{\theta}^{\otimes n}\right) < \zeta\text{
.}\label{b40comp1}\end{equation} 
  Here
\[Z_{{\theta}^n}=\Bigl\{{V'}_{{\theta}^n}\circ F^n(E(~|i)):
i\in\{1,\cdots,J_n\}\Bigr\}\text{ ,}\] where 
 ${V}_{\theta}'$ is the channel to the environment
defined by  $N_{\theta}$.

A non-negative number $R$ is an achievable  \bf secrecy
rate \it  with no CSI  at the encoder for the compound
wiretap quantum channel
$\{N_{\theta}: \theta \in \Theta\}$ if for every $\epsilon>0$, $\delta>0$,
$\zeta>0$ and sufficiently large $n$ there exist an  $(n, J_n)$
code $\mathcal{C} = \bigl(E, \{D_j : j = 1,\cdots J_n\}\bigr)$  such that $\frac{\log
J_n}{n}
> R-\delta$, and
\begin{equation} \max_{\theta \in \Theta}  1- \frac{1}{J_n} \sum_{j=1}^{J_n}
\mathrm{tr}(N_{\theta}^{\otimes n}\circ F^n(E(~|j))D_j) < \epsilon\text{ ,}\label{annian1comp2}\end{equation}
\begin{equation}\max_{\theta\in\Theta}
\chi\left(R_{uni};Z_{\theta}^{\otimes n}\right) < \zeta\text{
.}\label{b40comp2}\end{equation}

  The  supremum of all    secrecy rates with  CSI  at the encoder is called the   secrecy
capacity of $\{N_{\theta}: \theta \in \Theta\}$ with CSI, denoted by
$C_{s,csi} (\{N_{\theta}: \theta \in \Theta\})$.
  The  supremum of all    secrecy rates with no CSI  at the encoder is called the   secrecy
capacity of $\{N_{\theta}: \theta \in \Theta\}$ (with no CSI), denoted by
$C_{s} (\{N_{\theta}: \theta \in \Theta\})$.

\label{defofratecomp}
\end{definition}

\section{Main Results}
\label{mare}

Our main results 
are included in the following theorems and corollaries:\vspace{0.2cm}

\begin{theorem}\label{lntvttitbaa}
Let $\mathfrak{I}$ $=\{N_{\theta}: {\theta}\in\Theta\}$ be  an arbitrarily varying
quantum  channel.

\begin{enumerate}[1)]
\item If the
arbitrarily varying  quantum channel $\mathfrak{I}$
is not $L$-symmetrizable for some $L\in\mathbb{N}$, then 
 the deterministic secrecy capacity under the strong code 
concept      and  under the average error criterion     
  of $\{N_{\theta}:  {\theta}\in \Theta\}$ is given by
\begin{align} &
C_{s}(\mathfrak{I})  =  \mathfrak{C}(\mathfrak{L})\text{ .}
\end{align}

\item If $\mathfrak{I}$ is $L$-symmetrizable for all $L\in\mathbb{N}$,
then the deterministic secrecy capacity under strong code 
concept      and under the average error criterion       of 
$\{N_{\theta}:  {\theta}\in \Theta\}$ is equal to zero.\end{enumerate}
\end{theorem}\vspace{0.2cm}

\begin{corollary}\label{lntvttitbaalem}
Let $\mathfrak{I}$ $=\{N_{\theta}: {\theta}\in\Theta\}$ be  an arbitrarily varying
quantum  channel.
We  have
\begin{equation} 
C_{s}(\mathfrak{I},r) =
  \mathfrak{C}(\mathfrak{I})\text{ .}
\end{equation}
\end{corollary}\vspace{0.2cm}

\begin{lemma}\label{lntvttitbaalemmax}
Let $\mathfrak{I}$ $=\{N_{\theta}: {\theta}\in\Theta\}$ be  an arbitrarily varying
quantum  channel.
 We have $C_{s, max}(\mathfrak{I},r)$ $= C_{s}(\mathfrak{I},r)$.
In other words
\begin{equation} 
C_{s, max}(\mathfrak{I},r) =
\mathfrak{C}(\mathfrak{I})\text{ .}
\end{equation} 
\end{lemma}\vspace{0.2cm}

We distinguish here  message transmission under maximal error criterion and under
the average error criterion. 
The capacities under these two error criteria are
not equal for  classical arbitrarily varying channels,
but equal for arbitrarily varying classical-quantum channels.
The capacity formula of classical arbitrarily varying channels under
maximal error criterion is still an open problem
(cf. Remark \ref{iutcoavcqcum}).\vspace{0.2cm}

\begin{theorem}\label{lntvttitbaamax}
Let $\mathfrak{I}$ $=\{N_{\theta}: {\theta}\in\Theta\}$ be  an arbitrarily varying
quantum  channel.

\begin{enumerate}[1)]
\item If the
arbitrarily varying  quantum channel $\mathfrak{I}$
is not $L$-symmetrizable for some $L\in\mathbb{N}$, then 
 the deterministic secrecy capacity under strong code 
concept       and under the 
maximal error criterion      of $\{N_{\theta}:  {\theta}\in \Theta\}$ is given by
\begin{equation} 
C_{s,max}(\mathfrak{I}) =
  \mathfrak{C}(\mathfrak{I})\text{ .}
\end{equation}

\item If $\mathfrak{I}$ is $L$-symmetrizable for all $L\in\mathbb{N}$,
then the deterministic secrecy capacity under strong code 
concept and  the maximal error criterion of $\{N_{\theta}:  {\theta}\in \Theta\}$  is equal to zero.\end{enumerate}
\end{theorem}\vspace{0.2cm}

\begin{remark}
Theorem  \ref{lntvttitbaa} states that when an
arbitrarily varying  quantum channel 
is not $L$-symmetrizable for some $L\in\mathbb{N}$,
then secret message transmission is    possible
    even when the  coding scheme of
the legal transmitters is known by
the jammer and the eavesdropper.

Theorem  \ref{lntvttitbaamax}
states that when an
arbitrarily varying  quantum channel 
is not $L$-symmetrizable for some $L\in\mathbb{N}$,
then secret message transmission is    possible
     even when the  coding scheme of
the legal transmitters is known by
the jammer and the eavesdropper and, 
additionally, the jammer 
knows the  actual message
the legal channel users      want
to communicate 
     (cf. \ref{qotciiuttsatr}).
\end{remark}\vspace{0.2cm}

\begin{corollary}
	Let   $\Theta$ be a finite set and  $\mathfrak{J}$
	$=\{(\grave{W}_{\theta}, V_{\theta}): {\theta}\in \Theta\}$ be a
classical arbitrarily varying quantum wiretap 
channel.
If $\mathfrak{J}$
is not symmetrizable, 
then  
\begin{align}&C_{s,max}(\mathfrak{J})= \lim_{n\rightarrow \infty} \frac{1}{n} \max_{U\rightarrow A^n \rightarrow \{B_q^{\otimes n},Z_{{\theta}^n}:q,{\theta}^n\}}\left(
\min_{q \in P({\Theta})}  I(p_U,\grave{B}_q^n) -\max_{{\theta}^{n}\in \Theta^{n}}\chi (p_U;Z_{{\theta}^{n}})\right)\text{ .}\end{align}  
Here $\grave{B}_{\theta}$ are the resulting classical random variables at the output of the
legitimate receiver's channels and $Z_{{\theta}^n}$ are the resulting quantum states at the output of wiretap channels. The maximum is taken over all random
variables  that satisfy the Markov chain relationships:
$U\rightarrow A^n \rightarrow \{\grave{B}_q^{\otimes n},Z_{{\theta}^n}:q,{\theta}^n\}$ for every $\grave{B}_q \in Conv((\grave{B}_{\theta})_{{\theta}\in \Theta})$ and
$ {\theta}\in \Theta$.  $A$ is here a random
variable taking values on $\mathbf{A}$, and $U$ a random
variable taking values on some finite set $\mathbf{U}$
with probability  distribution $p_U$.
\label{lgwtvttitbacam}\end{corollary}\vspace{0.2cm}

In \cite{Bo/Sch/Po} it has been shown that
the secrecy capacity of
a classical arbitrarily
varying  channel
under  randomness assisted coding
is  continuous if
the eavesdropper's output can
be described by a finite alphabet set.

In Section \ref{iosccits} we will show that
the secrecy capacity of
an arbitrarily
varying quantum channel $\mathfrak{I}$
under  randomness assisted quantum coding
is  continuous if the receiver's system can be
described by a finite dimensional Hilbert space
in the following sense:\vspace{0.2cm}

\begin{corollary}

Let  $\{N_{\theta}: {\theta}\in\Theta\}$ be an arbitrarily varying
quantum  channel
$\mathcal{S}(H^\mathfrak{P}) \rightarrow \mathcal{S}(H^\mathfrak{Q})$.     Let
 $\dim H^\mathfrak{Q}$ be finite.      For 
a positive $\delta$, let
$\mathbf{C}_{\delta}$ be the set of all
arbitrarily
varying quantum channels
$\{\dot{N}_{\theta} : {\theta}\in \Theta\}$
such that
\[\max_{\rho \in \mathcal{S}(H^\mathfrak{P}) } \|N_{\theta}(\rho)- \dot{N}_{\theta}(\rho )\|_{1} <  \delta\]
for all ${\theta} \in \Theta$.      $\|\cdot\|_1$ denotes here the trace norm.

For any positive $\epsilon$ there is a positive $\delta$
such that for all  $\{\dot{N}_{\theta} : {\theta}\in \Theta\}$
$\in$ $\mathbf{C}_{\delta}$ we have
	\begin{equation} |C_{s}(\{N_{\theta}: {\theta}\in\Theta\},r)
	-C_{s}(\{(\dot{N}_{\theta}): {\theta} \in \Theta\},r)| \leq \epsilon
	\text{ .}\end{equation}
	\label{eetelctit}
\end{corollary}\vspace{0.2cm}

\begin{corollary}  Let $L\in\mathbb{N}$.
For    a finite set $\Theta$ and   an arbitrarily varying quantum channel
$\{N_{\theta}: {\theta}\in \Theta\}$, we define
	\begin{align}& \mathsf{F}_L(\{N_{\theta}: {\theta}\in \Theta\})\notag\\
	&:=\min_{\tau\in C\left(\Theta^L\mid \mathcal{S}(H^{\mathfrak{A}^L})\right) }\max_{\rho^L, {\rho^L}'}
\left\|\sum_{{\theta^L}\in\Theta^L}\tau(\theta^L\mid \rho^L)N_{\theta^L}({{\rho^L}'})
-\sum_{{\theta^L}\in\Theta^L}\tau(\theta^L\mid {{\rho^L}'})N_{\theta^L}(\rho^L)\right\|_1\text{ ,}\end{align}
where $C\left(\Theta^L\mid \mathcal{S}(H^{\mathfrak{A}^L})\right)$   is   the set of
parametrized distributions sets $\{\tau(\cdot\mid \rho^L):
 \rho^L\in \mathbf{A}^L\}$ on $\Theta^L$
    and
	\begin{equation} \mathsf{F}(\{N_{\theta}: {\theta}\in \Theta\}):=\sum_{L\in\mathbb{N}}\frac{1}{2^L}\mathsf{F}_L(\{N_{\theta}: {\theta}\in \Theta\})\text{ .}\end{equation}
The statement $\mathsf{F}_L(\{N_{\theta}: {\theta}\in \Theta\})=0$ is equivalent to $\{N_{\theta}: {\theta}\in \Theta\}$
being $L$-symmetrizable. The statement $\mathsf{F}(\{N_{\theta}: {\theta}\in \Theta\})=0$ is equivalent to $\{N_{\theta}: {\theta}\in \Theta\}$
being $L$-symmetrizable for every $L\in\mathbb{N}$.

 $C_{s}(\{N_{\theta}: \theta \})$,
the deterministic secrecy capacity of
arbitrarily
varying quantum channel
is discontinuous at
$\{N_{\theta}: {\theta}\in \Theta\}$, 
if
and only if the following hold:\\[0.15cm]
1)
The secrecy capacity of
$\{N_{\theta}: {\theta}\in \Theta\}$  under
 randomness assisted quantum coding
is positive;\\
2) $\mathsf{F}(\{N_{\theta}: \theta\})=0$, 
but for every  positive $\delta$ there is a
 $\{{N'}_{\theta}: {\theta}\in \Theta\}$
$\in$ $\mathbf{C}_{\delta}$ such that
$\mathsf{F}(\{{N'}_{\theta}: {\theta}\in \Theta\})>0$.\label{ftsdmiait}
\end{corollary}\vspace{0.2cm}

 \begin{remark}
The function  $\mathsf{F}$ is defined as a  uniformly convergent series of continuous functions on the set of arbitrarily
varying quantum channels.
Thus $\mathsf{F}$ itself is also a continuous function on the set of arbitrarily
varying quantum channels.
 \end{remark}\vspace{0.2cm}

Corollary \ref{ftsdmiait} completely characterizes
the continuity behavior of AVQC's secrecy capacity.
Particularly in Section \ref{appli} we  show 
that discontinuity occurs for AVQC's secrecy capacity.
Furthermore we will discuss its relation to 
the quantum capacity in Section \ref{pwatscc}.

The following Corollary \ref{ltbafsanttit} is a consequence of
Corollary \ref{ftsdmiait}. Corollary \ref{ltbafsanttit} shows
that the positive  values of 
the secrecy capacity of an  arbitrarily varying quantum  wiretap channel
are stable in the sense that when 
the secrecy capacity of an  arbitrarily varying quantum  wiretap channel
takes a positive value in an  interval of the domain,
then all  arbitrarily varying quantum  wiretap channels
in a sufficiently small neighborhood also have    positive
 secrecy capacities. \vspace{0.2cm}

\begin{corollary}
Let   $\Theta$ be a finite set and   $\{N_{\theta}: {\theta}\in \Theta\}$ be
an  arbitrarily varying quantum  wiretap channel.
When the secrecy capacity of $\{N_{\theta}: {\theta}\in \Theta\}$
is   positive,   then there is a  $\delta$
such that for all $\{{N'}_{\theta}: {\theta}\in \Theta\}$
$\in\mathbf{C}_{\delta}$ we have
\[C_{s}\left(\{{N'}_{\theta}: {\theta}\in \Theta\}\right)>0\text{ .}\]\label{ltbafsanttit}
\end{corollary}\vspace{0.2cm}

 \begin{corollary}
The deterministic secrecy capacity of
an arbitrarily
varying quantum channel
is in general not continuous.
\label{tscaavqq}\end{corollary}

\begin{corollary}
Super-activation occurs for secrecy capacities of arbitrarily varying quantum  
channels.\label{superactivationqqq}\end{corollary}
\vspace{0.2cm}

In \cite{Bo/Sch/Po} it has been shown that the 
secrecy capacity of a classical arbitrarily
varying wiretap channel under  randomness assisted coding is
continuous in the sense of the following  Lemma:

\begin{lemma}
For a classical arbitrarily
varying wiretap channel 
$\{(\mathsf{W}_{\theta},\mathsf{V}_{\theta}): \theta \in \Theta\}$, where
$\mathsf{W}_{\theta}$  $:$
$P(\mathbf{A}) \rightarrow P(\mathbf{B})$ and $\mathsf{V}_{\theta}$ $:$ $P(\mathbf{A})
\rightarrow P(\mathbf{C})$,
 and
a positive $\delta$, let
$\mathbf{C}_{\delta}$ be the set of all
classical arbitrarily
varying  wiretap channels
 $\{({\mathsf{W}'}_{\theta},{V}_{\theta}'): \theta \in \Theta\}$,
where
${\mathsf{W}'}_{\theta}$  $:$
$P(\mathbf{A}) \rightarrow P(\mathbf{B})$ and ${\mathsf{V}'}_{\theta}$ $:$ $P(\mathbf{A})
\rightarrow  P(\mathbf{C})$,
 such
that
\[\max_{ a\in \mathbf{A}} \|\mathsf{W}_{\theta}(a)- {\mathsf{W}'}_{\theta}(a)\|_1 <  \delta\]
and
\[\max_{ a\in \mathbf{A}} \|\mathsf{V}_{\theta}(a)- {\mathsf{V}'}_{\theta}(a)\|_1 <  \delta\]
for all $\theta \in \theta$, where $\| \cdot\|_1$ is 
 the $L^1$-Norm.

When $|\mathbf{A}|$ is finite, then
for any positive $\epsilon$ there is a positive $\delta$
such that for all  $\{({\mathsf{W}'}_{\theta},{\mathsf{V}'}_{\theta}): \theta \in \Theta\}$
$\in$ $\mathbf{C}_{\delta}$ we have
	\begin{equation} |C_s(\{(\mathsf{W}_{\theta},\mathsf{V}_{\theta}): \theta \in \Theta\};r)
	-C_s(\{(({\mathsf{W}'}_{\theta},{\mathsf{V}'}_{\theta}): \theta \in \Theta\};r)| \leq \epsilon
	\text{ .}\end{equation}
\label{facavwcmwtmvt}
\end{lemma}\vspace{0.2cm}

However, the proof for Lemma \ref{facavwcmwtmvt} in  \cite{Bo/Sch/Po}
requires  the outputs of the
 eavesdropper's channel to be include in a finite alphabet.
Since in general, the legal channel users do not have control over
the   the eavesdropper's channel, this proof 
is still  nonetheless	capable of     improvements.
We will give an alternative proof for
 Lemma \ref{facavwcmwtmvt}
such that the continuity does not depend on
the alphabet of the eavesdropper.\vspace{0.2cm}

In our earlier work
\cite{Bo/Ca/Ca/De} we determined the secrecy capacity of the
classical-quantum compound wiretap channel.
 Now we are going to
consider the secrecy capacity of  the compound wiretap quantum channel.\vspace{0.2cm}

\begin{corollary}
The secrecy capacity of the compound  quantum channel
 $\{N_{\theta}: {\theta}\in \Theta\}$ in the case with CSI
is given by
\begin{equation}  C_{S,CSI} (\{N_{\theta}: {\theta}\in \Theta\})=  \min_{{\theta}\in \Theta} 
\lim_{n\rightarrow \infty} \frac{1}{n}\max_{U\rightarrow A \rightarrow \{(BZ)_{\theta}:\theta\}}
\chi(p_U;B_{\theta}^n)-\chi(p_U;Z_{\theta}^n)
	\label{bibqicbttcomp} \text{ .}\end{equation}
	Here $B_{\theta}$ are the resulting
random variables at the output of legal receiver channels, and
$Z_{\theta}$ are the resulting random  quantum states at the output of wiretap
channels.

	The secrecy capacity of the compound  quantum channel
 $\{N_{\theta}: {\theta}\in \Theta\}$ in the case with CSI
is given by
\begin{equation}  C_{S} (\{N_{\theta}: {\theta}\in \Theta\})=
\lim_{n\rightarrow \infty} \frac{1}{n}\max_{U\rightarrow A \rightarrow \{(BZ)_{\theta}:\theta\}}
	\left(\min_{{\theta}\in \Theta} 	\chi(p_U;B_{\theta}^n)-\max_{{\theta}\in \Theta} \chi(p_U;Z_{\theta}^n)\right)
	\label{bibqicbttcomps} \text{ .}\end{equation}
		\label{bibqicbttcompg}
\end{corollary}\vspace{0.2cm}

\section{Complete Characterization of Secrecy Capacity of
Quantum Channels}\label{ccoscoqc}
\subsection{Previous Works, Strong Code Concept, and      Connection to Open Problems}\label{pwatscc}  

In our previous works \cite{Bo/Ca/De}, \cite{Bo/Ca/De2}, and \cite{Bo/Ca/De3}
we considered that
 the sender's
encoding is restricted to transmitting an  indexed finite set of
 quantum states $\{\rho_{x}: x\in \mathbf{A}\}\subset
\mathcal{S}(G)$ as a component of the channel. We
obtained a channel $\sigma_x := N(\rho_{x})$ with classical inputs $x\in \mathbf{A}$ and quantum outputs,
 which we call a classical-quantum
channel. We delivered the complete characterization of
arbitrarily varying classical-quantum
wiretap  channels' secure capacity in our previous works. For our result
in this work we use the results of those previous works in
multiple calculations.

In \cite{Bo/Ca/De}, the Ahlswede Dichotomy for arbitrarily varying classical-quantum
wiretap  channels has been established, 
i.e. either the deterministic 
capacity of an arbitrarily varying channel was zero or equal to its shared randomness assisted capacity.
Our proof was similar to the proof of the Ahlswede Dichotomy for  arbitrarily varying classical-quantum channels in
\cite{Ahl/Bli}: We   built   a two-part code word, the first part was used to
create  randomness for the sender and the legal receiver, the second part was used to transmit the message to the legal
receiver.      Here we use the weak code concept (cf. Definition \ref{stccwecc}).  \vspace{0.2cm}

\begin{lemma}[Ahlswede Dichotomy under weak code concept] Let $\mathfrak{K}$ $=\{(W_{\theta},{V}_{\theta}): {\theta} \in \Theta\}$
 be an arbitrarily
varying classical-quantum wiretap channel.
\begin{enumerate}
\item If the
arbitrarily varying  classical-quantum channel $\{W_{\theta} : {\theta} \in \Theta\}$
is not symmetrizable, then \begin{equation}
C_{s*}(\mathfrak{K})=C_{s}(\mathfrak{K};r)\text{ .}
\end{equation}\label{dichpart1a}
\item If $\{W_{\theta} : {\theta} \in \Theta\}$ is symmetrizable,
 \begin{equation}
C_{s*}(\mathfrak{K})=0\text{ .}
\end{equation} \label{dichpart1b}\end{enumerate}
\label{dichpart1}
\end{lemma}\vspace{0.2cm}

Here $C_{s*}(\mathfrak{K})$ and 
$C_{s}(\mathfrak{K};r)$
are the secrecy capacity under weak code concept
and  random   assisted   secrecy capacity of
$\mathfrak{K}$, respectively (cf. \cite{Bo/Ca/De}  and \cite{Bo/Ca/De3}).

We also analyzed the secrecy capacity     when   the sender and the
receiver used various resources. 
In \cite{Bo/Ca/De2} 
we determined the secrecy capacities under 
randomness assisted coding of arbitrarily varying classical-quantum
wiretap  channels.\vspace{0.2cm}

\begin{lemma}
Let   $\Theta$ $:=$ $\{1,\cdots,T\}$ be a finite index set.
	Let $\mathfrak{L}$ $=(W_{\theta},{V}_{\theta})_{ {\theta} \in \Theta}$ 
	be an arbitrarily varying classical-quantum wiretap channel.
	We have
\begin{align}&
	C_{s}(\mathfrak{L};r)= \mathfrak{C}(\mathfrak{L})
	\text{ .}\end{align}
	\label{commperm}
\end{lemma}\vspace{0.2cm}

 We also 
examined in \cite{Bo/Ca/De2}  when the secrecy
capacity is a continuous function of the system parameters.
  Furthermore, we proved the phenomenon
``super-activation'' for arbitrarily varying classical-quantum
wiretap channels.   
Combining the results of these two   papers      we obtain    the formula for
deterministic secrecy capacity of the arbitrarily varying classical-quantum
wiretap   channel.

 However, 
that formula is still 	capable of improvement 
when we explicitly allow the eavesdropper to 
have a small part of the code word 
be non-secure.
This reduces the generality of the code concept.
The code word we built was a composition of a public
code word to synchronize the second part and a  randomness assisted code word
to transmit the message.
We only    required    security for the last part.
As we    have   shown in 
 \cite{Bo/Ca/De3}, when
the jammer had access 
to the first part,
the code would be rendered
completely useless. Thus
the code concept only works when
the jammer is limited in his action, e.g.,
    when   
the   eavesdropper cannot
send messages towards the jammer.
Nevertheless, this code concept with weak criterion 
could be useful when   a small number   of public messages  
were   desired, e.g. when  the    receiver    used it
to estimate the channels.

 For classical arbitrarily varying wiretap channels
the authors of  \cite{No/Wi/Bo}
developed a new method to overcome
this problem: Applying 
a  technique introduced in \cite{Cs/Na},
they made the first part 
secure and used it to send the 
message instead of just the  randomness.
The code they constructed is thus 
 a one-part deterministic secure code.
    However, it is technically difficult to extend the random classical
code technique introduced in \cite{Cs/Na}
to classical-quantum  channels,
   thus we have to come up with a different construction.

Therefore, in  \cite{Bo/Ca/De3} we 
considered a general code concept
when we constructed
a code in such a way that every
part of it is secure.
We call
it the strong code concept  (cf. Definition \ref{stccwecc}) .
The main results of \cite{Bo/Ca/De3} 
are included in the following two lemmata:\vspace{0.2cm}

\begin{lemma}	
	Let $\mathfrak{L}$ $=(W_{\theta},{V}_{\theta})_{ {\theta} \in \Theta}$ 
	be an arbitrarily varying classical-quantum wiretap channel.
 If the
arbitrarily varying  classical-quantum channel $\{W_{\theta} : {\theta}\in \Theta\}$
is not symmetrizable, then
\begin{align}&C_{s}(\mathfrak{L}) =
\mathfrak{C}(\mathfrak{L})
	\label{bibqicbtt} \text{ ,}\end{align}
when we use a two-part code word
where both parts are secure.
	
Here $B_{\theta}$ are the resulting  quantum states at the output of the
legitimate receiver's channels. $Z_{{\theta}^n}$ are the resulting  quantum states  at
the output of wiretap channels.
 The maximum is taken over all random
variables  that satisfy the Markov chain relationships:
$U\rightarrow A \rightarrow B_q Z_{\theta}$ for every $B_q \in Conv((B_{\theta})_{{\theta}\in \Theta})$ and
$ {\theta}\in \Theta$.   Here $A$   is  a random
variable taking values on $\mathbf{A}$,   and  $U$ a random
variable taking values on some finite set $\mathbf{U}$
with probability  distribution $p_U$.
\label{pdwumsfsov}
\end{lemma}\vspace{0.2cm}

\begin{lemma}[Ahlswede Dichotomy under strong code concept] Let  $\Theta$ be a finite set and $\mathfrak{L}$ 
$=\{(W_{\theta},{V}_{\theta}): {\theta}\in \Theta\}$
 be an arbitrarily
varying classical-quantum wiretap channel.
\begin{enumerate}[1)]
\item If the
arbitrarily varying  classical-quantum channel $\{W_{\theta} : {\theta}\in \Theta\}$
is not symmetrizable, then \begin{equation}
C_{s}(\mathfrak{L})=C_{s}(\mathfrak{L};r)\text{ .}
\end{equation}
\item If $\{W_{\theta} : {\theta}\in \Theta\}$ is symmetrizable,   then 
 \begin{equation}
C_{s}(\mathfrak{L})=0\text{ .}
\end{equation} \end{enumerate}\label{dichpartstr}
\end{lemma}\vspace{0.2cm}

Here $C_{s}(\mathfrak{L})$ 
is the secrecy capacity under    the     strong code concept
(cf. \cite{Bo/Ca/De3}).\vspace{0.2cm}

These lemmata show that when the legal channel
is not symmetrizable, the sender can send
a small number of secure transmissions
which push the secure 
capacity  to the maximally attainable value.  
Thus, complete   security is granted.

The capacities of
various communication scenarios
(entanglement transmission, entanglement distillation, entanglement generating,
and  strong subspace transmission)
have been analyzed 
in \cite{Ahl/Bj/Bo/No}, \cite{Bo/No}, and \cite{Bo/No2}.
In these works the deterministic   capacity,
i.e., message capacity with no randomness  assisted 
code, of arbitrarily
varying quantum   channels for the case of positive deterministic capacity have been completely characterized.
Furthermore in  \cite{Ahl/Bj/Bo/No}
the randomness  assisted capacities
for message transmission, entanglement distillation, entanglement generating,
and  strong subspace transmission of
 arbitrarily
varying quantum   channels have been completely characterized.
Notably, it has been shown in  \cite{Ahl/Bj/Bo/No}
that the last three capacities are equal.
In addition, it has been shown that
for an  arbitrarily varying quantum channel $\{N_{\theta}: \theta\in\Theta\}$,
these capacities are equal to the respective
capacities of
the compound  quantum channel 
\begin{align*}&\{N_{p}: p\in P(\Theta)\}\\
&=\left\{ \text{completely positive,
trace preserving linear map } N: \mathcal{L}(H^\mathfrak{P}) \rightarrow \mathcal{L}(H^\mathfrak{Q}) : \exists ~ p\in P(\Theta) \text{ such that } N=\sum_{\theta} p(\theta)N_{\theta}\right\}\text{ ,}
\end{align*}
where $P(\Theta)$ is the set of
probability distributions on $\Theta$.
This implies that these three
 randomness  assisted capacities are 
continuous functions of the channel
system parameters.

In \cite{Ahl/Bj/Bo/No}, \cite{Bo/No}, and \cite{Bo/No2},
the so called ``Ahlswede Dichotomy'' for
 the
entanglement distillation capacity, the entanglement generating capacity,
and the strong subspace transmission capacity has
been shown, i.e., each of
these capacities is either zero or equal to the
respective capacity randomness assisted capacity.
The question if it can actually occur that
these capacities are not equal to the
respective capacities
under shared randomness assisted quantum coding
is still an open problem. However in  \cite{Ahl/Bj/Bo/No}
the authors strongly supposed that this case will not occur.
This conjecture
is still unsolved by now.

If this conjecture is true,
then these three capacities under 
deterministic coding are 
continuous functions of the channel
system parameters.
This means if this  conjecture is true
then for an AVQC,
 the behavior of these three capacities under 
deterministic coding and the
secure capacity under 
deterministic coding 
differ significantly.
This is insofar particularly interesting
since in \cite{De} Devetak used
the
secure capacity for the
characterization of
the above mentioned three capacities
for arbitrarily
varying quantum channels (the achievements
of the capacity formulas).

In this work, one of the major difficulties 
we have to overcome is the characterization
of the secure deterministic capacity
when the deterministic capacity is not
equal to zero. Additionally a important
result is that the 
continuity behavior of  the secure deterministic capacity
and the  secure capacity under 
deterministic coding 
differ significantly for AVQCs.

\subsection{Proof of the Main Results}\label{protmr}

Corollary \ref{lntvttitbaalem} follows
immediately from the definition of
the randomness  assisted secrecy capacity
and Lemma \ref{commperm}.\vspace{0.2cm}

The following proof of
Corollary \ref{lgwtvttitbacam} is almost
exact the same as  our proof of 
the capacity formula of classical arbitrarily
varying quantum wiretap channels
 in \cite{Bo/Ca/De2}, when we apply the coding technique of \cite{Bo/No}
for the maximal error criterion instead
of coding techniques for the average error criterion.\vspace{0.2cm}

\begin{proof}
 We fix a probability distribution
$p\in P(\mathbf{A})$
and choose an arbitrarily positive $\delta$.
Let 
\[J_n =  \lfloor  2^{\inf_{\grave{B}_q \in Conv((\grave{B}_{\theta'})_{{\theta'}\in \Theta})}I(p;\grave{B}_q^n)-\max_{{\theta}^n\in \Theta^n}\chi(p;Z_{{\theta}^n})-n\delta} \rfloor  \text{ ,}\]
and
\[L_{n} =  	\lceil 2^{\max_{{\theta}^n\in \Theta^n} (\chi(p;Z_{{\theta}^n})+n\delta)} \rceil 	\text{  .}\] 
Let $p' (x^n):= \begin{cases} \frac{p^{ n}(x^n)}{p^{ n}
(\mathcal{T}^n_{p,\delta})} & \text{if } x^n \in \mathcal{T}^n_{p,\delta}\text{ ;}\\
0 & \text{  else} \end{cases}$\\  
 and $X^n := \{X_{j,l}\}_{j \in
\{1, \cdots, J_n\}, l \in \{1, \cdots, L_{n}\}}$  be a family of
random matrices whose components are i.i.d. according to
$p'$.

By \cite{Bo/No}, \cite{Ahl/Bj/Bo/No} 
and the Alternative Covering Lemma in \cite{Bo/Ca/De2},
if $n$ is sufficiently
large
with a positive probability there exist
a realization $\{x_{j,l}\}_{j \in
\{1, \cdots, J_n\}, l \in \{1, \cdots, L_{n}\}}$ of $X^n$,
a set  of mutually disjoint sets 
$\{D_{j,l}: j \in
\{1, \cdots, J_n\}, l \in \{1, \cdots, L_{n}\}\}$ on $\mathbf{B}^n$ 
such that
 for all positive $\epsilon$, $\lambda$, all  ${\theta}^n \in \Theta^n$,
  $j \in \{1,\dots,J_n\}$,  and  $ l \in \{1, \cdots, L_{n}\}$
\begin{equation} \label{b3lnrimti} 
\grave{W}_{{\theta}^n}(D_{j,l}^c| x_{j,l})\leq \epsilon \text{ ,}\end{equation}
 and 
\begin{equation} \label{clruztr2wm}
\chi\left(R_{uni};Z_{{\theta}^n}\right)\leq \lambda
\text{ .}\end{equation}

\end{proof}\vspace{0.2cm}

 The following proof of
Corollary \ref{lntvttitbaalemmax} is 
similar to  our proof of 
the formula for randomness assisted capacity of classical arbitrarily
varying quantum wiretap channels.\vspace{0.2cm}

\begin{proof}

When we apply the coding technique of \cite{Bo/No}
where it has been shown that the  randomness  assisted  capacity of an
arbitrarily varying quantum channel under
the maximal error criterion
is equal to its  randomness  assisted  capacity  under
the average error criterion,
and the Alternative Covering Lemma in \cite{Bo/Ca/De2}
to our proof for 
the capacity formula of classical arbitrarily
varying quantum wiretap channels
 in \cite{Bo/Ca/De2},
we obtain:
If $n$ is sufficiently
large
with a positive probability there exist an
 $(n, J_n)$
code $\mathcal{C}^{\gamma} = \bigl(E^{\gamma}, \{D_{j,l}^{\gamma} : j \in
\{1, \cdots, J_n\}, l \in \{1, \cdots, L_{n}\}\}\bigr)$ for $\gamma = 1, \cdots, n^3$ and a $F$ 
such that
 for all positive $\epsilon$, $\lambda$, all  ${\theta}^n \in \Theta^n$,
  $j \in \{1,\dots,J_n\}$,  and  $ l \in \{1, \cdots, L_{n}\}$
\begin{equation}1-\frac{1}{n^3}\sum_{\gamma=1}^{n^3}
{N}_{\theta^{n}}\circ {F}^n (D_{j,l}^{\gamma}|E^{\gamma}(\cdot |j,l))\leq  \varepsilon\text{ ,}\end{equation} 
 and 
\begin{equation} \label{clruztr2wmqq}\frac{1}{n^3}\sum_{\gamma=1}^{n^3}
\chi\left(R_{uni};Z_{\mathcal{C}^{\gamma},{\theta}^n}\right)\leq \lambda
\text{ .}\end{equation}

 Here for
$\theta^n\in\Theta^n$ and $\mathcal{C}^{\gamma}=\{(w(j)^{n,\gamma},D_j^{\gamma}):
 j=1,\ldots,J_n\}$,
\[Z_{\mathcal{C}^{\gamma},\theta^n} :=\left\{{V}_{\theta^{n}}(w(1)^{n,\gamma}),
{V}_{\theta^{n}}(w(2)^{n,\gamma}),\ldots,
{V}_{\theta^{n}}(w(n)^{n,\gamma})\right\}\text{ .}\]
\end{proof}\vspace{0.2cm}

Now we will deliver  the proof for
Theorem \ref{lntvttitbaa}:
\vspace{0.2cm}

\begin{proof}
\it i)  Proof of   2) \rm \vspace{0.2cm}

In \cite{Bo/No2}, the following was shown:
When $\mathfrak{I}$ is $L$-symmetrizable for all $L\in\mathbb{N}$,
then for  any
chosen $\mathbf{A}$ and
$F:$ $\mathbf{A}$ $\rightarrow \mathcal{S}(H^{\mathfrak{P}})$,
the arbitrarily varying classical-quantum channel $\{N_{\theta}\circ F: {\theta}\in\Theta\}$
is symmetrizable. By Lemma \ref{dichpartstr}, 
the secrecy capacity 
concept of $\{N_{\theta}:  {\theta}\in \Theta\}$ with any classical
input is equal to zero. Thus
2) holds.\vspace{0.3cm}

\it ii)  Converse for   1) \rm \vspace{0.2cm}

Now let us assume that there is a  $L'\in\mathbb{N}$ such that
 $\mathfrak{I}$ is $L'$-symmetrizable.\vspace{0.2cm}

We fix  a  finite set $\mathbf{A}$, and a
map $F:$ $\mathbf{A}$ $\rightarrow \mathcal{S}(H^{\mathfrak{P}})$.
By Lemma \ref{dichpartstr},
the  secrecy capacity under strong code for the arbitrarily varying classical-quantum wiretap channel
$\{(N_{\theta}\circ F,{V}_{\theta}'\circ F):  {\theta}\in \Theta\}$ under strong code concept 
cannot exceed   $\limsup_{n\rightarrow \infty}$ $\frac{1}{n}\sup_{p}
\inf_{q \in P({\Theta})} $ $ (\chi(P,B_q^n) -\max_{{\theta}^{n}\in \Theta^{n}}\chi (p,Z_{{\theta}^{n}}))$.
Thus
\begin{equation}C_{s}(\mathfrak{I}) \leq
  \limsup_{n\rightarrow \infty} \frac{1}{n} \sup_{\mathbf{A}, F, P}\left(
\inf_{q \in P({\Theta})}  \chi(P,B_q^n) -\max_{{\theta}^{n}\in \Theta^{n}}\chi (P,Z_{{\theta}^{n}})\right)\text{ .}\label{csclntvt}\end{equation}\vspace{0.3cm}

\it iii)  Achievement for   1) \rm \vspace{0.2cm}

We assume that there is a  $L'\in\mathbb{N}$ such that
 $\mathfrak{I}$ is not $L'$-symmetrizable.\vspace{0.2cm}

When 
 \[\limsup_{n\rightarrow \infty} \frac{1}{n} \sup_{\mathbf{A}, F, P}
(\inf_{q \in P({\Theta})}  \chi(P,B_q^n) -\max_{{\theta}^{n}\in \Theta^{n}}\chi (P,Z_{{\theta}^{n}}) )\leq 0\]
holds, we have $C_{s}(\mathfrak{I}) = 0$
by (\ref{csclntvt}), and there is nothing to prove.\vspace{0.2cm}

Now we assume there is a finite set of letter $\mathbf{A}'$, a map $F'$ 
$: \mathbf{A}' \rightarrow \mathcal{S}(H)$, and a distribution  $p'$
on $\{ {F}'(x) : x\in\mathbf{A}'\}$ such that for sufficiently large $n$ we have
\begin{equation}\frac{1}{n}\Bigl(\inf_{{B'}_q \in Conv(({B'}_{\theta})_{{\theta}\in \Theta})}\chi(p';{B'}_q^{\otimes n})
-\max_{{\theta}^n\in \Theta^n}\chi(p';{Z'}_{{\theta}^n})\Bigr) > 0\text{ ,}\label{finbibqicbt}\end{equation}
where ${B'}_{\theta}$ and ${Z'}_{\theta}$
are the resulting quantum states at the outputs
of ${N}_{\theta}\circ F'$ and ${V}_{\theta}'\circ F'$, 
respectively.

In \cite{Bo/No2}
a technique has been introduced to
construct an arbitrarily varying classical-quantum channel by means of
an  arbitrarily varying quantum channel with
positive deterministic capacity.
However this technique does not work
for the classical arbitrarily varying quantum wiretap 
channel since it cannot provide security.
We have to find a more sophisticated way.\vspace{0.2cm}

We have  used a trick  for the proof of Lemma
\ref{pdwumsfsov} in \cite{Bo/Ca/De3} to 
define a set of classical-quantum channels
with positive deterministic secure capacity in sense of the strong
code concept. However, the
construction of an arbitrarily varying classical-quantum wiretap channel by means of
an  arbitrarily varying quantum wiretap channel
with positive deterministic secure capacity in sense of the strong
code concept is even harder. We have to combine
the technique in \cite{Bo/Ca/De3}, the technique
of \cite{Bo/No2} and additionally, the concept of
block coding for our proof, where we get a 
deterministic code word which is a
composition of a
$((\log \log n)^3,(\log n)^3)$ secret deterministic code word and a $((\log n)^2-(\log \log n)^3,n^3)$ 
secret random code word (cf. Remark \ref{ntecwot}).

\it a) Definition of an arbitrarily varying classical-quantum channel   which is  not   symmetrizable \rm\vspace{0.2cm}

When 
$\{N_{\theta}:  {\theta} \in \Theta\}$
is not ${L'}$-symmetrizable, then it is also not
${L}$-symmetrizable
for all $L > L'$ (cf. \cite{Bo/No2}).
We choose a $L \geq L'$ such that
\begin{equation}\frac{1}{L}\Bigl(\inf_{{B'}_q \in Conv(({B'}_{\theta})_{{\theta}\in \Theta})}\chi(p';{B'}_q^{\otimes L})
-\max_{{\theta}^L\in \Theta^L}\chi(p';{Z'}_{{\theta}^L})\Bigr)  = C_1> 0\text{ .}\label{bibqicbttit}\end{equation}
We now combine the idea of \cite{Bo/No2} and the
concept of block coding to 
define a set of classical-quantum channels.\vspace{0.2cm}

We denote the dimension of $H^{\mathfrak{A}}$ by $d$. 
For every $\breve{L}\in\mathbb{N}$ we
 choose arbitrarily $d^{2\breve{L}}$
density operators
$\bar{M}_i$, $i=1, \cdots, {d^{\breve{L}}}^2$ which span the space of Hermitian operators on
 $H^{\mathfrak{A}^{\breve{L}}}$. 
 We define a new set of letters
  $\mathbf{A}_1$      $:= {\mathbf{A}'} \cup \{1, \cdots, d^{2}\}$ 
  and a map $F$ $: \mathbf{A}_1 \rightarrow \mathcal{S}(H)$
  by \[F(a) = \begin{cases} {{F}'}(a)& \text{if } a\in {\mathbf{A}'} \text{ ;}\\
  \bar{M}_a& \text{if } a\in \{1, \cdots, d^{2}\} \end{cases}\text{ .}\]\vspace{0.2cm}

When $\{N_{{\theta}^L} \circ \breve{F} : {\theta}^L \in \Theta^L\}$
is symmetrizable then
there is a  $\{\tau(\cdot\mid a^L):
 a^L\in \mathbf{U}^L\}$ on $\Theta^L$ such that 
\[\sum_{{\theta}^L\in\Theta^L}\tau({\theta}^L\mid a^L)N_{{\theta}^L} \circ {F}^L ({a'}^L)
=\sum_{{\theta}^L\in\Theta^L}\tau({\theta}^L\mid {a'}^L)N_{{\theta}^L} \circ {F}^L (a^L)\]
for all $a, a'\in\mathbf{A}_1$.

This implies that
\[\sum_{{\theta}^L\in\Theta^L}\tau({\theta}^L\mid i^L)N_{{\theta}^L} (\bar{M}_i)
=\sum_{{\theta}^L\in\Theta^L}\tau({\theta}^L\mid {i'}^L)N_{{\theta}^L} (\bar{M}_{i'})\]
for all $i,i'\in\{1,\cdots, {d^L}^2\}$. 

Since 
$\{\bar{M}_i: i=1, \cdots, {d^L}^2\}$ spans  the space of Hermitian operators on    $H^{\mathfrak{A}^L}$,    
 $\{N_{{\theta}^L} : {\Theta}^L \in \Theta^L\}$ is $L$-symmetrizable.
This is a contradiction to our assumption.
Therefore 
$\{N_{{\theta}^L}  \circ F^L : {\theta}^L \in \Theta^L\}$
is not symmetrizable.\vspace{0.2cm}

\it b) The secure transmission of  the message with a deterministic code\rm\vspace{0.2cm}

We define a distribution  $p$
on $\{ {F}^L(a^L) : a^L\in\mathbf{A}_1^L\}$ by
\[p\left({F}^L(a^L)\right) = \begin{cases} p'\left({F}^L(a^L)\right)& \text{if } a^L\in \mathbf{A}'^L \text{ ;}\\
0& \text{else } \end{cases}\text{ .}\]

By (\ref{bibqicbttit}) 
\begin{align}&\frac{1}{L} \Bigl(\inf_{{B'}_q \in Conv(({B'}_{\theta})_{{\theta}\in \Theta})}\chi(p;{B'}_q^{\otimes L})
-\max_{{\theta}^L\in \Theta^L}\chi(p;{Z'}_{{\theta}^L})\Bigr)\notag\\
&=\frac{1}{L}\Bigl(\inf_{{B'}_q \in Conv(({B'}_{\theta})_{{\theta}\in \Theta})}\chi(p';{B'}_q^{\otimes L})
-\max_{{\theta}^L\in \Theta^L}\chi(p';{Z'}_{{\theta}^L})\Bigr)\notag\\
& = C_1 \text{ ,}\label{0bltzpclt}\end{align}
where ${B}_{\theta}$ and ${Z}_{\theta}$
are the resulting quantum states at the outputs
of ${N}_{\theta}\circ F$ and ${V}_{\theta}'\circ F$, 
respectively.\vspace{0.2cm}

We apply block coding and
regarding $L$ uses of the quantum channel as a single
quantum channel.
 Since $\{N_{{\theta}^L}  \circ {F}^L : {\theta}^L \in \Theta^L\}$
is not symmetrizable, 
by Lemma \ref{pdwumsfsov},  (\ref{0bltzpclt}) implies
that 

\[C_{s}\left(\{\left(N_{{\theta}^L}  \circ {F}^L,  {V'}_{{\theta}^L}  \circ {F}^L\right) : {\theta}^L \in \Theta^L\}\right) = C_1 >0\text{ .}\]

By Corollary \ref{lntvttitbaalemmax},
for any $(n, J_n)$  randomness  assisted quantum code
 there exists an $(n, J_n)$   randomness assisted quantum code
using randomness on a set of polynomial order of $n$ (actually a set of size of $n^3$)
with the same secrecy rate (cf.  \cite{Bo/Ca/De}  for the reduction 
of  randomness).

Since $(\log n)^2 \gg 3\frac{\log n}{C_1}$
we can build a $((\log n)^2,n^3)$
code $(\grave{E}^{L(\log n)^2},\{\grave{D}_i^{L(\log n)^2}$ $:i\in\{1,\cdots,n^3\}\})$
such that for any positive
$\varepsilon$ when $n$ is sufficiently large we have
\begin{align}& 1- \min_{{\theta}^{L(\log n)^2} \in \Theta^{L(\log n)^2}} \min_{i\in\{1,\cdots,n^3\}} 
N_{{\theta}^{L(\log n)^2}}\circ {F}^{L(\log n)^2} \left( \grave{E}^{L(\log n)^2} \left(\cdot \mid i\right)\right) \grave{D}_i^{L(\log n)^2}\notag\\
&\leq  \varepsilon\label{f1n3mtwtnr2A}\end{align}
  and  
\begin{equation}
\max_{ {\theta}^{L(\log n)^2} \in \Theta^{L(\log n)^2}}\left\|{V'}_{{\theta}^{L(\log n)^2}}\circ {F}^{L(\log n)^2}
\left( \grave{E}^{L(\log n)^2} \left(\cdot \mid i\right)\right)- \Xi_{{\theta}^{L(\log n)^2}}\right\|_{1}
\leq \varepsilon\text{ ,}\label{mt6l2itA}
\end{equation}
where
$\Xi_{{\theta}^{L(\log n)^2}}$ is  a quantum state which is independent of $i$ (cf. \cite{Bo/Ca/De3}).\vspace{0.2cm}

\begin{remark}Note that each code word of this  $((\log n)^2,n^3)$ secret deterministic code itself is actually a composition of a
$((\log \log n)^3,(\log n)^3)$ secret deterministic code word and a $((\log n)^2-(\log \log n)^3,n^3)$ 
secret random code word (cf. \cite{Bo/Ca/De3}).\label{ntecwot}\end{remark}\vspace{0.2cm}

\it iv) The secure transmission of both the message and the randomization index \rm\vspace{0.2cm}

Recall that by (\ref{0bltzpclt}),
\begin{align*}& \frac{1}{n}\Bigl(\inf_{{B}_q \in Conv(({B}_{\theta})_{{\theta}\in \Theta})}\chi(p;{B}_q^{\otimes n})
-\max_{{\theta}^n\in \Theta^n}\chi(p;{Z}_{{\theta}^n})\Bigr)\\
&=\frac{1}{n}\Bigl(\inf_{{B'}_q \in Conv(({B'}_{\theta})_{{\theta}\in \Theta})}\chi(p';{B'}_q^{\otimes n})
-\max_{{\theta}^n\in \Theta^n}\chi(p';{Z'}_{{\theta}^n})\Bigr) \end{align*} 
where ${B}_{\theta}$ and ${Z}_{\theta}$
are the resulting quantum states at the outputs
of ${N}_{\theta}\circ F$ and ${V}_{\theta}'\circ F$, 
respectively.

We choose an arbitrary  positive $\delta$. Let
\[J_n = \frac{1}{n}  \left(\inf_{{B}_q \in Conv(({B}_{\theta})_{{\theta}\in \Theta})} 
\chi(p;{B}_q^{\otimes n}) - \max_{{\theta}^n\in \Theta^n}\chi(p;{Z}_{{\theta}^n})\right) -\delta\text{ .}\]

By Lemma \ref{commperm} (cf. \cite{Bo/Ca/De2} for details),
if $n$ is sufficiently large  we can construct an $(n, J_n)$
 randomness assisted quantum code
$\Bigl\{\left(E^n, \{ D_{j,i}^n : j\in\{1,\cdots,J_n\}\}\right):$ $i\in\{1,\cdots,n^3\}\Bigr\}$, and positive $\lambda$, $\zeta$, 
such that for  all ${\theta}^n\in\Theta^n$ 
\begin{align}&
\frac{1}{n!}\frac{1}{J_n} \sum_{\pi\in\mathsf{S}_n}  \sum_{j=1}^{J_n}  
\mathrm{tr}\left(N_{{\theta}^n}\circ {{F}'}^{n}\left(E^n(\pi(\cdot|j))\right) P_{\pi} D_{j}^n P_{\pi}^{\dagger} \right)
\geq 1-2^{-n^{1/16}\lambda} \text{ ,}\end{align}
and for all ${\theta}^n\in\Theta^n$, $j\in\{1,\cdots,J_n\}$ and all $i\in\{1,\cdots,n^3\}$ 
\begin{align}&\lVert
{V'}_{{\theta}^n}\circ {{F}'}^{n}\left(\pi(E^n(\cdot|j))\right) - P_{\pi}
\Xi_{\pi^{-1}({\theta}^n)} P_{\pi}^{\dagger}  \rVert_{1}
< 2^{-\sqrt{n}\zeta}\text{ ,}\end{align} 
where $\Xi_{{\theta}^n}\in\mathcal{S}(H^n)$ is  a quantum state which is independent of $j$ and $i$ (cf. \cite{Bo/Ca/De3}).

Using   the    technique 
of Lemma \ref{dichpart1}
(cf.  \cite{Bo/Ca/De}  for details) to reduce the amount
of 
randomness if $n$ is sufficiently   large,  
we can find  a set $\{\pi_1,\cdots,\pi_{n^3}\} \subset \mathsf{S}_n$ 
such that
\begin{align}&
\frac{1}{n^3}\frac{1}{J_n} \sum_{i=1}^{n^3} \sum_{j=1}^{J_n}  
\mathrm{tr}\left(N_{{\theta}^n}\circ {{F}'}^{n}\left(\pi_i(E^n(\cdot|j))\right)P_{\pi_i}D_{j}^n P_{\pi_i}^{\dagger}\right) 
\geq 1-2^{-n^{1/16}\lambda} \text{ ,}\label{mtnitn1n3f1jnA}\end{align}
and for all ${\theta}^n\in\Theta^n$, $j\in\{1,\cdots,J_n\}$ and all $i\in\{1,\cdots,n^3\}$ 
\begin{align}&\lVert
{V'}_{{\theta}^n}\circ {{F}'}^{n}\left(\pi_i(E^n(\cdot|j))\right) -
 P_{\pi_i}
\Xi_{\pi_i^{-1}({\theta}^n)} P_{\pi_i}^{\dagger} \rVert_{1}
< 2^{-\sqrt{n}\zeta}\text{ ,}\label{balf1lnsl13A}\end{align} 
where $\Xi_{{\theta}^n}\in\mathcal{S}(H^n)$ is  a quantum state which is independent of $j$ and $i$ (cf. \cite{Bo/Ca/De3}).\vspace{0.2cm}

 	Furthermore, 
	by the permutation-invariance of $p$ we also have  
$\Xi_{{\theta}^n} = P_{\pi}
\Xi_{\pi({\theta}^n)} P_{\pi}^{\dagger}$ for all $\pi\in\mathsf{S}_n$.\vspace{0.2cm}

We  now define our 
secure $(L(\log n)^2+n, J_n+n^3)$ code
for $\{({N}_{\theta}\circ F, {V}_{\theta}'\circ F): {\theta}\in\Theta\}$
by
	\begin{equation}E^{L(\log n)^2+n}(a^{L(\log n)^2+n}\mid i,j) :=\grave{E}^{L(\log n)^2} (a^{L(\log n)^2}|i)\cdot
E^n(\pi_i(a^n)|j)
\text{ ,}\end{equation}
for every $a^{{(\log n)^3}+n}= (a^{(\log n)^3},a^{n})\in \mathbf{U}^{{(\log n)^3}+n}$
and 
	\begin{equation}D_{i,j}^{L(\log n)^2+n} :=\grave{D}_i^{L(\log n)^2} \otimes (P_{\pi_i}D_j^n P_{\pi_i}^{\dagger})
\text{ .}\end{equation}

By (\ref{f1n3mtwtnr2A}) and   (\ref{mtnitn1n3f1jnA}),  
for every   ${\theta}^{L(\log n)^2+n}= ({\theta}^{L(\log n)^2},{\theta}^n)\in{\Theta}^{L(\log n)^2+n}$,   we have
\begin{align}& \frac{1}{n^3}\frac{1}{J_n}\sum_{i=1}^{n^3} \sum_{j=1}^{J_n} 
\mathrm{tr}\left(N_{{\theta}^{L(\log n)^2}+n}\circ {F}^{L(\log n)^2+n} (E^{L(\log n)^2+n}(\cdot\mid i,j)) D_{i,j}^{L(\log n)^2+n}\right) \notag\\
&=\frac{1}{n^3}\frac{1}{J_n}\sum_{i=1}^{n^3} \sum_{j=1}^{J_n} 
\mathrm{tr}\Biggl(\left[N_{{\theta}^{L(\log n)^2}}\circ {F}^{L(\log n)^2}  (\grave{E}^{L(\log n)^2} (\cdot|i))\otimes
 \left(
N_{{\theta}^n}\circ {F}^{n}(\pi_i( E^n(\cdot|j)))\right)\right]\allowdisplaybreaks\notag\\
&\left[ \grave{D}_i^{L(\log n)^2} 
\otimes (P_{\pi_i}D_j^n P_{\pi_i}^{\dagger})\right]\Biggr) \allowdisplaybreaks\notag\\
&=\frac{1}{n^3}\sum_{i=1}^{n^3}  
\mathrm{tr}\Biggl(\left[N_{{\theta}^{L(\log n)^2}}\circ {F}^{L(\log n)^2}  (\grave{E}^{L(\log n)^2} (\cdot|i))
\grave{D}_i^{L(\log n)^2}\right]\allowdisplaybreaks\notag\\
& \otimes\left[ \frac{1}{J_n}\sum_{j=1}^{J_n}\left(
N_{{\theta}^n}\circ {F}^{n}(\pi_i(E^n(\cdot|j)))\right)
 P_{\pi_i}D_j^n P_{\pi_i}^{\dagger}\right]\Biggr) \allowdisplaybreaks\notag\\
&=\frac{1}{n^3}\sum_{i=1}^{n^3}  \Biggl(
\mathrm{tr}\left(N_{{\theta}^{L(\log n)^2}}\circ {F}^{L(\log n)^2}  (\grave{E}^{L(\log n)^2} (\cdot|i))
\grave{D}_i^{L(\log n)^2}\right)\allowdisplaybreaks\notag\\
& \cdot  \mathrm{tr}\left( \frac{1}{J_n}\sum_{j=1}^{J_n}\left(
N_{{\theta}^n}\circ {F}^{n}(\pi_i(E^n(\cdot|j)))\right)
 P_{\pi_i}D_j^n P_{\pi_i}^{\dagger}\right) \Biggr)\notag\\
&\geq 1-\frac{1}{n^{1/16}}2^{\lambda}-2\cdot 2^{-n^{1/16}\lambda}\notag\\
&\geq 1-  \varepsilon\label{l21gnbrtippnjd}
\end{align}
for any positive 
$\varepsilon$   when $n$ is sufficiently large.  

By (\ref{mt6l2itA}) and   (\ref{balf1lnsl13A}),  
for every ${\theta}^{L(\log n)^2+n}= ({\theta}^{L(\log n)^2},{\theta}^n)\in {\theta}^{L(\log n)^2+n}$ 
and $i \in \{1,\cdots, n^3\}$ and   $j\in \{1,\cdots, J_n\}$,   we have

\begin{align}& \lVert
{V'}_{{\theta}^{L(\log n)^2+n}}\circ {F}^{L(\log n)^2+n} (E^{L(\log n)^2+n}(\cdot\mid i,j)) -
\Xi_{{\theta}^{L(\log n)^2}} \otimes
\Xi_{{\theta}^n}  \rVert_{1}\notag\\
&= \lVert {V'}_{{\theta}^{L(\log n)^2}\circ {F}^{L(\log n)^2}}\left( \grave{E}^{L(\log n)^2}  \left(\cdot \mid i\right)\right) \otimes
 {V'}_{{\theta}^n}\circ {F}^{n}\left(\pi(E^n(\cdot|j))\right) -
\Xi_{{\theta}^{L(\log n)^2}} \otimes
\Xi_{{\theta}^n}  \rVert_{1}\notag\\
&< \frac{1}{\sqrt{n}}2^{\zeta}+ 2\cdot 2^{-\sqrt{n}\zeta}
\text{ .}\end{align}\vspace{0.2cm}

\begin{lemma}[Fannes-Audenaert  Ineq.,
 cf. \cite{Fa}, \cite{Au}]\label{eq_9}  
Let $\Phi$ and $\Psi$ be two  quantum states in a
$d$-dimensional complex Hilbert space and
$\|\Phi-\Psi\| \leq \mu < \frac{1}{e}$, then
\begin{equation} |S(\Phi)-S(\Psi)| \leq \mu \log (d-1)
+ h(\mu)\text{ ,}\label{faaudin}\end{equation}where $h(\nu) := -\nu \log \nu - (1- \nu) \log (1-\nu)$
for $\nu\in [0,1]$.\end{lemma}\vspace{0.15cm}

 The Fannes  Inequality was first introduced in \cite{Fa}, where it has been
shown that $|S(\mathfrak{X})-S(\mathfrak{Y})| \leq \mu \log d - \mu
\log \mu $. In \cite{Au} the result of \cite{Fa} has been
improved, and (\ref{faaudin}) has been proved.\vspace{0.15cm}

Let 
 $R_{n^3}$ be the 
uniform distribution on $\{1,\cdots,n^3\}$. We define
a random variable $R_{n^3,uni}$ on the
set $\{1,\cdots,n^3\} \times \{1,\cdots,R_n\}$
 by $R_{n^3,uni}:=R_{n^3} \times R_{uni}$.
Applying Lemma \ref{eq_9}
we obtain

\begin{align}&
\max_{ {\theta}^{L(\log n)^2+n} \in {\theta}^{L(\log n)^2+n}}\chi\left(R_{n^3,uni}; Z_{{\theta}^{L(\log n)^2+n}}\right)\notag\\
&\leq\max_{ {\theta}^{L(\log n)^2} \in {\theta}^{L(\log n)^2}}\chi\left(R_{n^3}; Z_{{\theta}^{L(\log n)^2}}\right)\notag\\
&~+\frac{1}{n^3}\sum_{i=1}^{n^3}\max_{ {\theta}^{n} \in {\theta}^{n}}\chi\left(R_{uni}; {V'}_{{\theta}^{L(\log n)^2}} \circ {F}^{L(\log n)^2}
(\grave{E}^{L(\log n)^2} (\cdot|i))\otimes Z_{{\theta}^{n},\pi_i}\right)\allowdisplaybreaks\notag\\
&=  \max_{ {\theta}^{L(\log n)^2} \in {\theta}^{L(\log n)^2}}\Biggl[ S\left(\frac{1}{n^3}
\frac{1}{J_n}\sum_{i=1}^{n^3} \sum_{j=1}^{J_n}{V'}_{{\theta}^{L(\log n)^2+n}}\circ {F}^{L(\log n)^2+n} (E^{L(\log n)^2+n}(\cdot\mid i,j)) \right)\allowdisplaybreaks\notag\\
&~-\frac{1}{n^3}\sum_{i=1}^{n^3} S\left(
\frac{1}{J_n}\sum_{j=1}^{J_n}{V'}_{{\theta}^{L(\log n)^2+n}}\circ {F}^{L(\log n)^2+n} (E^{L(\log n)^2+n}(\cdot\mid i,j)) \right)
\Biggr]\allowdisplaybreaks\notag\\
&~ + \frac{1}{n^3}\sum_{i=1}^{n^3}\max_{ {\theta}^{n} \in {\theta}^{n}}\Biggl[ S\biggl(
\frac{1}{J_n} \sum_{j=1}^{J_n}{V'}_{{\theta}^{L(\log n)^2}}\circ {F}^{L(\log n)^2} (\grave{E}^{L(\log n)^2} (\cdot|i))
\otimes
 \left(
{V'}_{{\theta}^n}\circ {F}^{n}(\pi_i( E^n(\cdot|j)))\right) \biggr)\allowdisplaybreaks\notag\\
&~-  \frac{1}{J_n}\sum_{j=1}^{J_n}S\left(
{V'}_{{\theta}^{L(\log n)^2}}\circ {F}^{L(\log n)^2} (\grave{E}^{L(\log n)^2} (\cdot|i))\otimes
 \left(
{V'}_{{\theta}^n}\circ {F}^{n}(\pi_i( E^n(\cdot|j)))\right) \right)\Biggr]\allowdisplaybreaks\notag\\
&\leq \max_{ {\theta}^{L(\log n)^2} \in {\theta}^{L(\log n)^2}}\Biggl[ ~~\Biggl\vert S\left(\frac{1}{n^3}
\frac{1}{J_n}\sum_{i=1}^{n^3} \sum_{j=1}^{J_n}{V'}_{{\theta}^{L(\log n)^2+n}}\circ {F}^{L(\log n)^2+n} (E^{L(\log n)^2+n}(\cdot\mid i,j)) \right)\allowdisplaybreaks\notag\\
&~- S\left(\Xi_{{\theta}^{L(\log n)^2}} \otimes
\Xi_{{\theta}^n}\right) \Biggr\vert\allowdisplaybreaks\notag\\
&~+\Biggl\vert S\left(\Xi_{{\theta}^{L(\log n)^2}} \otimes
\Xi_{{\theta}^n}\right)-\frac{1}{n^3}\sum_{i=1}^{n^3} S\left(
\frac{1}{J_n}\sum_{j=1}^{J_n}{V'}_{{\theta}^{L(\log n)^2+n}}\circ {F}^{L(\log n)^2+n} (E^{L(\log n)^2+n}(\cdot\mid i,j)) \right)\Biggr\vert
~~\Biggr]\allowdisplaybreaks\notag\\
&~ + \max_{ {\theta}^{n} \in {\theta}^{n}}\frac{1}{n^3}\sum_{i=1}^{n^3}\Biggl[ ~~\Biggl\vert S\biggl(
\frac{1}{J_n} \sum_{j=1}^{J_n}{V'}_{{\theta}^{L(\log n)^2}}\circ {F}^{L(\log n)^2} (\grave{E}^{L(\log n)^2} (\cdot|i))
\otimes
 \left(
{V'}_{{\theta}^n}\circ {F}^{n}(\pi_i( E^n(\cdot|j)))\right) \biggr)\allowdisplaybreaks\notag\\
&~- S\left( {V'}_{{\theta}^{L(\log n)^2}}\circ {F}^{L(\log n)^2} (\grave{E}^{L(\log n)^2} (\cdot|i))\otimes\Xi_{{\theta}^n} \right)\Biggr\vert\allowdisplaybreaks\notag\\
&~+\Biggl\vert S\left( {V'}_{{\theta}^{L(\log n)^2}}\circ {F}^{L(\log n)^2} (\grave{E}^{L(\log n)^2} (\cdot|i))\otimes \Xi_{{\theta}^n}\right)\allowdisplaybreaks\notag\\
&~ -  \frac{1}{J_n}\sum_{j=1}^{J_n}S\left(
{V'}_{{\theta}^{L(\log n)^2}}\circ {F}^{L(\log n)^2} (\grave{E}^{L(\log n)^2} (\cdot|i))\otimes
 \left(
{V'}_{{\theta}^n}\circ {F}^{n}(\pi_i( E^n(\cdot|j)))\right) \right)\Biggr\vert~~\Biggr]\allowdisplaybreaks\notag\\
&\leq(\frac{1}{\sqrt{n}}2^{\zeta}+ 2\cdot 2^{-\sqrt{n}\zeta})\log (d^{L(\log n)^2} -1)+
h(\frac{1}{\sqrt{n}}2^{\zeta}+ 2\cdot 2^{-\sqrt{n}\zeta}) \notag\\
&~+2\cdot 2^{-\sqrt{n}\zeta}\log (d^n-1) +h(2\cdot 2^{-\sqrt{n}\zeta})\notag\\
&\leq \varepsilon
\end{align}
for any positive 
$\varepsilon$ when $n$ is sufficiently large. Here 
$Z_{i,{\theta}^{n}}$    denotes      the resulting quantum state at ${V'}_{{\theta}^n}\circ {F}^{n}$
after $i\in\{1,\cdots,n^3\}$ has been sent with $E^{L(\log n)^2}$.

For any positive $\delta$, if $n$ is   sufficiently large,   we have $\frac{1}{n}\log J_n -\frac{1}{L(\log n)^2 +n}\log J_n$
$\leq \delta$.
Thus 
\begin{equation}C_{s}(\mathfrak{I}) \geq
  \limsup_{n\rightarrow \infty} \frac{1}{n} \sup_{\mathbf{A}, F, P}\left(
\inf_{q \in P({\Theta})}  \chi(P,B_q^n) -\max_{{\theta}^{n}\in \Theta^{n}}\chi (P,Z_{{\theta}^{n}})\right)-2\delta\text{ .}\label{csclntvta}\end{equation}

\end{proof}\vspace{0.2cm}

Next we will deliver  the proof for
Theorem  \ref{lntvttitbaamax}.\vspace{0.2cm}

\begin{proof}

Let   $\Theta$ $:=$ $\{1,\cdots,T\}$ be a finite index set.
Let $\{(N_{\theta},{V}_{\theta}') : {\theta}\in \Theta\}$
be an arbitrarily varying wiretap quantum channel
defined by an arbitrarily varying
quantum  channel $\mathfrak{I}$ $=\{N_{\theta}: {\theta}\in\Theta\}$.
\begin{equation}C_{s}(\mathfrak{I}) \geq C_{s, max}(\mathfrak{I})\label{cscntitg}\end{equation}
follows directly from the definitions.
Thus the converse is trivial.

Now we are going to prove the achievability.
When
 $\mathfrak{I}$ is  $L$-symmetrizable
for all  $L\in\mathbb{N}$, then by 
Theorem \ref{lntvttitbaa}, we have $C_{s}(\mathfrak{I}) =0$.
When 
 \[\limsup_{n\rightarrow \infty} \frac{1}{n} \sup_{\mathbf{A}, F, P}
(\inf_{q \in P({\Theta})}  \chi(P,B_q^n) -\max_{{\theta}^{n}\in \Theta^{n}}\chi (P,Z_{{\theta}^{n}}) )\leq 0\]
holds, we also have $C_{s}(\mathfrak{I}) = 0$.
By (\ref{cscntitg}), there is nothing to prove in this case.\vspace{0.2cm}

We assume that there is a  $L'\in\mathbb{N}$ such that
 $\mathfrak{I}$ is not $L'$-symmetrizable, and
 \[\limsup_{n\rightarrow \infty} \frac{1}{n} \sup_{\mathbf{A}, F, P}
(\inf_{q \in P({\Theta})}  \chi(P,B_q^n) -\max_{{\theta}^{n}\in \Theta^{n}}\chi (P,Z_{{\theta}^{n}}) )> 0\]
holds.\vspace{0.2cm}

Now we assume there is a finite set of letters $\mathbf{A}'$, a map $F'$ 
$: \mathbf{A}' \rightarrow \mathcal{S}(H)$, and a distribution  $p'$
on $\{ {F}'(x) : x\in\mathbf{A}'\}$ such that for sufficiently large $n$ and some $p'\in P(\mathbf{A}')$  we have
\begin{equation}\frac{1}{n}\Bigl(\inf_{{B'}_q \in Conv(({B'}_{\theta})_{{\theta}\in \Theta})}\chi(p';{B'}_q^{\otimes n})
-\max_{{\theta}^n\in \Theta^n}\chi(p';{Z'}_{{\theta}^n})\Bigr) > c\label{finbibqicbtmax}\end{equation}
for a positive $c$. Here ${B'}_{\theta}$ and ${Z'}_{\theta}$
are the resulting quantum states at the outputs
of ${N}_{\theta}\circ F'$ and ${V}_{\theta}'\circ F'$, 
respectively.

We choose an arbitrary positive $\mu$ and define
\[{J'}_m := \lfloor \exp(\inf_{q \in P({\Theta})}  \chi(P,B_q^m) -2m\mu) \rfloor\text{ .}\]
By Corollary \ref{lntvttitbaalemmax}, if $m$ is sufficiently large we can find a 
$(m, {J'}_m)$   randomness assisted
quantum code  $\Bigl\{\bigl({E'}^{\gamma} , \{{D'}_{j}^{\gamma} : j \in \{1, \cdots,{J'}_m\}\}\bigr):\gamma \in \Gamma\Bigr\}$ and  positive  $\lambda$, $\zeta$
such that  for all ${\theta}^m\in \Theta^m$ 
	\begin{equation}
\min_{j\in \{1,\cdots, {J'}_m)\}}\frac{1}{|\Gamma|}\sum_{\gamma \in \Gamma}
\mathrm{tr}\left( N_{{\theta}^m}\circ {F'}^m  ( {E'}^{\gamma} (~|j)  ){D'}_{j}^{\gamma}\right)\geq 1-2^{-m^{1/16}\lambda}\text{ .}
\label{1f1jmspimax1}\end{equation}

When 
 $\mathfrak{I}$ is not $L'$-symmetrizable
 for some $L'\in\mathbb{N}$,
by \cite{Ahl/Bj/Bo/No} if  $m$ is sufficiently large, there exist 
quantum states $\sigma_1$, $\sigma_2$ $\in  \mathcal{S}(H^m)$
and positive-semidefinite operators 
$\dot{D}_1$, $\dot{D}_2$ on ${H}^{\otimes m}$, where  
$\dot{D}_1 + \dot{D}_2$ $=1_{{H}^{\otimes m}}$,
such that for all ${\theta}^m\in \Theta^m$ and $i\in\{1,2\}$
\begin{equation}\label{mtlntmsi}
\mathrm{tr}\left( N_{{\theta}^m}(\sigma_i)\dot{D}_i\right) > \frac{1}{2}\text{ .}\end{equation}

 We define a new set of letters
 $\mathbf{A}_2$ by 
  $\mathbf{A}_2$ $:= \mathbf{A}' \cup \{1, 2\}$
  and for every ${\theta}^m \in \Theta^m$ a 
	classical-quantum channel $\dot{N}_{{\theta}^m}$ $:\mathbf{A}_2
	\rightarrow \mathcal{S}(H^m)$ by
 \[\dot{N}_{{\theta}^m}(a) = \begin{cases} N_{{\theta}^m}({F'}^m(a))& \text{if } a\in \mathbf{A}' \text{ ;}\\
 N_{{\theta}^m}(\sigma_a)& \text{if } a\in \{1, 2\} \end{cases}\text{ .}\]\vspace{0.2cm}

We define a
$p\in P(\mathbf{A}_2)$
by 
 \[p(a) = \begin{cases} p'(a)& \text{if } a\in \mathbf{A}' \text{ ;}\\
0& \text{if } a\in \{1, 2\} \end{cases}\text{ .}\]\vspace{0.2cm}
By (\ref{finbibqicbtmax}) and (\ref{1f1jmspimax1})
it holds
\begin{equation}\frac{1}{m}\Bigl(\inf_{\dot{B}_q \in Conv((\dot{B}_{\theta})_{{\theta}\in \Theta})}\chi(p;\dot{B}_q^{\otimes m})
-\max_{{\theta}^m\in \Theta^m}\chi(p;{Z'}_{{\theta}^m})\Bigr) > c\text{ ,}\label{finbibqicbtmax2}\end{equation}
where $\dot{B}_{{\theta}^m}$
are the resulting quantum states at the outputs
of $\dot{N}_{{\theta}^m}$.

By Theorem 1 of   \cite{Bo/Ca/De2},   if $m'$ is sufficiently large we can find a 
$(m', 2)$   code $\bigl(\dot{E} , \{\ddot{D}_{j} : j \in \{3, 4\}\}\bigr)$ and  positive  $\lambda$, $\zeta$
such that for all ${\theta}^{mm'}\in \Theta^{mm'}$ 
	\begin{equation}
\min_{j\in \{3,4\}}
\mathrm{tr}\left( \dot{N}_{{\theta}^{mm'}} ( \dot{E} (~|j))\ddot{D}_{j}\right)\geq 1-2^{-{m'}^{1/16}\lambda}
\label{1f1jmspimax2}\end{equation}
and 
\begin{align}&\max_{j\in \{3,4\}}\lVert
{V'}_{{\theta}^{mm'}}\circ{F'}^{mm'} \left(\dot{E} (~|j))\right) -
\Xi_{{\theta}^{mm'}} \rVert_{1}
< 2^{-\sqrt{m'}\zeta}
\label{balf1lnsl1max2}\end{align}   
for a $\Xi_{{\theta}^{mm'}}\in\mathcal{S}(H^{mm'})$ which is independent of $j$.\vspace{0.2cm}

Now for every ${\theta}^{mm'}\in \Theta^{mm'}$ we
construct a classical channel $\ddot{N}_{{\theta}^{mm'}}$
$:P(\{1,2,3,4\})$ $\rightarrow$  $P(\{1,2,3,4\})$ by

\begin{equation}
\ddot{N}_{{\theta}^{mm'}}(a\mid b) :=\begin{cases} \frac{1}{2}\mathrm{tr}\left(  N_{{\theta}^{mm'}}(\sigma_b^{\otimes m'})\dot{D}_a^{\otimes m'}\right)  & \text{if } a\in \{1, 2\} \text{ ;}\\
 \frac{1}{2}\mathrm{tr}\left(  N_{{\theta}^{mm'}}( \dot{E}^{m'} (~|b))\ddot{D}_{a}\right)  & \text{if } a\in \{3, 4\} \end{cases}\text{ .}\end{equation}
Since $ \frac{1}{2}(\dot{D}_1^{\otimes m'}+\dot{D}_2^{\otimes m'}+\ddot{D}_{3}+\ddot{D}_{4})=1_{H^{mm'}}$, 
for every $a\in \{1,2,3,4\}$ we have $\sum_{b=1}^{4}\ddot{N}_{{\theta}^{mm'}}(a\mid b) =1$, thus
this definition is valid.

By (\ref{mtlntmsi}), the deterministic capacity  of the classical arbitrarily
varying  channel
$\{\ddot{N}_{{\theta}^{mm'}}: {\theta}^{mm'}\in \Theta^{mm'}\}$ under
the maximal error criterion is positive.



By Corollary \ref{lgwtvttitbacam},
the deterministic secrecy
capacity of the
classical arbitrarily
varying quantum wiretap channel $\{\left(\ddot{N}_{{\theta}^{mm'}},{V'}_{{\theta}^{mm'}}\circ{F'}^{mm'} \right): {\theta}^{mm'}\in \Theta^{mm'}\}$
under the maximal error criterion
is positive.

A secure deterministic code 
 $\bigl({E} , \{S_{j} : j\}\bigr)$ 
for the classical arbitrarily
varying wiretap channel
$\{\left(\ddot{N}_{{\theta}^{mm'}},{V'}_{{\theta}^{mm'}}\circ{F'}^{mm'} \right): {\theta}^{mm'}\in \Theta^{mm'}\}$
defines a secure deterministic code 
 $\bigl({E} , \{{D}_{j} : j\}\bigr)$ 
for
$\{\left({N}_{{\theta}^{mm'}}\circ{F'}^{mm'},{V'}_{{\theta}^{mm'}}\circ{F'}^{mm'} \right): {\theta}^{mm'}\in \Theta^{mm'}\}$
when we define $\ddot{D}_{j} := \dot{D}_j^{\otimes m'}$ for $j\in\{1,2\}$ and
set \[\frac{1}{2}\sum_{a^m\in S_{j}} \ddot{D}_{a^m} \text{ .}\]
Here, for $a^{mm'}=(a_1,\cdots,a_{mm'})$, we set $\ddot{D}_{a^m}=\ddot{D}_{a_1}\otimes \cdots \otimes \ddot{D}_{a_{mm'}}$.
Thus the deterministic secrecy
capacity of the
 arbitrarily
varying classical-quantum wiretap channel $\{\left(N_{{\theta}^{mm'}}\circ{F'}^{mm'},{V'}_{{\theta}^{mm'}}\circ{F'}^{mm'} \right): {\theta}^{mm'}\in \Theta^{mm'}\}$
under the maximal error criterion
is positive.

Similar to  the proof of
Theorem \ref{lntvttitbaa},
we can  construct a  code
with two-part code words for $\{(N_{\theta},{V}_{\theta}'): {\theta} \in \Theta\}$. 
The first parts of the code words  are
 used to send the randomization index
to the legal receiver such that the eavesdropper  knows 
nothing about the  randomness
such that the maximal probability of the decoding error
can be kept arbitrarily small. The second parts are 
  randomness assisted secure
code words  transmitting the actual secret message
such that the maximal probability of the decoding error
can be kept arbitrarily small.
By Corollary \ref{lntvttitbaalemmax} there exists a 
randomness assisted secure
code such that its
secure deterministic capacity
under strong code concept and the 
maximal error criterion is large
or equal to 
$\frac{1}{n}\Bigl(\inf_{{B'}_q \in Conv(({B'}_{\theta})_{{\theta}\in \Theta})}\chi(p';{B'}_q^{\otimes n})
-\max_{{\theta}^n\in \Theta^n}\chi(p';{Z'}_{{\theta}^n})\Bigr)$.  
This shows the direct part of 
Theorem  \ref{lntvttitbaamax}, 1).
\end{proof}\vspace{0.2cm}

\cite{Bo/No}
shows that 
the capacities of  arbitrarily varying classical-quantum channel under
maximal error criterion and under the average error criterion are equal.
Theorem  \ref{lntvttitbaamax}
shows that 
the secrecy capacities of  arbitrarily varying quantum channel under
maximal error criterion and under the average error criterion
are equal.\vspace{0.2cm}

\begin{remark}
In
\cite{Ahl/Bj/Bo/No},
where  arbitrarily varying quantum channels  were analyzed, actually the following
result has been shown:

The deterministic capacity of an arbitrarily varying quantum channel
$\{N_{\theta} : {\theta} \in \Theta\}$  under the average error criterion
is zero if and only if for every $L\in\mathbb{N}$
and every finite set $\{\rho_1^L, \cdots, \rho_K^L  \}$
 $\subset$  $\mathcal{S}(H^{\mathfrak{P}^L})$
 there exists a $\tau$ such that for all 
$\rho^L$, ${\rho'}^L$  $\in\{\rho_1^L, \cdots, \rho_K^L  \}$,we have
\begin{equation}\sum_{\theta^L\in\Theta^L}\tau(\rho^L)(\Theta^L) {N}_{\theta^L}({\rho'}^L)
=\sum_{\theta^L\in\Theta^L}\tau({\rho'}^L)(\theta^L) {N}_{\theta^L}(\rho^L)
\text{ .}\label{stlitltrltl}\end{equation}

The deterministic capacity of an arbitrarily varying quantum channel
$\{N_{\theta} : {\theta} \in \Theta\}$  under the maximal error criterion
is zero if and only if for every  $L\in\mathbb{N}$
and every $\{\rho_1^L,  \rho_2^L  \}$
 $\subset$  $\mathcal{S}(H^{\mathfrak{P}^L})$ we have
\begin{equation}
Conv\left(\left\{N_{{\theta}^L}(\rho_1^L) : {\theta}^L\in \Theta^L\right\}\right)
\cap Conv\left(\left\{N_{{\theta}^L}(\rho_1^L) : {\theta}^L\in \Theta^L\right\}\right) \not= \emptyset
\text{ .}\label{cllntlr1l}\end{equation}

In
\cite{Ahl/Bj/Bo/No} the authors then showed that the zero-capacity-condition
for $\{N_{\theta} : {\theta} \in \Theta\}$  under the maximal error
(\ref{cllntlr1l}) and the $L$-symmetrizability in sense of
Definition \ref{symmetL} are equivalent.

In
\cite{Bo/No} the authors  showed that the zero-capacity-condition
for $\{N_{\theta} : {\theta} \in \Theta\}$  under the maximal error criterion 
(\ref{cllntlr1l}) and the $L$-symmetrizability in the sense of
Definition \ref{symmetL} is equivalent (cf. Remark \ref{icbntlswdaf}).
This is one of the key statements
by the authors of \cite{Bo/No}
to show the equality of the  capacity of an
arbitrarily varying quantum channel under
the maximal error criterion and
 its   capacity  under
the average error criterion.
\label{icwavqcwa}\end{remark}\vspace{0.2cm}

\begin{remark}
Interestingly,  
unlike for arbitrarily varying classical-quantum channels,  
for classical arbitrarily varying channels, the capacities under
maximal error criterion and under
the average error criterion are not equal.
See \cite{Ahl1} for a discussion of
the zero-capacity-condition for classical arbitrarily
varying channel under maximal error criterion and under the average error
criterion.

The capacity formula of classical arbitrarily varying channels under
maximal error criterion is still an open problem.
In \cite{Ahl0} the equality of the zero-error capacity of a
classical arbitrarily varying channel and its capacity
under maximal error criterion has been demonstrated.
In \cite{Bj/Bo/Ja/No} it has been shown that this
equality does not hold for 
arbitrarily varying classical-quantum channels.
\label{iutcoavcqcum}\end{remark}\vspace{0.2cm}

\section{Applications}
\label{appli}

\subsection{Investigation of Secrecy Capacity's Continuity}\label{iosccits}

In this section we give the proof to Corollary
\ref{eetelctit}. This shows the continuity of
the secrecy capacity of
an arbitrarily
varying quantum channel $\mathfrak{I}$
under  randomness assisted quantum coding
when the receiver's system can be
described by a finite dimensional Hilbert space.\vspace{0.2cm}

\begin{proof} 

\cite{St} shows that for a noisy quantum channel $N_{\theta}$, 
when we consider interactions with the environment as an external system,
it is possible to express it as $\mathrm{tr}_{\mathfrak{E}}\left(U_{N} \rho U_{N}^*\right)$,
 where $U_{N}$ is a  linear operator
$\mathcal{S}(H^{\mathfrak{P}})$ $\rightarrow$
$\mathcal{S}(H^{\mathfrak{QE}})$ such that $U_{N}^*U_{N}=\mathrm{id}_{H^\mathfrak{P}}$, and $\mathfrak{E}$ is the quantum
system of  the environment. By our assumption, where we follow \cite{De} using a most general protocol, 
the environment  is completely under the control of the eavesdropper.
A continuity theorem
for Stinespring Dilation has been shown in \cite{Kre/Sch/Wer}. This
continuity theorem is one of the tools our proof is based on.

By
\cite{Kre/Sch/Wer}, when
\[\max_{\rho \in \mathcal{S}(H^\mathfrak{P}) } \|N_{\theta}(\rho)- \dot{N}_{\theta}(\rho )\|_{1} <  \delta\text{ ,}\]
there exist 
Stinespring dilations $U_{N_{\theta}}$ and $U_{\dot{N}_{\theta}}$ 
for $N_{\theta}$ and $\dot{N}_{\theta}$
 such
that 
\[\max_{ \rho \in \mathcal{S}(H^\mathfrak{P}) } \|U_{N_{\theta}}(\rho \otimes id^\mathfrak{P})
- U_{\dot{N}_{\theta}}(\rho \otimes id^\mathfrak{P})\|_{1} <  \delta\]
for all ${\theta} \in \Theta$.

When there are
Stinespring dilations $U_{N_{\theta}}$ and $U_{\dot{N}_{\theta}}$ 
for $N_{\theta}$ and $\dot{N}_{\theta}$
 such
that 
\[\max_{ \rho \in \mathcal{S}(H^\mathfrak{P}) } \|U_{N_{\theta}}(\rho \otimes id^\mathfrak{P})
- U_{\dot{N}_{\theta}}(\rho \otimes id^\mathfrak{P})\|_{1} <  \delta\text{ ,}\]
by the triangle inequality,
we then have
\[\max_{ \rho \in \mathcal{S}(H^\mathfrak{P}) } \|{V}_{\theta}'(\rho)- {\dot{V}'}_{\theta}(\rho )\|_{1} <  \delta\text{ .}\]

We fix  a  finite set $\mathbf{A}$, and a
map $F:$ $\mathbf{A}$ $\rightarrow \mathcal{S}(H^{\mathfrak{P}})$.
By \cite{Bo/Ca/De2},
the secrecy capacity of
 $\{(N_{\theta}\circ F,{V}_{\theta}'\circ F):  {\theta}\in \Theta\}$ 
under  randomness assisted quantum coding is
	\[\lim_{n\rightarrow \infty} \frac{1}{n}
\max_{U\rightarrow A \rightarrow \{B_q^{\otimes n},{Z'}_{{\theta}^n}:q,{\theta}_n\}}\Bigl(\inf_{B_q \in Conv((B_{\theta})_{{\theta}\in \Theta})}\chi(p_U;B_q^{\otimes n})-
 \max_{{\theta}^n\in \Theta^n}\chi(p_U;{Z'}_{{\theta}^n})\Bigr)
	\text{ ,}\] 
and for every  $\{(\dot{N}_{\theta}\circ F,{\dot{V}'}_{\theta}\circ F):  {\theta}\in \Theta\}$
$\in$ $\mathbf{C}_{\delta}$, its secrecy capacity
 under  randomness assisted quantum coding is
	\[ \lim_{n\rightarrow \infty} \frac{1}{n}
\max_{U\rightarrow A \rightarrow \{{\dot{B}}_q^{\otimes n},{\dot{Z}'}_{{\theta}^n}:q,{\theta}_n\}}
\Bigl(\inf_{{\dot{B}}_q \in Conv(({\dot{B}}_{\theta})_{{\theta}\in \Theta})}\chi(p_U;{\dot{B}}_q^{\otimes n})
-\max_{{\theta}^n\in \Theta^n}\chi(p_U;{\dot{Z}'}_{{\theta}^n})\Bigr)
	\text{ ,}\] 
where ${\dot{B}}_{\theta}$ is the resulting  quantum state at the output of
${\dot{N}}_{\theta}\circ F$ and ${\dot{Z}'}_{\theta}$ is the resulting  quantum state  at
the output of ${\dot{V}'}_{\theta}\circ F$.\vspace{0.2cm}

To analyze $\vert\chi(p;{Z'}_{{\theta}^n})-\chi(p;{\dot{Z}'}_{{\theta}^n})\vert$,
we use the technique introduced in
\cite{Le/Sm} and apply the following lemma given in \cite{Al/Fa}.\vspace{0.2cm}

\begin{lemma}[Alicki-Fannes Inequality] Suppose we have
a composite system $\mathfrak{PQ}$ with components
 $\mathfrak{P}$ and $\mathfrak{Q}$. Let $G^\mathfrak{P}$ and $G^\mathfrak{Q}$ be the
 Hilbert space of $\mathfrak{P}$ and $\mathfrak{Q}$, respectively.  
Suppose we have two
bipartite
quantum states $\phi^\mathfrak{PQ}$ and
 $\sigma^\mathfrak{PQ}$ in $\mathcal{S}(G^\mathfrak{PQ})$ such that
$\|\phi^\mathfrak{PQ}-\sigma^\mathfrak{PQ}\|_{1} = \epsilon <1$,  it holds
\begin{equation}S(\mathfrak{P}\mid\mathfrak{Q})_{\rho}- S(\mathfrak{P}\mid\mathfrak{Q})_{\sigma}
\leq 4 \epsilon \log(d-1) - 2h(\epsilon)\text{ ,}\end{equation}
where $d$ is the dimension of $G^\mathfrak{P}$ and $h(\epsilon)$ is defined as in Lemma \ref{eq_9}.\label{AFLswhacs}
\end{lemma}\vspace{0.2cm}

We fix an $n\in\mathbb{N}$ and a ${\theta}^n$ $= $ $({\theta}_1,\cdots {\theta}_n)$ $\in\Theta^n$.
For any $a^n\in \mathbf{A}^n$ we have
\begin{align*}&\left|S\left({V'}_{{\theta}^n}( F^n(a^n))\right)-S\left({\dot{V}'}_{{\theta}^n}( F^n(a^n))\right)\right|\\
&= \biggl|\sum_{k=1}^{n} S\left({V'}_{({\theta}_1,\cdots {\theta}_{k-1})}\otimes{\dot{V}'}_{({\theta}_{k},\cdots {\theta}_n)}( F^n(a^n ))\right)
-S\left({V'}_{({\theta}_1,\cdots {\theta}_k)}\otimes{\dot{V}'}_{({\theta}_{k+1},\cdots {\theta}_n)}( F^n(a^n))\right)\biggr|\\
&\leq \sum_{k=1}^{n} \biggl|S\left({V'}_{({\theta}_1,\cdots {\theta}_{k-1})}\otimes {\dot{V}'}_{({\theta}_{k},\cdots {\theta}_n)}( F^n(a^n))\right)
-S\left({V'}_{({\theta}_1,\cdots {\theta}_k)}\otimes{\dot{V}'}_{({\theta}_{k+1},\cdots {\theta}_n)}( F^n(a^n))\right)\biggr|\text{ .}
\end{align*}\vspace{0.2cm}

For a $k\in\{1,\cdots, n\}$ and $a^n$ $= $ $(a_1,\cdots a_n)$ $\in \mathbf{A}^n$, by Lemma \ref{AFLswhacs}
we have
\begin{align}&\biggl|S\left({V'}_{({\theta}_1,\cdots {\theta}_{k+1})}\otimes {\dot{V}'}_{({\theta}_{k},\cdots {\theta}_n)}( F^n(a^n))\right)
-S\left({V'}_{({\theta}_1,\cdots {\theta}_{k+1})}\otimes{\dot{V}'}_{({\theta}_{k+1},\cdots {\theta}_n)}( F^n(a^n)\right))\biggr|\notag\\
&=\biggl|S\left({V'}_{({\theta}_1,\cdots {\theta}_k)}\otimes {\dot{V}'}_{({\theta}_{k},\cdots {\theta}_n)}( F^n(a^n))\right)\allowdisplaybreaks\notag\\
&-S\left({V'}_{({\theta}_1,\cdots {\theta}_{k-1})}\otimes {\dot{V}'}_{({\theta}_{k+1},\cdots {\theta}_n)}( F^{n-1}((a_1,\cdots a_{k-1}, a_{k+1},\cdots a_n)))\right)\allowdisplaybreaks\notag\\
&-S\left({V'}_{({\theta}_1,\cdots {\theta}_k)}\otimes{\dot{V}'}_{({\theta}_{k+1},\cdots {\theta}_n)}( F^n(a^n))\right)\allowdisplaybreaks\notag\\
&+ S\left({V'}_{({\theta}_1,\cdots {\theta}_{k-1})}\otimes {\dot{V}'}_{({\theta}_{k+1},\cdots {\theta}_n)}( F^{n-1}((a_1,\cdots a_{k-1}, a_{k+1},\cdots a_n)))\right)\biggr|\allowdisplaybreaks\notag\\
&= \biggl| S\left( {\dot{V}'}_{{\theta}_{k}}(a_k)\mid  {V'}_{({\theta}_1,\cdots {\theta}_{k-1})}\otimes {\dot{V}'}_{({\theta}_{k+1},\cdots {\theta}_n)}
( F^{n-1}((a_1,\cdots a_{k-1}, a_{k+1},\cdots a_n)))  \right)\allowdisplaybreaks\notag\\
&-S\left( {V'}_{{\theta}_{k}}(a_k)\mid  {V'}_{({\theta}_1,\cdots {\theta}_{k-1})}\otimes {\dot{V}'}_{({\theta}_{k+1},\cdots {\theta}_n)}
( F^{n-1}((a_1,\cdots a_{k-1}, a_{k+1},\cdots a_n)))  \right) \biggr|\allowdisplaybreaks\notag\\
&\leq  4 \delta \log(\dim H^\mathfrak{E}-1) - 2\cdot h(\delta) \notag\\
&=  4 \delta \log(\dim H^\mathfrak{Q}-1) - 2\cdot h(\delta)\text{ .}
\label{bslvt1ctk}\end{align}\vspace{0.2cm}

Thus, 
\begin{equation}\left|S\left({V'}_{{\theta}^n}( F^n(a^n))\right)-
S\left({\dot{V}'}_{{\theta}^n}( F^n(a^n))\right)\right|\leq  4 n\delta \log(\dim H^\mathfrak{Q}-1) - 2n\cdot h(\delta) \text{ .}
\label{lswtnanr}\end{equation}

For any probability distribution $p\in P(\mathbf{A})$, $n\in\mathbb{N}$, and ${\theta}^n\in\Theta^n$ we have
\begin{align}&
\vert \chi(p;{Z'}_{{\theta}^n}) - \chi(p;{\dot{Z}'}_{{\theta}^n}) \vert \notag\\
&=\Bigl\vert S(\sum_{a}p(a){V'}_{{\theta}^n}( F^n(a)))- \sum_{a}p(a)S({V'}_{{\theta}^n}( F^n(a)))  \notag\\
&- S(\sum_{a}p(a){\dot{V}'}_{{\theta}^n}( F^n(a))) + S(\sum_{a}p(a){\dot{V}'}_{{\theta}^n}( F^n(a))) \Bigr\vert \notag\\
&\leq\Bigl\vert S(\sum_{a}p(a){V'}_{{\theta}^n}( F^n(a)))- S(\sum_{a}p(a){\dot{V}'}_{{\theta}^n}(F^n(a))) \Bigr\vert \notag\\
&+ \Bigl\vert \sum_{a}p(a)S({\dot{V}'}_{{\theta}^n}( F^n(a)))  - \sum_{a}p(a)S({\dot{V}'}_{{\theta}^n}( F^n(a))) \Bigr\vert \notag\\
&\leq 8 n\delta \log(\dim H^\mathfrak{Q}-1) - 4n\cdot h(\delta)\text{ .}\label{lbndln14n}
\end{align}

\vspace{0.2cm}

We fix a  probability distribution $q$ on $\Theta$, a probability distribution $p\in P(\mathbf{A})$,
  and an $n\in\mathbb{N}$. By Lemma \ref{eq_9} 
we have

\begin{align}&
\vert \chi(p;B_q) - \chi(p;{\dot{B}}_q) \vert \notag\\
&=\Bigl\vert \sum_{\theta}q(t)S(\sum_{a}p(a)N_{\theta}(F(a)))- \sum_{\theta}\sum_{a}q(t)p(a)S(N_{\theta}(F(a)))  \notag\\
&- \sum_{\theta}q(t)S(\sum_{a}p(a){\dot{N}}_{\theta}(F(a))) + S(\sum_{\theta}\sum_{a}q(t)p(a){\dot{N}}_{\theta}(F(a))) \Bigr\vert \notag\\
&\leq\Bigl\vert \sum_{\theta}q(t)S(\sum_{a}p(a)N_{\theta}(F(a)))- \sum_{\theta}q(t)S(\sum_{a}p(a){\dot{N}}_{\theta}(F(a))) \Bigr\vert \notag\\
&+ \Bigl\vert \sum_{\theta}\sum_{a}q(t)p(a)S(N_{\theta}(F(a)))  - S(\sum_{\theta}\sum_{a}q(t)p(a){\dot{N}}_{\theta}(F(a))) \Bigr\vert \notag\\
&\leq 8 \delta \log(\dim H^\mathfrak{Q}-1) - 4\cdot h(\delta)\text{ .}
\end{align}\vspace{0.2cm}

Thus for any probability distribution $q$ on $\Theta$, 
$n\in\mathbb{N}$,  $p\in P(\mathbf{A})$, ${\theta}^n\in\Theta^n$ we have for
all $\{{\dot{N}}_{\theta}: {\theta} \in \Theta\}$
$\in$ $\mathbf{C}_{\delta}$ 
\begin{align}&
\Bigl\vert  (\chi(p;B_q)-\frac{1}{n}\chi(p;{Z'}_{{\theta}^n}))
-  (\chi(p;{\dot{B}}_q)-\frac{1}{n}\chi(p;{\dot{Z}'}_{{\theta}^n}))\Bigr\vert\notag\\
&\leq 
16 \delta \log(\dim H^\mathfrak{Q}-1) -
 8\cdot h(\delta)
	\text{ .}\end{align}
	
For any positive $\epsilon$ we can find a positive $\delta$
such that $16 \delta \log(\dim H^\mathfrak{Q}-1)$ 
 $-$ $8 \cdot h(\delta)$ 
$\leq$ $\epsilon$.

When $\inf_{q \in P(\theta)}\chi(p;B_q)- \max_{{\theta}^n\in \Theta^n}\chi(p;{Z'}_{{\theta}^n})$
achieves its maximum in $F$ and $p$, and when
$\inf_{q' \in P(\theta)}\chi(\acute{p};{\acute{\dot{B}}}_{q'})
-\frac{1}{n}\max_{{{\theta}^n}'\in \Theta^n}\chi(\acute{p};{\acute{\dot{Z}}'}_{{{\theta}^n}'})$
achieves its maximum in $\acute{F}$ and $\acute{p}$ for some
map $\acute{F}:$ $\mathbf{A}$ $\rightarrow \mathcal{S}(H^{\mathfrak{P}})$ and a $\acute{p}\in P(\mathbf{A})$,
where ${\acute{\dot{B}}}_{\theta}$ is the resulting  quantum state at the output of
${\dot{N}}_{\theta}\circ \acute{F}$ and ${\acute{\dot{Z}}'}_{\theta}$ is the resulting  quantum state  at
the output of ${\dot{V}'}_{\theta}\circ \acute{F}$, the following inequality holds.

For all $n\in\mathbb{N}$ and any positive $\epsilon$ we can find a positive $\delta$
such that for
all $\{{\dot{N}}_{\theta}: {\theta} \in \Theta\}$
$\in$ $\mathbf{C}_{\delta}$ 
	\begin{align}&\Bigl\vert
\max_{p}(\inf_{q \in P(\theta)}\chi(p;B_q)- \max_{{\theta}^n\in \Theta^n}\chi(p;{Z'}_{{\theta}^n}))\notag\\
&	- \max_{\acute{p}}(\inf_{q' \in P(\theta)}\chi(\acute{p};{\acute{\dot{B}}}_{q'})
-\frac{1}{n}\max_{{{\theta}^n}'\in \Theta^n}\chi(\acute{p};{\acute{\dot{Z}}'}_{{{\theta}^n}'}))\Bigr\vert\notag\\
&\leq 	3\epsilon\text{ ,}\label{bv1nmbibq1}\end{align}
since elsewise we would have

\begin{align*}&\inf_{q \in P(\theta)}\chi(p;{\dot{B}}_q)
-\frac{1}{n}\max_{{\theta}^n\in \Theta^n}\chi(p;{\dot{Z}'}_{{\theta}^n})\\
&>\inf_{q' \in P(\theta)}\chi(\acute{p};{\acute{\dot{B}}}_{q'})
-\frac{1}{n}\max_{{{\theta}^n}'\in \Theta^n}\chi(\acute{p};;{\acute{\dot{Z}}'}_{{{\theta}^n}'})\text{ .}\end{align*}
	(\ref{bv1nmbibq1}) shows Corollary \ref{eetelctit}.
\end{proof}\vspace{0.2cm}

\begin{remark}
In
\cite{Bo/Sch/Po} the continuity of
the secrecy capacity of a classical arbitrarily varying channel 
under  randomness assisted quantum coding has been shown
in a similar way (cf.  Lemma \ref{facavwcmwtmvt}). That proof is only valid when the alphabet that
the eavesdropper uses is finite. Similarly, our proof of Corollary \ref{eetelctit} only
holds when the quantum system of the environment as eavesdropper has a finite dimension.
However, the quantum system of the environment as eavesdropper can be always
chosen such that its dimension is upper bounded by the dimension of the receiver's system
(cf. \ref{bslvt1ctk}).\label{ictcotscoacavc}
\end{remark}\vspace{0.2cm}

Since for the classical case we cannot always expect that
the eavesdropper is restrict to a finite alphabet,
we may use another way to overcome the problem (cf. Remark \ref{ictcotscoacavc})  by
giving an alternative proof to
 Lemma \ref{facavwcmwtmvt}:\vspace{0.2cm}

\begin{proof} 
For sets $\mathbf{A}$, $\mathbf{B}$, $\mathbf{C}$ and a finite index set
 $\Theta$, let $\{(\mathsf{W}_{\theta},\mathsf{V}_{\theta}): \theta \in \Theta\}$
be a classical arbitrarily
varying wiretap channel. 
Here
$\mathsf{W}_{\theta}$  $:$
$P(\mathbf{A}) \rightarrow P(\mathbf{B})$ and $\mathsf{V}_{\theta}$ $:$ $P(\mathbf{A})
\rightarrow P(\mathbf{C})$.
By \cite{Bj/Bo/So} we have
\begin{equation}
C_s(\{(\mathsf{W}_{\theta},\mathsf{V}_{\theta}): \theta \in \Theta\};r)
=\max_{\mathcal{U} \rightarrow A
\rightarrow (BZ)_t} \min_{\theta\in\Theta} I(p_U,B_{\theta}) -  \max_{\theta\in\Theta} I(p_U,Z_{\theta})  \text{ .}
 \label{tzuptitmbtpi}\end{equation}
The maximum is taken over all random
variables  that satisfy the Markov chain relationships: $U \rightarrow A
\rightarrow (BZ)_t$. Here $B_t$ are the resulting
random variables at the output of legal receiver channels, and
$Z_t$ are the resulting random variables at the output of wiretap
channels. $U$ is a random
variable taking values on some finite set $\mathbf{U}$
with probability  distribution $p_U$.

 Assume we have a classical arbitrarily
varying  wiretap channel
 $\{({\mathsf{W}'}_{\theta},{\mathsf{V}'}_{\theta}): \theta \in \Theta\}$, where
${\mathsf{W}'}_{\theta}$  $:$
$P(\mathbf{A}) \rightarrow P(\mathbf{B})$ and ${\mathsf{V}'}_{\theta}$ $:$ $P(\mathbf{A})
\rightarrow  P(\mathbf{C})$, and a positive $\delta$
 such
that
\[\max_{ a\in \mathbf{A}} \|\mathsf{W}_{\theta}(a)- {\mathsf{W}'}_{\theta}(a)\|_1 <  \delta\]
and
\[\max_{ a\in \mathbf{A}} \|\mathsf{V}_{\theta}(a)- {\mathsf{V}'}_{\theta}(a)\|_1 <  \delta\]
for all $\theta \in \theta$.

We fix a  $\theta\in\Theta$ and a random variable $X$ distributed on $\mathbf{A}$ according to a probability distribution $p_x$.
Let $Z_{\theta}$ and ${Z'}_{\theta}$, distributed on $\mathbf{Z}$, be the resulting random variables  at the output of ${\mathsf{V}}_{\theta}$ and
${\mathsf{V}'}_{\theta}$, respectively.
 We have
\begin{align}&|I(X,Z_{\theta})-I(X,{Z'}_{\theta})|\notag\\
&= |H(X)-H(X|Z_{\theta})-H(X)+H(X|{Z'}_{\theta})|\notag\\
&=|H(X|Z_{\theta})-H(X|{Z'}_{\theta})|\notag\\
&= \sum_{z\in\mathbf{Z}} \left\vert {\mathsf{V}}_{\theta}(z\mid p_X)- {\mathsf{V}'}_{\theta}(z\mid p_X)\right\vert \cdot H(X|Z_{\theta}=z)\notag\\
&\leq \epsilon \max_{z\in\mathbf{Z}} H(X|Z_{\theta}=z)\notag\\
&\leq \epsilon \log|\mathbf{A}|
\text{ .}\label{ixztixztnhx}\end{align}

Let $B_{\theta}$ and ${B'}_{\theta}$ be the resulting random variables  at the output of ${\mathsf{W}}_{\theta}$ and
${\mathsf{W}'}_{\theta}$, respectively.
Similar to (\ref{ixztixztnhx}), we have 
\begin{equation}|I(X,B_{\theta})-I(X,{B'}_{\theta})| \leq \epsilon \log|\mathbf{A}|\label{ixztixztnhx2}\text{ .}\end{equation}

Combining
(\ref{tzuptitmbtpi}) with (\ref{ixztixztnhx}) and (\ref{ixztixztnhx2})
we have
\begin{equation}|C_s(\{(\mathsf{W}_{\theta},\mathsf{V}_{\theta}): \theta \in \Theta\};r)
	-C_s(\{(({\mathsf{W}'}_{\theta},{\mathsf{V}'}_{\theta}): \theta \in \Theta\};r)|\leq  2\epsilon \log|\mathbf{A}|\text{ .}\end{equation}

This shows that in the classical case also, the 
continuity does not depend on some unknown strategy
of the eavesdropper.
\end{proof}\vspace{0.2cm}


Corollary \ref{ftsdmiait}
and Corollary \ref{ltbafsanttit}
are direct consequences of 
Corollary \ref{eetelctit}  when
we apply the techniques in 
\cite{Bo/Ca/De3} for the discontinuity points of classical-quantum channels
to Theorem \ref{lntvttitbaa} and
Corollary \ref{eetelctit}.

Now we are going to prove
Corollary  \ref{ftsdmiait}.\vspace{0.2cm}

\begin{proof}
At first we assume that the secrecy capacity of 
$\{N_{\theta}: {\theta}\in \Theta\}$  under
 randomness assisted quantum coding
is positive and $\mathsf{F}(\{N_{\theta}: \theta\})=0$.
We choose a positive $\epsilon$
such that $C_s(\{N_{\theta}: \theta\};r)$
$-\epsilon$ $:=C$ $>0$.
By Corollary \ref{eetelctit},  
the secrecy capacity  under
 randomness assisted quantum coding
is continuous. Thus
there exist a 
positive $\delta$ such that  
for all 
$\{N_{\theta}': {\theta}\in \Theta\}$
$\in$ $\mathbf{C}_{\delta}$ we have
\[ C_s\left(\{N_{\theta}': {\theta}\in \Theta\};r\right)\geq 
C_s\left (\{N_{\theta}: \theta\};r\right)-\epsilon \text{ .}\]\vspace{0.15cm}

Now we assume that there is a
 $\{N_{\theta}'': {\theta}\in \Theta\}$
$\in$ $\mathbf{C}_{\delta}$ such that
$\mathsf{F}(\{N_{\theta}'': \theta\})>0$. This means that
$\{N_{\theta}'': \theta\}$ is not  symmetrizable.
By Theorem \ref{lntvttitbaa}
it holds
\[ C_s\left(\{N_{\theta}'': {\theta}\in \Theta\}\right)
= C_s\left (\{N_{\theta}'': \theta\};r\right)\geq C > 0 \text{ .}\]

$\{{N}_{\theta}: \theta\}$ is symmetrizable since $\mathsf{F}(\{N_{\theta}: \theta\})=0$.
 By Theorem \ref{lntvttitbaa},  
 \[C_s(\{N_{\theta}: {\theta}\in \Theta\})=0\text{ .}\] Therefore 
the deterministic secrecy capacity
is discontinuous at
$\{N_{\theta}: {\theta}\in \Theta\}$ when 1) and 2) hold.\vspace{0.2cm}

Now let us consider the case when  the deterministic secrecy capacity
is discontinuous at
$\{N_{\theta}: {\theta}\in \Theta\}$.\vspace{0.15cm}

We fix a  $L\in\mathbb{N}$, a
$\tau\in C(\Theta^L\mid \mathcal{S}({H}^{\mathfrak{A}^L}))$,
and $\rho^L$, ${\rho^L}'$ $\in\mathcal{S}({H}^{\mathfrak{A}^L})$.
The map  
\[\{N_{\theta}: {\theta}\in \Theta\}\rightarrow 
\frac{1}{2^L}\left\|\sum_{{\theta^L}\in\Theta^L}\tau(\theta^L\mid \rho^L)N_{\theta^L}({{\rho^L}'})-\sum_{{\theta^L}\in\Theta^L}\tau(\theta^L\mid {{\rho^L}'})N_{\theta^L}(\rho^L)\right\|_1\]
is continuous
in the following  sense:
When  
\[\frac{1}{2^L}\Bigl\|\sum_{{\theta^L}\in\Theta^L}\tau(\theta^L\mid \rho^L)
N_{\theta^L}({{\rho^L}'}) -\sum_{{\theta^L}\in\Theta^L}\tau(\theta^L\mid {{\rho^L}'})N_{\theta^L}(\rho^L)\Bigr\|_1=C\] 
  holds,   then
for every positive $\delta$ and any
$\{N_{\theta}': {\theta}\in \Theta\}$
$\in$ $\mathbf{C}_{\delta}$ we have
\[\left| \frac{1}{2^L}\|\sum_{{\theta^L}\in\Theta^L}\tau(\theta^L\mid \rho^L)N_{\theta^L}'({{\rho^L}'})-\sum_{{\theta^L}\in\Theta^L}\tau(\theta^L\mid {{\rho^L}'})N_{\theta^L}'(\rho^L)\|_1 -C\right|\leq 2\delta\text{ .}\]
Thus if for a $\tau\in C(\Theta^L\mid \mathbf{A})$
we have  
\[\frac{1}{2^L}\Bigl\|\sum_{{\theta^L}\in\Theta^L}\tau(\theta^L\mid a^L)N_{\theta^L}({{\rho^L}'})
-\sum_{{\theta^L}\in\Theta^L}\tau(\theta^L\mid {{\rho^L}'})N_{\theta^L}(\rho^L)\Bigr\|_1 =C >0\] 
for all $\rho^L$, ${\rho^L}'$ $\in\mathbf{A}$, 
we also have 
\[\frac{1}{2^L}\left\|\sum_{{\theta^L}\in\Theta^L}\tau(\theta^L\mid \rho^L)N_{\theta^L}'({{\rho^L}'})-\sum_{{\theta^L}\in\Theta^L}\tau(\theta^L\mid {{\rho^L}'}){N}_{\theta^L}'(\rho^L)\right\|_1 \geq C-2\delta\text{ .}\]
When  $\mathsf{F}(\{N_{\theta}: \theta\})>0$ holds,
then there is at least one $L\in \mathbb{N}$ such that 
 $\mathsf{F}_L(\{N_{\theta}: \theta\})>0$.
We can find a 
 positive $\delta$ such that $\mathsf{F}_L(\{{W'}_{\theta}: \theta\})>0$, and thus
$\mathsf{F}(\{{W'}_{\theta}: \theta\})>0$) holds for all
$\{N_{\theta}': {\theta}\in \Theta\}$
$\in$ $\mathbf{C}_{\delta}$.
By  Theorem \ref{lntvttitbaa},   it
holds
\[ C_s\left(\{N_{\theta}': {\theta}\in \Theta\}\right)
= C_s\left (\{N_{\theta}': \theta\};r\right)\geq C > 0 \text{ .}\]
By Corollary \ref{eetelctit},   $ C_s\left (\{N_{\theta}': \theta\};r\right)$
is continuous. 

Therefore, when  the deterministic secrecy capacity
is discontinuous at
$\{N_{\theta}: {\theta}\in \Theta\}$, $\mathsf{F}(\{N_{\theta}: \theta\})$ cannot be positive.\vspace{0.15cm}

We consider now that
$\mathsf{F}(\{N_{\theta}: \theta\})=0$ holds. By  Theorem \ref{lntvttitbaa},  
\[C_s(\{N_{\theta}: {\theta}\in \Theta\} =0\text{ .}\]
When for every  $\{N_{\theta}': {\theta}\in \Theta\}$
$\in$ $\mathbf{C}_{\delta}$ we have $\mathsf{F}(\{{N}_{\theta}': \theta\})=0$,
then by  Theorem \ref{lntvttitbaa}
 \[C_s(\{N_{\theta}': {\theta}\in \Theta\})=0\text{ ,}\]
and the deterministic secrecy capacity
is thus continuous at
$\{N_{\theta}: {\theta}\in \Theta\}$. 

Therefore, when  the deterministic secrecy capacity
is discontinuous at
$\{N_{\theta}: {\theta}\in \Theta\}$,  for every  positive $\delta$ there is a
 $\{N_{\theta}': {\theta}\in \Theta\}$
$\in$ $\mathbf{C}_{\delta}$ such that
$\mathsf{F}(\{{N}_{\theta}': \theta\})>0$.\vspace{0.15cm}

When for every  positive $\delta$ there is a
 $\{N_{\theta}': {\theta}\in \Theta\}$
$\in$ $\mathbf{C}_{\delta}$ such that
$\mathsf{F}(\{{N}_{\theta}': \theta\})>0$ and $C_s(\{N_{\theta}: {\theta}\in \Theta\},r)$ $=0$
holds,
then by  Theorem \ref{lntvttitbaa} we have
\[C_s(\{N_{\theta}': {\theta}\in \Theta\})=C_s(\{N_{\theta}': {\theta}\in \Theta\},r)\text{ ,}\]
and the deterministic secrecy capacity
is  continuous at
$\{N_{\theta}: {\theta}\in \Theta\}$. 

Therefore, when  the deterministic secrecy capacity
is discontinuous at
$\{N_{\theta}: {\theta}\in \Theta\}$, $C_s(\{N_{\theta}: {\theta}\in \Theta\},r)$
must be positive.
\end{proof}\vspace{0.2cm}

Corollary \ref{tscaavqq} states that 
the  deterministic secrecy capacity of an arbitrarily varying  quantum  channel is, in general, not continuous.
We show Corollary \ref{tscaavqq} by giving an example.
\vspace{0.2cm}

 \begin{example}\label{lt1122awtit}

Let $\Theta:=\{1^{(+)},1^{(-)},2^{(+)},2^{(-)}\}$ and  $\{W_{\theta}: {\theta}\in \Theta\}$
 be defined as in Example \ref{ltsqsbphma} in Section \ref{aewtdscoaavqwc}.
Let $\{|0\rangle^{\mathfrak{A}}, |1\rangle^{\mathfrak{A}} \}$ be a set of orthonormal vectors
on ${H}^{\mathfrak{A}}$. Let $\{|0\rangle^{\mathfrak{B}},$ $|1\rangle^{\mathfrak{B}}, $ $
|2\rangle^{\mathfrak{B}},$ $|3\rangle^{\mathfrak{B}},$ $|4\rangle^{\mathfrak{B}},$ $|5\rangle^{\mathfrak{B}}
|6\rangle^{\mathfrak{B}}\}$ be a set of orthonormal vectors
on ${H}^{\mathfrak{B}}$. Let $\lambda$ be $\in [0,1]$.
 We define a  map $W_{1^{(+)}}^{\lambda}$:
$\mathcal{S}({H}^{\mathfrak{A}})$ $\rightarrow$ $\mathcal{S}({H}^{\mathfrak{B}})$
by\begin{align*}& c_1|0\rangle\langle 0|^{\mathfrak{A}} + c_2|1\rangle\langle 1|^{\mathfrak{A}}
+ c_3|1\rangle\langle 0|^{\mathfrak{A}} + \overline{c_3}|0\rangle\langle 1|^{\mathfrak{A}}\\
& \rightarrow (1-\lambda) c_1|0\rangle\langle 0|^{\mathfrak{B}} + (1-\lambda) c_2|1\rangle\langle 1|^{\mathfrak{B}}\\
&~~~~+ (1-\lambda) c_3|1\rangle\langle 0|^{\mathfrak{B}} + (1-\lambda) \overline{c_3}|0\rangle\langle 1|^{\mathfrak{B}}
 + \lambda(c_1+c_2)|3\rangle\langle 3|^{\mathfrak{B}} \\
&= (1-\lambda) W_{1^{(+)}} \Bigl(c_1|0\rangle\langle 0|^{\mathfrak{A}} + c_2|1\rangle\langle 1|^{\mathfrak{A}}
+ c_3|1\rangle\langle 0|^{\mathfrak{A}} + \overline{c_3}|0\rangle\langle 1|^{\mathfrak{A}}\Bigr) + \lambda(c_1+c_2) |3\rangle\langle 3|^{\mathfrak{B}}\text{ ,}\end{align*}
 a  map $W_{1^{(-)}}^{\lambda}$:
$\mathcal{S}({H}^{\mathfrak{A}})$ $\rightarrow$ $\mathcal{S}({H}^{\mathfrak{B}})$
by\begin{align*}& c_1|0\rangle\langle 0|^{\mathfrak{A}} + c_2|1\rangle\langle 1|^{\mathfrak{A}}
+ c_3|1\rangle\langle 0|^{\mathfrak{A}} + \overline{c_3}|0\rangle\langle 1|^{\mathfrak{A}}\\
& \rightarrow  (1-\lambda) c_1|0\rangle\langle 0|^{\mathfrak{B}} +  (1-\lambda) c_2|1\rangle\langle 1|^{\mathfrak{B}}\\
&~~~~-  (1-\lambda) c_3|1\rangle\langle 0|^{\mathfrak{B}} -  (1-\lambda) \overline{c_3}|0\rangle\langle 1|^{\mathfrak{B}}+ \lambda(c_1+c_2) |4\rangle\langle 4|^{\mathfrak{B}}\\
&= (1-\lambda) W_{1^{(-)}} \Bigl(c_1|0\rangle\langle 0|^{\mathfrak{A}} + c_2|1\rangle\langle 1|^{\mathfrak{A}}
+ c_3|1\rangle\langle 0|^{\mathfrak{A}} + \overline{c_3}|0\rangle\langle 1|^{\mathfrak{A}}\Bigr) + \lambda(c_1+c_2) |4\rangle\langle 4|^{\mathfrak{B}} \text{ ,}\end{align*}
 a map $W_{2^{(+)}}^{\lambda}$:
$\mathcal{S}({H}^{\mathfrak{A}})$ $\rightarrow$ $\mathcal{S}({H}^{\mathfrak{B}})$
by\begin{align*}& c_1|0\rangle\langle 0|^{\mathfrak{A}} + c_2|1\rangle\langle 1|^{\mathfrak{A}}
+ c_3|1\rangle\langle 0|^{\mathfrak{A}} + \overline{c_3}|0\rangle\langle 1|^{\mathfrak{A}}\\
& \rightarrow  (1-\lambda) c_1|1\rangle\langle 1|^{\mathfrak{B}} + (1-\lambda)  c_2|2\rangle\langle 2|^{\mathfrak{B}}\\
&~~~~+  (1-\lambda) c_3|2\rangle\langle 1|^{\mathfrak{B}} +  (1-\lambda) \overline{c_3}|1\rangle\langle 2|^{\mathfrak{B}}+\lambda  (c_1+c_2)|5\rangle\langle 5|^{\mathfrak{B}} \\
&= (1-\lambda) W_{2^{(+)}} \Bigl(c_1|0\rangle\langle 0|^{\mathfrak{A}} + c_2|1\rangle\langle 1|^{\mathfrak{A}}
+ c_3|1\rangle\langle 0|^{\mathfrak{A}} + \overline{c_3}|0\rangle\langle 1|^{\mathfrak{A}}\Bigr) +\lambda  (c_1+c_2)|5\rangle\langle 5|^{\mathfrak{B}} \text{ ,}\end{align*}
and a  map $W_{2^{(-)}}^{\lambda}$:
$\mathcal{S}({H}^{\mathfrak{A}})$ $\rightarrow$ $\mathcal{S}({H}^{\mathfrak{B}})$
by\begin{align*}& c_1|0\rangle\langle 0|^{\mathfrak{A}} + c_2|1\rangle\langle 1|^{\mathfrak{A}}
+ c_3|1\rangle\langle 0|^{\mathfrak{A}} + \overline{c_3}|0\rangle\langle 1|^{\mathfrak{A}}\\
& \rightarrow  (1-\lambda) c_1|1\rangle\langle 1|^{\mathfrak{B}} +  (1-\lambda) c_2|2\rangle\langle 2|^{\mathfrak{B}}\\
&~~~~-  (1-\lambda) c_3|2\rangle\langle 1|^{\mathfrak{B}} -  (1-\lambda) \overline{c_3}|1\rangle\langle 2|^{\mathfrak{B}} + \lambda (c_1+c_2)|6\rangle\langle 6|^{\mathfrak{B}}\\
&= (1-\lambda) W_{2^{(-)}} \Bigl(c_1|0\rangle\langle 0|^{\mathfrak{A}} + c_2|1\rangle\langle 1|^{\mathfrak{A}}
+ c_3|1\rangle\langle 0|^{\mathfrak{A}} + \overline{c_3}|0\rangle\langle 1|^{\mathfrak{A}}\Bigr) + \lambda (c_1+c_2)|6\rangle\langle 6|^{\mathfrak{B}} \text{ .}\end{align*}\vspace{0.2cm}

These maps are obviously linear and trace-preserving. 
They are also completely positive: When 
$\begin{pmatrix}
c_1 & c_3\\
\overline{c_3} & c_2
 \end{pmatrix}$ is positive semidefinite then all its leading principal minors
are non-negative, therefore $c_1$ and $c_2$ are non-negative.
\[\begin{pmatrix}
(1-\lambda)c_1 & (1-\lambda)c_3&0\\
(1-\lambda)\overline{c_3} & (1-\lambda)c_2&0\\
0&0&\lambda (c_1+c_2)
 \end{pmatrix}\] is a matrix with positive semidefinite matrices on
its diagonal, and thus, also positive semidefinite.\vspace{0.2cm}

\it i)
If $\lambda$ is not equal to zero, then $\{W_{\theta}^{\lambda}:\theta\in\Theta\}$ is not symmetric. 
\rm\vspace{0.2cm}

Let $\upsilon$ and $\phi$ be two arbitrary quantum states $\in \mathcal{S}({H}^{\mathfrak{A}})$.
We suppose that there are two  distributions $p_{\upsilon}$ and $p_{\phi}$
 on $\Theta$ such that
\[\sum_{\theta\in\Theta} p_{\phi}(\theta)\cdot W_{\theta}^{\lambda} (\upsilon) = \sum_{\theta\in\Theta} p_{\upsilon}(\theta) \cdot W_{\theta}^{\lambda} (\phi)\text{ .}\]
We have
\begin{align}& (1-\lambda)\sum_{\theta\in\Theta} p_{\upsilon}(\theta)W_{\theta}(\phi) + \lambda [ p_{\upsilon}(1^{(+)})|3\rangle\langle 3|^{\mathfrak{B}}
+ p_{\upsilon}(1^{(-)})|4\rangle\langle 4|^{\mathfrak{B}}\notag\\
& ~+ p_{\upsilon}(2^{(+)})|5\rangle\langle 5|^{\mathfrak{B}} 
+ p_{\upsilon}(2^{(-)})|6\rangle\langle 6|^{\mathfrak{B}}]\notag\\
&= (1-\lambda)\sum_{\theta\in\Theta}p_{\phi} (\theta) W_{\theta}(\upsilon) + \lambda [ p_{\phi}(1^{(+)})|3\rangle\langle 3|^{\mathfrak{B}}
+ p_{\phi}(1^{(-)})|4\rangle\langle 4|^{\mathfrak{B}}\notag\\
& ~+ p_{\phi}(2^{(+)})|5\rangle\langle 5|^{\mathfrak{B}} 
+ p_{\phi}(2^{(-)})|6\rangle\langle 6|^{\mathfrak{B}}]\text{ .}\label{1lspujw}\end{align}

If $\lambda\not= 0$, (\ref{1lspujw})  implies that 
\[p_{\phi}(\theta)=p_{\upsilon}(\theta)\]
for all $\theta\in\Theta$.

If  $\{W_{\theta}^{\lambda}:\theta\in\Theta\}$ was  symmetric, i.e., if there existed a
parametrized set of distributions $\{p_{\rho}(\cdot):
 \rho \in \mathcal{S}({H}^{\mathfrak{A}})\}$ on $\Theta$ such that for all $\rho$, ${\rho'}\in \mathcal{S}({H}^{\mathfrak{A}})$,
and if $\lambda\not= 0$
$\sum_{\theta\in\Theta}p_{\rho'}(\theta) W_{\theta}^{\lambda}(\rho)$ $=$ $\sum_{\theta\in\Theta}p_{\rho} (\theta) W_{\theta}^{\lambda}(\rho')$,
by (\ref{1lspujw}) there would be a distribution $p'$ on $\Theta$ such that
\[p'(\theta)=p_{\rho}(\theta)\]
 for all $\theta\in\Theta$ and $\rho\in \mathcal{S}({H}^{\mathfrak{A}})$.

But there is 
clearly no such distribution $p'$ such that 
$\sum_{\theta\in\Theta}p'(\theta) W_{\theta}^{\lambda}(\rho)$ $=$ $\sum_{\theta\in\Theta}p'(\theta) W_{\theta}^{\lambda}(\rho')$.
For example 
\[\sum_{\theta\in\Theta}p'(\theta) W_{\theta}^{\lambda}(|0\rangle\langle 0|^{\mathfrak{A}})
=\sum_{\theta\in\Theta}p'(\theta) W_{\theta}^{\lambda}(|1\rangle\langle 1|^{\mathfrak{A}})\] implies
that \[p'(1^{(+)})=-p'(1^{(-)})\] and \[p'(2^{(+)})=-p'(2^{(-)})\text{ ,}\] but
\begin{align*}&p'(1^{(+)}) W_{1^{(+)}}^{\lambda}(|0\rangle\langle 0|^{\mathfrak{A}}) -
p'(1^{(+)}) W_{1^{(-)}}^{\lambda}(|0\rangle\langle 0|^{\mathfrak{A}})\\
&~+p'(2^{(+)}) W_{2^{(+)}}^{\lambda}(|0\rangle\langle 0|^{\mathfrak{A}}) -
p'(2^{(+)}) W_{2^{(-)}}^{\lambda}(|0\rangle\langle 0|^{\mathfrak{A}})\\
&=p'(1^{(+)}) W_{1^{(+)}}^{\lambda}(\frac{1}{2}[|0\rangle\langle 0|^{\mathfrak{A}} + |1\rangle\langle 1|^{\mathfrak{A}}
+ |1\rangle\langle 0|^{\mathfrak{A}} + |0\rangle\langle 1|^{\mathfrak{A}}]) \\
&-p'(1^{(+)}) W_{1^{(-)}}^{\lambda}(\frac{1}{2}[|0\rangle\langle 0|^{\mathfrak{A}} + |1\rangle\langle 1|^{\mathfrak{A}}
+ |1\rangle\langle 0|^{\mathfrak{A}} + |0\rangle\langle 1|^{\mathfrak{A}}])\\
&~+p'(2^{(+)}) W_{2^{(+)}}^{\lambda}(\frac{1}{2}[|0\rangle\langle 0|^{\mathfrak{A}} + |1\rangle\langle 1|^{\mathfrak{A}}
+ |1\rangle\langle 0|^{\mathfrak{A}} + |0\rangle\langle 1|^{\mathfrak{A}}]) \\
&~-p'(2^{(+)}) W_{2^{(-)}}^{\lambda}(\frac{1}{2}[|0\rangle\langle 0|^{\mathfrak{A}} + |1\rangle\langle 1|^{\mathfrak{A}}
+ |1\rangle\langle 0|^{\mathfrak{A}} + |0\rangle\langle 1|^{\mathfrak{A}}])\end{align*}
implies
that \[p'(1^{(+)})=-p'(1^{(-)})=p'(2^{(+)})=-p'(2^{(-)})=0\text{ .}\]
This is of course a contradiction to the assumption  that $p'$  is a distribution on $\Theta$.
Thus $\{W_{\theta}^{\lambda}:\theta\in\Theta\}$ is not symmetric.\vspace{0.2cm}

\it ii) There is a positive $\eta$ and a positive $\lambda'$
such that for any positive $\gamma\leq \lambda'$ 
the   randomness assisted secrecy capacity of  
 $\{W_{\theta}^{\lambda}:\theta\in\Theta\}$ is larger than $\eta$.  
\rm \vspace{0.2cm}

Let the sender's
encoding be restricted to $\{|0\rangle^{\mathcal{A}}$, $|1\rangle^{\mathcal{A}}\}$.
We have a classical-quantum
channel. By \cite{Bl/Ca}, the classical-quantum  secrecy capacity of 
 $\{W_{\theta}^{\lambda}:\theta\in\Theta\}$ can be lower bounded by
\[\max_{P\text{ on } \{0,1\} }
\left(\min_{Q\in\mathcal{Q}}\chi\left(P,W_{Q}^{\lambda} \right) -
\lim_{n\rightarrow\infty}\max_{{\theta}^n\in\Theta^n}\frac{1}{n}\chi(P^n,V_{{\theta}^n}^{\lambda})\right)\text{
,}\] where $\{V_{\theta}^{\lambda}:\theta\in\Theta\}$: $\mathcal{S}({H}^{\mathfrak{A}})$ $\rightarrow$
 $\mathcal{S}({H}^{\mathfrak{E}})$ are the channels where
the wiretapper observes  the outputs, and  $\mathcal{Q}$ is the set
of  distributions on $\Theta$.\vspace{0.2cm}

The  
 Kraus representations and the  Stinespring dilation of $W_{\theta}^{\lambda}$,
$\theta\in\Theta$ are simple.
It is easy to verify that the following fact holds:

When the sender sends the state $\rho$, and when the legal channel to
the receiver can be represented by  Kraus matrices $\left( A_1, A_2,\cdots\right)$,
the quantum state which the environment obtains can be written in matrix form
\[\left(A_i^*\rho A_j\right)_{i,j}\text{ .}\]
Thus for all ${\theta}\in\Theta$, $V_{\theta}^{\lambda}\left(\begin{pmatrix}
c_1 & c_3\\
\overline{c_3} & c_2
 \end{pmatrix}\right)$ can be written as (minus the zero rows and columns, and 
to another set of basis vectors for each ${\theta}\in\Theta$, of course)

\begin{align*}&
\begin{pmatrix}
(1-\lambda)(c_1+c_2)&0&0\\
0&\lambda c_1 & \lambda c_3\\
0&\lambda\overline{c_3} & \lambda c_2
\end{pmatrix}
\text{ .}\end{align*}

We assume that $\theta$, the channel information, is known by    a eavesdropper who has access to 
the complete environment.     Thus
we may assume that
this eavesdropper knows to which basis its output can be represented in the
above matrix form, i.e., 
when $c_1|0\rangle\langle 0|^{\mathcal{A}}$ $+$ $c_3|1\rangle\langle  0|^{\mathcal{A}}$ $+$ 
$\overline{c_3}|0\rangle\langle 1|^{\mathcal{A}}$ $+$ $(1-c_1)|1\rangle\langle 1|^{\mathcal{A}}$
is transmitted, the  eavesdropper can 
find a set of orthonormal vectors $\{|0\rangle^{\mathcal{E}}$, $|1\rangle^{\mathcal{E}}$,
$|2\rangle^{\mathcal{E}}\}$ on ${H}^{\mathfrak{E}}$ and
perform a transformation to his output such that
it obtains
$\lambda c_1|0\rangle\langle 0|^{\mathcal{E}}$ $+$ $\lambda c_3|1\rangle\langle  0|^{\mathcal{E}}$ $+$ 
$\lambda\overline{c_3}|0\rangle\langle 1|^{\mathcal{E}}$ $+$ $\lambda (1-c_1)|1\rangle\langle 1|^{\mathcal{E}}$
$(1-\lambda)|2\rangle\langle 2|^{\mathcal{E}}$. If it is not able to do this, the amount of information
it gathers will be, of course, less.

Let $P$ be a probability distribution  on the set  $\{|0\rangle^{\mathcal{A}}$, $|1\rangle^{\mathcal{A}}\}$,
such that $P(|0\rangle^{\mathcal{A}}) = a_0$ and $P(|1\rangle^{\mathcal{A}}) = a_1$,
$a_0 + a_1 = 1$.
When the sender's
encoding is restricted to $\{|0\rangle^{\mathcal{A}}$, $|1\rangle^{\mathcal{A}}\}$,
then  for any ${\theta}^n\in\Theta^n$ we have

\begin{align}&
\frac{1}{n}\chi(P^n,V_{{\theta}^n}^{\lambda})\notag\\
&=-\frac{1}{n}\sum_{\begin{subarray}{l}0\leq k_1,k_2,k_3 \leq n:\\
k_1+k_2+k_3=n\end{subarray}}\binom{n}{k_1,k_2,k_3} 
(1-\lambda)^{k_1}\lambda^{k_2}a_0^{k_2}\lambda^{k_3}a_1^{k_3}\log((1-\lambda)^{k_1}\lambda^{k_2}a_0^{k_2}\lambda^{k_3}a_1^{k_3})\notag\\
&+\frac{1}{n}\sum_{\begin{subarray}{l}0\leq k_1,k_2,k_3 \leq n:\\
k_1+k_2+k_3=n\end{subarray}}\binom{n}{k_1,k_2,k_3} 
(1-\lambda)^{k_1}\lambda^{k_2}a_0^{k_2}\lambda^{k_3}a_1^{k_3}\log((1-\lambda)^{k_1}\lambda^{k_2}\lambda^{k_3})\notag\\
&=-\frac{1}{n}\sum_{\begin{subarray}{l}0\leq k_1,k_2,k_3 \leq n:\\
k_1+k_2+k_3=n\end{subarray}}\binom{n}{k_1,k_2,k_3} 
(1-\lambda)^{k_1}\lambda^{k_2}a_0^{k_2}\lambda^{k_3}a_1^{k_3}\log(a_0^{k_2}a_1^{k_3})\notag\\
&\leq -\lambda\frac{1}{n}\sum_{\begin{subarray}{l}0\leq k_1,k_2,k_3 \leq n:\\
k_1+k_2+k_3=n\end{subarray}}\binom{n}{k_1,k_2,k_3} 
a_0^{k_2}a_1^{k_3}\log(a_0^{k_2}a_1^{k_3})\notag\\
&\leq\lambda\frac{1}{n}H(P^n)\notag\\
&=\lambda
\text{ .}\end{align}

We consider a positive $\lambda_1$ such that for any $0<\lambda\leq\lambda_1$,
we have 
\[\lim_{n\rightarrow\infty}\max_{{\theta}^n\in\Theta^n}\frac{1}{n}\chi(P^n,V_{{\theta}^n}^{\lambda})
\leq \frac{1}{4}C\text{ ,}\]
where $C$ is the random  capacity of the arbitrarily varying classical-quantum channel 
$(W_{\theta}^{0})_{{\theta}\in\Theta}$, which is positive by Example \ref{ltsqsbphma} in Section
\ref{aewtdscoaavqwc}.\vspace{0.2cm}

The function $[0,1]\rightarrow \mathbb{R}$:
$\lambda \rightarrow \min_{Q\in\mathcal{Q}}\chi\left(P,W_{Q}^{\lambda} \right)$ is
continuous. Thus there is a positive $\lambda_2$ such that for any $0<\lambda\leq\lambda_2$,
we have 
\[\min_{Q\in\mathcal{Q}}\chi\left(P,W_{Q}^{\lambda}\right)\geq \frac{3}{4}C\text{ .}\]\vspace{0.2cm}

For all $0< \lambda \leq \min\{\lambda_1,\lambda_2\}$, 
we have
\begin{equation}\max_{P\text{ on } \{0,1\} }
\left(\min_{Q\in\mathcal{Q}}\chi\left(P,W_{Q}^{\lambda} \right) -
\lim_{n\rightarrow\infty}\max_{{\theta}^n\in\Theta^n}\frac{1}{n}\chi(P^n,V_{{\theta}^n}^{\lambda})\right)\geq \frac{1}{4}C\text{
.}\label{c21glntvnp}\end{equation}

Since the random security capacity of the  arbitrarily varying quantum channel $\{W_{\theta}^{\lambda}:\theta\in\Theta\}$ is large or
equal to the random security  capacity of the  arbitrarily varying classical-quantum channel when
 the sender's
encoding is restricted to $\{|0\rangle^{\mathcal{A}}$, $|1\rangle^{\mathcal{A}}\}$,
 (\ref{c21glntvnp}) proves ii).\vspace{0.2cm}

By i) and ii), when $\lambda\in]0,1]$ approaches  
zero, the deterministic secrecy capacity of
$\{W_{\theta}^{\lambda}:\theta\in\Theta\}$ does not approach  zero, which
is   the deterministic secrecy capacity of
$\{W_{\theta}^{0}:\theta\in\Theta\}$ by Example \ref{ltsqsbphma} in Section \ref{aewtdscoaavqwc}.
Thus the deterministic secrecy capacity of
$\{W_{\theta}^{\lambda}:\theta\in\Theta\}$ is not continuous at zero.

 \end{example}\vspace{0.2cm}

Now we are going to prove Corollary \ref{ltbafsanttit}.\vspace{0.2cm}

\begin{proof}
Suppose we have $C_s(\{N_{\theta}: {\theta}\in \Theta\})$ $>0$.
Then $\{N_{\theta}: {\theta}\in \Theta\}$
is not symmetrizable, which means that
$\mathsf{F}(\{N_{\theta}: \theta\})$ is positive.
In the proof of Corollary  \ref{ftsdmiait} we show that
$F$ is continuous. Thus there is a positive $\delta'$ such that
$\mathsf{F}(\{N_{\theta}': \theta\})$ $>0$ for all $\{N_{\theta}': {\theta}\in \Theta\}$
$\in\mathsf{C}_{\delta'}$ .
When $\{N_{\theta}: {\theta}\in \Theta\}$
is not symmetrizable then we have $C_s(\{N_{\theta}: {\theta}\in \Theta\},r)$ $=$
$C_s(\{N_{\theta}: {\theta}\in \Theta\})$ $>0$.
By  Corollary 5.1 in \cite{Bo/Ca/De2}, 
the secrecy capacity  under
 randomness assisted quantum coding is continuous.
Thus there is a positive $\delta''$ such that
 $C_s(\{N_{\theta}': {\theta}\in \Theta\},r)$ $>0$
for all $\{N_{\theta}': {\theta}\in \Theta\}$
$\in\mathsf{C}_{\delta''}$ .
We define $\delta$ $:=$ $\min (\delta',\delta'')$ and the
Corollary is shown.
\end{proof}

\subsection{Further Notes and Super-Activation}\label{aewtdscoaavqwc}

Theorem \ref{lntvttitbaa} states that either the deterministic
security capacity of an arbitrarily varying quantum
 channel is zero, or it equals its
randomness assisted
 security  capacity.
There are actually arbitrarily varying quantum
 channels which have zero deterministic  security capacity
 but achieve a positive security capacity, if the sender and the
 legal receiver can use a resource (cf. Example \ref{ltsqsbphma}).
 This shows that the Ahlswede Dichotomy
 is indeed a ``dichotomy'', and how helpful  a resource can be for
 the robust and secure message transmission.

Now we give
an example when the  deterministic secrecy capacity  of an arbitrarily varying quantum channel
 is not equal to its randomness assisted secrecy capacity.

As already mentioned in Section \ref{pwatscc}, Example \ref{ltsqsbphma}
shows a case when the deterministic secrecy capacity of an arbitrarily varying quantum channel is 
equal to zero while 
 its randomness assisted secrecy capacity is positive.

If the previously mentioned conjecture in  \cite{Ahl/Bj/Bo/No}
is true, i.e., if this behavior does not occur for
 the
entanglement distillation capacity, the entanglement generating capacity,
and the strong subspace transmission capacity,
then it is not possible to find any respective
examples for those capacities.

 \begin{example}
\label{ltsqsbphma}

Let the sender's quantum system be presented by ${H}^{\mathfrak{A}}$ $=$ $\mathbb{C}^{2}$.
Let the receiver's quantum system be presented by ${H}^{\mathfrak{B}}$ $=$ $\mathbb{C}^{3}$.
We choose an orthonormal basis $\{|0\rangle^{\mathfrak{A}}, |1\rangle^{\mathfrak{A}}\}$ on
${H}^{\mathfrak{A}}$ and an orthonormal basis $\{|0\rangle^{\mathfrak{B}}, |1\rangle^{\mathfrak{B}},
|2\rangle^{\mathfrak{B}}\}$ on
${H}^{\mathfrak{B}}$.

\it i) Definition of an arbitrarily varying quantum channel \rm\vspace{0.2cm}

Let $\Theta:=\{1^{(+)},1^{(-)},2^{(+)},2^{(-)}\}$ be  a set of indices.
 We      describe      an arbitrarily varying quantum channel $\{W_{\theta}: {\theta}\in \Theta\}$ in the following way:
 We define a  map $W_{1^{(+)}}$:
$\mathcal{S}({H}^{\mathfrak{A}})$ $\rightarrow$ $\mathcal{S}({H}^{\mathfrak{B}})$
by\begin{align*}& c_1|0\rangle\langle 0|^{\mathfrak{A}} + c_2|1\rangle\langle 1|^{\mathfrak{A}}
+ c_3|1\rangle\langle 0|^{\mathfrak{A}} + \overline{c_3}|0\rangle\langle 1|^{\mathfrak{A}}\\
& \rightarrow c_1|0\rangle\langle 0|^{\mathfrak{B}} + c_2|1\rangle\langle 1|^{\mathfrak{B}}
+ c_3|1\rangle\langle 0|^{\mathfrak{B}} + \overline{c_3}|0\rangle\langle 1|^{\mathfrak{B}} \text{ ,}\end{align*}
 a  map $W_1^{(-)}$:
$\mathcal{S}({H}^{\mathfrak{A}})$ $\rightarrow$ $\mathcal{S}({H}^{\mathfrak{B}})$
by\begin{align*}& c_1|0\rangle\langle 0|^{\mathfrak{A}} + c_2|1\rangle\langle 1|^{\mathfrak{A}}
+ c_3|1\rangle\langle 0|^{\mathfrak{A}} + \overline{c_3}|0\rangle\langle 1|^{\mathfrak{A}}\\
& \rightarrow c_1|0\rangle\langle 0|^{\mathfrak{B}} + c_2|1\rangle\langle 1|^{\mathfrak{B}}
- c_3|1\rangle\langle 0|^{\mathfrak{B}} - \overline{c_3}|0\rangle\langle 1|^{\mathfrak{B}} \text{ ,}\end{align*}
 a map $W_{2^{(+)}}$:
$\mathcal{S}({H}^{\mathfrak{A}})$ $\rightarrow$ $\mathcal{S}({H}^{\mathfrak{B}})$
by\begin{align*}& c_1|0\rangle\langle 0|^{\mathfrak{A}} + c_2|1\rangle\langle 1|^{\mathfrak{A}}
+ c_3|1\rangle\langle 0|^{\mathfrak{A}} + \overline{c_3}|0\rangle\langle 1|^{\mathfrak{A}}\\
& \rightarrow c_1|1\rangle\langle 1|^{\mathfrak{B}} + c_2|2\rangle\langle 2|^{\mathfrak{B}}
+ c_3|2\rangle\langle 1|^{\mathfrak{B}} + \overline{c_3}|1\rangle\langle 2|^{\mathfrak{B}} \text{ ,}\end{align*}
and a  map $W_{2^{(-)}}$:
$\mathcal{S}({H}^{\mathfrak{A}})$ $\rightarrow$ $\mathcal{S}({H}^{\mathfrak{B}})$
by\begin{align*}& c_1|0\rangle\langle 0|^{\mathfrak{A}} + c_2|1\rangle\langle 1|^{\mathfrak{A}}
+ c_3|1\rangle\langle 0|^{\mathfrak{A}} + \overline{c_3}|0\rangle\langle 1|^{\mathfrak{A}}\\
& \rightarrow c_1|1\rangle\langle 1|^{\mathfrak{B}} + c_2|2\rangle\langle 2|^{\mathfrak{B}}
- c_3|2\rangle\langle 1|^{\mathfrak{B}} - \overline{c_3}|1\rangle\langle 2|^{\mathfrak{B}}  \text{ .}\end{align*}

It is easy to verify that all these maps are linear and trace preserving,
it is also easy to verify that $W_{1^{(+)}}$ and $W_{2^{(+)}}$ are
complete positive.
$W_{1^{(-)}}$ is
 complete positive, since
for any $\rho \in \mathcal{S}({H}^{\mathfrak{A}})$,
 $W_{1^{(+)}}(\rho)$ and  $W_{1^{(-)}}(\rho)$ are unitarily equivalent.
Similarly, $W_{2^{(-)}}(\rho)$ is unitarily equivalent to $W_{2^{(+)}}(\rho)$
for every $\rho \in \mathcal{S}({H}^{\mathfrak{A}})$ and therefore $W_{2^{(-)}}$ is
complete positive, too.\vspace{0.2cm}

 $W_{1^{(+)}}$,  $W_{1^{(-)}}$, $W_{2^{(+)}}$, and $W_{2^{(-)}}$ are unitary transformations,
they have very simple 
 Kraus representations (consist of only one matrix) and very simple Stinespring dilations (a  identity matrix 
on a subspace).
It is easy to verify that there is a vector $|0\rangle\langle 0|^{\mathfrak{E}}$ on
$H^{\mathfrak{E}}$, the environment space, such that the channel $V_{\theta}$ from the sender to 
the environment can be described as 
\[ c_1|0\rangle\langle 0|^{\mathfrak{A}} + c_2|1\rangle\langle 1|^{\mathfrak{A}}
+ c_3|1\rangle\langle 0|^{\mathfrak{A}} + \overline{c_3}|0\rangle\langle 1|^{\mathfrak{A}}\rightarrow
  (c_1 + c_2)|0\rangle\langle 0|^{\mathfrak{E}} \]
for all four ${\theta}\in\Theta$.\vspace{0.2cm}

\it ii)    $L$-symmetrizability for all $L\in \mathbb{N}$ \rm\vspace{0.2cm}

We fix an arbitrary $L\in \mathbb{N}$.
For every $\rho^L\in \mathcal{S}({H}^{\mathfrak{A}^L})$
we define a parametrized set of distributions $\tilde{q}_{\rho^L}$
on $\{\Theta^L: \theta\in\{0,1\}\}$ by
\[\tilde{q}_{\rho^L}(\theta^L) :=\left(\otimes_{i=1}^L \langle \theta_i |^{\mathfrak{A}}\right) \rho^L \left(\otimes_{i=1}^L | \theta_i \rangle^{\mathfrak{A}}\right)\]
for $j^L =(j_1, \cdots,  j_L)$.

We define a map $g:$ $\Theta^L \rightarrow \{0,1\}^L$ in the following way.
For every ${\theta}^L$
$=(k_1^{(s_1)}, k_2^{(s_2)}, \cdots, k_L^{(s_L)})$ $\in \Theta^L$,
$k_i \in \{1,2\}$, $s_i \in \{+,-\}$, we define
\[g({\theta}^L) := (k_1-1,k_2-1,\cdots k_L-1)\text{ .}\]

We define
a parametrized set of distributions $\{q_{\rho^L}(\cdot):
 \rho \in \mathcal{S}({H}^{\mathfrak{A}^L})\}$ on $\Theta^L$ by
\[q_{\rho^L}({\theta}^L) := \frac{1}{2^L}\tilde{q}_{\rho}^L(g({\theta}^L))\text{ .}\]

For all $\rho^L$, ${\rho'}^L\in \mathcal{S}({H}^{\mathfrak{A}^L})$ it holds
\begin{align*}&\sum_{{\theta}^L\in\Theta^L}q_{{\rho'}^L}({\theta}^L)W_{{\theta}^L}(\rho^L)\\
&=\sum_{{\theta}^L\in\Theta^L} q_{{\rho'}^L}\sum_{{j'}^L\in \{0,1\}^L}
\left(\otimes_{i=1}^L\langle j_i' |^{\mathfrak{A}}\right)  {\rho^L} \left(\otimes_{i=1}^L | j_i' \rangle^{\mathfrak{A}}\right)
W_{{\theta}^L}\left(\otimes_{i=1}^L | j_i' \rangle\langle  j_i' |^{\mathfrak{A}}\right)\\
&=\frac{1}{2^L}\sum_{{\theta}^L\in\Theta^L} \left(\otimes_{i=1}^L \langle g({\theta}^L)_i|^{\mathfrak{A}}\right) {\rho'}^L \left(\otimes_{i=1}^L | g({\theta}^L)_i \rangle^{\mathfrak{A}}\right)\\
&~~\cdot \left[\sum_{{j'}^L\in \{0,1\}^L }\left(\otimes_{i=1}^L\langle j_i' |^{\mathfrak{A}}\right)  {\rho^L} \left(\otimes_{i=1}^L | j_i' \rangle^{\mathfrak{A}}\right)
\left(\otimes_{i=1}^L| j_i'+g({\theta}^L)_i \rangle\langle j_i'+g({\theta}^L)_i |^{\mathfrak{B}}\right)\right]\\
&=\sum_{j^L\in \{0,1\}^L} \left(\otimes_{i=1}^L \langle j_i|^{\mathfrak{A}}\right) {\rho'}^L \left(\otimes_{i=1}^L | j_i \rangle^{\mathfrak{A}}\right)\\
&~~\cdot \left[\sum_{{j'}^L\in \{0,1\}^L }\left(\otimes_{i=1}^L\langle j_i' |^{\mathfrak{A}}\right)  {\rho^L} \left(\otimes_{i=1}^L | j_i' \rangle^{\mathfrak{A}}\right)
\left(\otimes_{i=1}^L| j_i'+j_i \rangle\langle j_i'+j_i |^{\mathfrak{B}}\right)\right]\\
&=\sum_{m^L \in  \{0,1,2\}^L} ~~\sum_{j^L, {j'}^L \in  \{0,1\}^L: j^L+{j'}^L = m^L} 
\left(\otimes_{i=1}^L\langle j_i |^{\mathfrak{A}}\right)  {\rho'}^L \left(\otimes_{i=1}^L | j_i \rangle^{\mathfrak{A}}\right)\\
&~~\cdot \left(\otimes_{i=1}^L \langle j_i' |^{\mathfrak{A}}\right) \rho^L \left(\otimes_{i=1}^L | j_i' \rangle^{\mathfrak{A}}\right)
\left[\otimes_{i=1}^L| m_i \rangle\langle m_i |^{\mathfrak{B}}\right]\\
&=\sum_{{j'}^L \in  \{0,1\}^L} \left(\otimes_{i=1}^L \langle j_i'|^{\mathfrak{A}}\right) \rho^L \left(\otimes_{i=1}^L | j_i' \rangle^{\mathfrak{A}}\right)\\
&~~\cdot \left[\sum_{{j'}^L\in \{0,1\}^L }\left(\otimes_{i=1}^L\langle j_i' |^{\mathfrak{A}}\right)  {\rho^L} \left(\otimes_{i=1}^L | j_i' \rangle^{\mathfrak{A}}\right)
\left(\otimes_{i=1}^L| j_i'+j_i \rangle\langle j_i'+j_i |^{\mathfrak{B}}\right)\right]\\
&=\frac{1}{2^L}\sum_{{\theta}^L\in\Theta^L}\left(\otimes_{i=1}^L \langle g({\theta}^L)_i |^{\mathfrak{A}}\right) {\rho}^L \left(\otimes_{i=1}^L | g({\theta}^L)_i \rangle^{\mathfrak{A}}\right)\\
&~~\cdot\left[\sum_{j^L\in \{0,1\}^L}\left(\otimes_{i=1}^L\langle j_i |^{\mathfrak{A}}\right)  {\rho'}^L \left(\otimes_{i=1}^L | j_i \rangle^{\mathfrak{A}}\right)
\left(\otimes_{i=1}^L| j_i+g({\theta}^L)_i \rangle\langle j_i+g({\theta}^L)_i |^{\mathfrak{B}}\right)\right]\\
&=\sum_{{\theta}^L\in\Theta^L} q_{\rho^L}\sum_{j^L\in \{0,1\}^L}
\left(\otimes_{i=1}^L\langle j_i |^{\mathfrak{A}}\right)  {\rho'}^L \left(\otimes_{i=1}^L | j_i \rangle^{\mathfrak{A}}\right)
W_{{\theta}^L}\left(\otimes_{i=1}^L | j_i \rangle\langle  j_i |^{\mathfrak{A}}\right)\\
&=\sum_{{\theta}^L\in\Theta^L}q_{\rho^L}({\theta}^L)W_{{\theta}^L}({\rho'}^L)\text{ ,}\end{align*}
where for $j^L=(j_1,\cdots,j_L)$ and ${j'}^L=(j_1',\cdots,j_L')$ $\in  \{0,1\}^L$ 
we define $ j^L+{j'}^L := (j_1+j_1',\cdots,j_L+j_L')$ $\in  \{0,1,2\}^L$.
\vspace{0.2cm}

Thus 
  $\{W_{\theta}: {\theta}\in \Theta\}$ is $L$-symmetrizable.\vspace{0.2cm}

\it iii)   Positivity of the random security capacity  \rm\vspace{0.2cm}

Let $\Theta':=\{1,2\}$ be  a finite set of indices. We define a complete positive trace preserving map $W_1$:
$\mathcal{S}({H}^{\mathfrak{A}})$ $\rightarrow$ $\mathcal{S}({H}^{\mathfrak{B}})$
by\begin{align*}& c_1|0\rangle\langle 0|^{\mathfrak{A}} + c_2|1\rangle\langle 1|^{\mathfrak{A}}
+ c_3|1\rangle\langle 0|^{\mathfrak{A}} + \overline{c_3}|0\rangle\langle 1|^{\mathfrak{A}}\\
& \rightarrow c_1|0\rangle\langle 0|^{\mathfrak{B}} + c_2|1\rangle\langle 1|^{\mathfrak{B}} \text{ ,}\end{align*}
and a complete positive trace preserving map $W_2$:
$\mathcal{S}({H}^{\mathfrak{A}})$ $\rightarrow$ $\mathcal{S}({H}^{\mathfrak{B}})$
by\begin{align*}& c_1|0\rangle\langle 0|^{\mathfrak{A}} + c_2|1\rangle\langle 1|^{\mathfrak{A}}
+ c_3|1\rangle\langle 0|^{\mathfrak{A}} + \overline{c_3}|0\rangle\langle 1|^{\mathfrak{A}}\\
& \rightarrow c_1|1\rangle\langle 1|^{\mathfrak{B}} + c_2|2\rangle\langle 2|^{\mathfrak{B}} \text{ .}\end{align*}

Let $\mathcal{T}$ and $\mathcal{T}'$
be the  set of distributions on
the index set $\Theta$ and $\Theta'$, respectively.
We consider the case when the sender's encoding is restricted to transmitting
the indexed finite set  of orthogonal quantum states $\{|0\rangle\langle 0|^{\mathfrak{A}}, 
|1\rangle\langle 1|^{\mathfrak{A}}\}$
 on $\mathcal{S}({H}^{\mathfrak{A}})$.
For any $\sigma\in \{|0\rangle\langle 0|^{\mathfrak{A}}, 
|1\rangle\langle 1|^{\mathfrak{A}}\}$ we have
$W_{1}(\sigma)=W_{1^{(+)}}(\sigma)=W_{1^{(-)}}(\sigma)$ and
$W_{2}(\sigma)=W_{2^{(+)}}(\sigma)=W_{2^{(-)}}(\sigma)$.\vspace{0.2cm}

For any $p\in\mathcal{T}$ we define a distribution $q_p \in\mathcal{T}'$
by $q_p(1)=p(1^{(+)})+p(1^{(-)})$ and $q_p(2)=p(2^{(+)})+p(2^{(-)})$.
For any $\sigma\in \{|0\rangle\langle 0|^{\mathfrak{A}}, 
|1\rangle\langle 1|^{\mathfrak{A}}\}$ we have
\begin{align*}&p(1^{(+)})W_{1^{(+)}}(\sigma)+p(1^{(-)})W_{1^{(-)}}(\sigma)\\
&+p(2^{(+)})W_{2^{(+)}}(\sigma)+p(2^{(-)})W_{2^{(-)}}(\sigma)\\
&=\left(p(1^{(+)})+p(1^{(-)})\right)W_{1}(\sigma)+\left(p(2^{(+)})+p(2^{(-)})\right)W_{2}(\sigma)\\
&=q_p(1)W_{1}(\sigma)+ q_p(2)W_{2}(\sigma)\text{ .}\end{align*}

For any $q \in\mathcal{T}'$ we define a distribution $p_q \in\mathcal{T}$
by $p_q(1^{(+)})=q(1)$,  $p_q(2^{(+)})=q(2)$, and $p_q(1^{(-)})=p_q(2^{(-)})=0$.
For any $\sigma\in \{|0\rangle\langle 0|^{\mathfrak{A}}, 
|1\rangle\langle 1|^{\mathfrak{A}}\}$ 
it always holds  trivially
\[ q(1)W_{1}(\sigma)+ q(2)W_{2}(\sigma) =  p_q(1^{(+)})W_{1^{(+)}}(\sigma) + p_q(2^{(+)})W_{2^{(+)}}(\sigma) \text{ .}\]

Thus 
\[\min_{s\in \mathcal{T}}\max_{p\in \mathcal{P}}\chi(p,B_{s}) =
\min_{s'\in \mathcal{T}'}\max_{p\in \mathcal{P}}\chi(p,B_{s}')\text{ .}\]
Here $\mathcal{P}$ is the  set of distributions on the input set $\{|0\rangle\langle 0|^{\mathfrak{A}}, 
|1\rangle\langle 1|^{\mathfrak{A}}\}$, $B_{\theta}'$ is the quantum output of  $(W_{\theta})_{{\theta}\in \Theta'}$, and
$B_{\theta}$ is the quantum output of  $\{W_{\theta}: {\theta}\in \Theta\}$.

We denote by $p'$ the distribution on $\{|0\rangle\langle 0|^{\mathfrak{A}}, 
|1\rangle\langle 1|^{\mathfrak{A}}\}$ such that $p'(|0\rangle\langle 0|^{\mathfrak{A}})=p'(|1\rangle\langle 1|^{\mathfrak{A}})=\frac{1}{2}$.
Let $q\in[0,1]$.  We define $Q(1)=q$,   $Q(2)=1-q$. We have
\begin{align*}
&\chi\left(p',\{W_Q(a): a\in {\mathbf{A}}\}\right)\\
&=-\frac{1}{2}q\log \frac{1}{2}q + \frac{1}{2}(1-q)\log\frac{1}{2}(1-q)
-\frac{1}{2}\log\frac{1}{2}\\
&+q\log q + (1-q)\log(1-q)\text{ .}\end{align*}
By the differentiation  by $q$,
we obtain
\begin{align*}&\frac{1}{\log e}\biggl(-\frac{1}{2}\log \frac{1}{2}q - \frac{1}{2}+ 
\frac{1}{2}\log\frac{1}{2}(1-q)+\frac{1}{2}
+\log q +1- \log(1-q)-1\biggr)\\
&=\frac{1}{2\log e}\left(\log q - \log(1-q)\right)\text{ .}\end{align*}
This term is equal to zero if and only if $q=\frac{1}{2}$. By further calculation,
one can show
that  $\chi\Bigl(p',\{W_Q(\rho): \rho\in \{|0\rangle\langle 0|^{\mathfrak{A}}, 
|1\rangle\langle 1|^{\mathfrak{A}}\}\}\Bigr)$ achieves its minimum when  $q=\frac{1}{2}$.  This
minimum is equal to $-\frac{1}{2}\log\frac{1}{4}+\frac{1}{2}\log\frac{1}{2}$ $=$ $\frac{1}{2}$ $>0$.
Thus \[\max_p\min_{q} \chi\left(p,B_q\right)\geq\frac{1}{2}\text{ .}\]
\vspace{0.2cm}

 For any $n\in\mathbb{N}$, an arbitrary quantum state $\rho^{\otimes n}$ in $\mathcal{S}({H}^{\mathfrak{A}^{n}})$,
and an arbitrary sequence ${\theta}^n \in \Theta^n$,
we have $V_{{\theta}^n}(\rho^{\otimes n})$ $=$ $|0\rangle\langle 0|^{\mathfrak{E}^{n}}$. This can be easily shown by induction: 
For any $m\in\mathbb{N}$, we assume $\rho^{\otimes m}$ is an arbitrary quantum state  in $\mathcal{S}({H}^{\mathfrak{A}^{m}})$
and ${\theta}^m=({\theta}_1,\cdots,{\theta}_m)$ is an arbitrary sequence in $\Theta$.
It holds $V_{{\theta}^m}(\rho^{\otimes m})$ $=$ $|0\rangle\langle 0|^{\mathfrak{E}} \otimes V_{{\theta}^{m-1}}(\rho^{\otimes m-1})$, where
$\rho^{\otimes m-1}$ is a quantum state in  $\mathcal{S}({H}^{\mathfrak{A}^{m-1}})$  and ${\theta}^{m-1}=({\theta}_2,\cdots,{\theta}_m)$.
Thus 
for every $p\in
\mathcal{P}$ and ${\theta}^n\in \Theta^n$,
$\chi(p^n,Z_{{\theta}^n})$ cannot  exceed $1 \log 1 - 1 \log 1 = 0$.\vspace{0.2cm}

Thus 
\[C_{s}(\{(W_{\theta}, V_{\theta}): {\theta}\in \Theta\},r)\geq\frac{1}{2} > 0=C_{s}(\{(W_{\theta}, V_{\theta}): {\theta}\in \Theta\},r)\text{ .}\]
This     gives    an example of
 an arbitrarily varying quantum channel such that
 its deterministic  capacity  is zero, but its random  capacity  is positive.

 \end{example}

 Thus, a ``useless'' arbitrarily varying  quantum  channel,
 i.e. with zero deterministic  secrecy capacity,  allows
 secure transmission if the sender and the legal receiver have the possibility
 to use a resource.\vspace{0.3cm}

A common property of     most      classical channels is that where the capacity of two sub-channels, when
used together,  is the sum of the
their capacities.      Particularly,
 a channel system consisting of two orthogonal classical
channels, where both are ``useless''  in the sense  that they both have zero
capacity for message transmission, the capacity  for message transmission
of the whole system is zero as well (``$0 + 0 = 0$''). 
 For the definition of   ``two orthogonal
 channels'' in classical systems please see  \cite{Fa/Ka}.

In contrast to the classical information  theory, it is known that
the capacities of quantum  channels can be super-additive, i.e.,
there are cases in which the capacity of the product $W_1\otimes W_2$
 of
 two  quantum  channels $W_1$
 and $W_2$
 are
larger than the sum of the capacity of $W_1$ and
 the capacity of $W_2$ (cf. \cite{Li/Wi/Zou/Guo} and \cite{Gi/Wo}).
``The whole is greater than the sum of its parts'' - Aristotle.

 Particularly in quantum information  theory, there are examples of two quantum
channels, $W_1$ and $W_2$,  with zero capacity, which
 allow     positive      transmission if they are used together, i.e.,  the   capacity of
their product $W_1\otimes W_2$ is positive, (cf. \cite{Smi/Yar},
\cite{Smi/Smo/Yar}, \cite{Opp} and also \cite{Bo/Wy} for a rare case
result when this phenomenon occurs using two classical arbitrarily
varying wiretap channels). This is due to the fact that there are
different reasons why a quantum channel can have zero capacity. If
we have two channels which have zero capacity for different reasons,
they can ``remove'' their weaknesses from each other, or in
other words, ``activate'' each other. We
call this phenomenon   ``super-activation'' (``$0+0 >0$'').

In our previous work 
\cite{Bo/Ca/De} we proved super-activation for
 arbitrarily varying classical-quantum  wiretap
channels.

 Now we will prove Theorem  \ref{superactivationqqq}, i.e., super-activation of  secrecy
capacities for
arbitrarily varying quantum 
channels,
by giving an example.\vspace{0.2cm}

\begin{example}

Let the sender's quantum system be presented by ${H}^{\hat{\mathfrak{A}}}$ $=$ $\mathbb{C}^{4}$ and
the receiver's quantum system be presented by ${H}^{\hat{\mathfrak{B}}}$ $=$ $\mathbb{C}^{4}$.
Let $\hat{\Theta}:=\{1,2\}$ be  a finite set of indices.

Choosing an orthonormal basis $\{|0\rangle^{\hat{\mathfrak{A}}}, |1\rangle^{\hat{\mathfrak{A}}}, 
|2\rangle^{\hat{\mathfrak{A}}}, |3\rangle^{\hat{\mathfrak{A}}}\}$ on
${H}^{\hat{\mathfrak{A}}}$ and an orthonormal basis $\{|0\rangle^{\hat{\mathfrak{B}}}, |1\rangle^{\hat{\mathfrak{B}}},
|2\rangle^{\hat{\mathfrak{B}}}, |3\rangle^{\hat{\mathfrak{B}}}\}$ on
${H}^{\hat{\mathfrak{B}}}$,
 we define a complete positive trace-preserving map ${\hat{W}}_1$:
$\mathcal{S}({H}^{\hat{\mathfrak{A}}})$ $\rightarrow$ $\mathcal{S}({H}^{\hat{\mathfrak{B}}})$
with respect to these bases
by
\begin{align*}&
\begin{pmatrix}c_{00}&c_{01}&c_{02}&c_{03}\\
\overline{c_{01}}&c_{11}&c_{12}&c_{13}\\
\overline{c_{02}}&\overline{c_{12}}&c_{22}&c_{23}\\
\overline{c_{03}}&\overline{c_{13}}&\overline{c_{23}}&c_{33}\\
\end{pmatrix}
\rightarrow
\begin{pmatrix}c_{00}&c_{01}&0&0\\
\overline{c_{01}}&c_{11}&0&0\\
0&0&c_{22}+c_{33}&0\\
0&0&0&0\\
\end{pmatrix}
 \text{ ,}\end{align*}

and a complete positive trace-preserving map ${\hat{W}}_2$:
$\mathcal{S}({H}^{\hat{\mathfrak{A}}})$ $\rightarrow$ $\mathcal{S}({H}^{\hat{\mathfrak{B}}})$
with respect to these bases
by
\begin{align*}&
\begin{pmatrix}c_{00}&c_{01}&c_{02}&c_{03}\\
\overline{c_{01}}&c_{11}&c_{12}&c_{13}\\
\overline{c_{02}}&\overline{c_{12}}&c_{22}&c_{23}\\
\overline{c_{03}}&\overline{c_{13}}&\overline{c_{23}}&c_{33}\\
\end{pmatrix}
\rightarrow
\begin{pmatrix}c_{00}+c_{11}&0&0&0\\
0&0&0&0\\
0&0&c_{22}&c_{23}\\
0&0&\overline{c_{23}}&c_{33}\\
\end{pmatrix}
 \text{ .}\end{align*}\vspace{0.15cm}

With respect to 
an orthonormal basis $\{|0\rangle^{\hat{\mathfrak{E}}}, |1\rangle^{\hat{\mathfrak{E}}},
|2\rangle^{\hat{\mathfrak{E}}}, |3\rangle^{\hat{\mathfrak{E}}}\}$ on
${H}^{\hat{\mathfrak{E}}}$,
${\hat{W}}_{1}$ has the Stinespring matrix
\[\left ( \begin{array} {rrrrrrrrrrrrrrrr}
1&0&0&0& 0&0&0&0& 0&0&0&0& 0&0&0&0\\
0&1&0&0& 0&0&0&0& 0&0&0&0& 0&0&0&0\\
0&0&0&0& 0&0&0&0& 0&0&1&0& 0&0&0&0\\
0&0&0&0& 0&0&0&0& 0&0&0&0& 0&0&1&0\\
0&0&0&0& 1&0&0&0& 0&0&0&0& 0&0&0&0\\
0&0&0&0& 0&1&0&0& 0&0&0&0& 0&0&0&0\\
0&0&0&0& 0&0&1&0& 0&0&0&0& 0&0&0&0\\
0&0&0&0& 0&0&0&1& 0&0&0&0& 0&0&0&0\\
0&0&0&0& 0&0&0&0& 1&0&0&0& 0&0&0&0\\
0&0&0&0& 0&0&0&0& 0&1&0&0& 0&0&0&0\\
0&0&1&0& 0&0&0&0& 0&0&0&0& 0&0&0&0\\
0&0&0&0& 0&0&0&0& 0&0&0&0& 0&0&0&1\\
0&0&0&0& 0&0&0&0& 0&0&0&0& 1&0&0&0\\
0&0&0&0& 0&0&0&0& 0&0&0&0& 0&1&0&0\\
0&0&0&1& 0&0&0&0& 0&0&0&0& 0&0&0&0\\
0&0&0&0& 0&0&0&0& 0&0&0&1& 0&0&0&0\\
 \end{array}\right )\text{ .}\]

The quantum state which the environment obtains is
\begin{align*}&
\begin{pmatrix}c_{00}&c_{01}&c_{02}&c_{03}\\
\overline{c_{01}}&c_{11}&c_{12}&c_{13}\\
\overline{c_{02}}&\overline{c_{12}}&c_{22}&c_{23}\\
\overline{c_{03}}&\overline{c_{13}}&\overline{c_{23}}&c_{33}\\
\end{pmatrix}
\rightarrow
\begin{pmatrix}c_{00}+c_{11}&0&0&0\\
0&0&0&0\\
0&0&c_{22}&c_{23}\\
0&0&\overline{c_{23}}&c_{33}\\
\end{pmatrix}
\end{align*} with respect to 
the orthonormal basis $\{|0\rangle^{\hat{\mathfrak{E}}}, |1\rangle^{\hat{\mathfrak{E}}},
|2\rangle^{\hat{\mathfrak{E}}}, |3\rangle^{\hat{\mathfrak{E}}}\}$ on
${H}^{\hat{\mathfrak{E}}}$.\vspace{0.15cm}

With respect to 
an orthonormal basis $\{|0\rangle^{\hat{\mathfrak{E}}}, |1\rangle^{\hat{\mathfrak{E}}},
|2\rangle^{\hat{\mathfrak{E}}}, |3\rangle^{\hat{\mathfrak{E}}}\}$ on
${H}^{\hat{\mathfrak{E}}}$,
${\hat{W}}_{2}$ has the Stinespring matrix
\[\left ( \begin{array} {rrrrrrrrrrrrrrrr}
0&0&0&0& 0&0&0&0& 1&0&0&0& 0&0&0&0\\
0&0&0&0& 0&0&0&0& 0&1&0&0& 0&0&0&0\\
0&0&1&0& 0&0&0&0& 0&0&0&0& 0&0&0&0\\
0&0&0&0& 0&0&1&0& 0&0&0&0& 0&0&0&0\\
0&0&0&0& 1&0&0&0& 0&0&0&0& 0&0&0&0\\
0&0&0&0& 0&1&0&0& 0&0&0&0& 0&0&0&0\\
0&0&0&1& 0&0&0&0& 0&0&0&0& 0&0&0&0\\
0&0&0&0& 0&0&0&0& 0&0&0&0& 0&0&0&1\\
1&0&0&0& 0&0&0&0& 0&0&0&0& 0&0&0&0\\
0&1&0&0& 0&0&0&0& 0&0&0&0& 0&0&0&0\\
0&0&0&0& 0&0&0&0& 0&0&1&0& 0&0&0&0\\
0&0&0&0& 0&0&0&0& 0&0&0&1& 0&0&0&0\\
0&0&0&0& 0&0&0&0& 0&0&0&0& 1&0&0&0\\
0&0&0&0& 0&0&0&0& 0&0&0&0& 0&1&0&0\\
0&0&0&0& 0&0&0&0& 0&0&0&0& 0&0&1&0\\
0&0&0&0& 0&0&0&1& 0&0&0&0& 0&0&0&0\\
 \end{array}\right )\text{ .}\]
The quantum state which the environment obtains is
\begin{align*}&
\begin{pmatrix}c_{00}&c_{01}&c_{02}&c_{03}\\
\overline{c_{01}}&c_{11}&c_{12}&c_{13}\\
\overline{c_{02}}&\overline{c_{12}}&c_{22}&c_{23}\\
\overline{c_{03}}&\overline{c_{13}}&\overline{c_{23}}&c_{33}\\
\end{pmatrix}
\rightarrow
\begin{pmatrix}c_{00}&c_{01}&0&0\\
\overline{c_{01}}&c_{11}&0&0\\
0&0&c_{22}+c_{33}&0\\
0&0&0&0\\
\end{pmatrix}
\end{align*}
 with respect to 
a orthonormal basis $\{|0\rangle^{\hat{\mathfrak{E}}}, |1\rangle^{\hat{\mathfrak{E}}},
|2\rangle^{\hat{\mathfrak{E}}}, |3\rangle^{\hat{\mathfrak{E}}}\}$ on
${H}^{\hat{\mathfrak{E}}}$.\vspace{0.2cm}

When we interpret  the environment as a wiretapper,
then  the  arbitrarily varying  wiretap quantum channel $\{({\hat{W}}_{\theta'},{\hat{V}}_{\theta'}): {\theta'}\in\hat{\Theta}\}$
has the following property: ${\hat{W}}_1$ and ${\hat{V}}_2$ are identical with respect to
the corresponding orthonormal bases on ${H}^{\hat{\mathfrak{B}}}$
and ${H}^{\hat{\mathfrak{E}}}$,      while      ${\hat{W}}_2$ and ${\hat{V}}_1$ are identical with respect to
the corresponding orthonormal bases on ${H}^{\hat{\mathfrak{B}}}$
and ${H}^{\hat{\mathfrak{E}}}$.

Thus for any input set of quantum states and any probability distribution $p$
on this set we always have
\begin{equation}\min_{i\in\hat{\Theta}}\chi(p,{\hat{W}}_{\theta'})- \max_{{\theta''}\in\hat{\Theta}}\chi(p,{\hat{V}}_{\theta''})=
\min_{{\theta'}\in\hat{\Theta}}\chi(p,{\hat{W}}_{\theta'})- \max_{{\theta''}\in\hat{\Theta}}\chi(p,{\hat{W}}_{\theta''})\leq 0\text{ .}\label{miitcpwi}\end{equation}
The secrecy capacity of this arbitrarily varying wiretap
quantum channel is thus always zero:
 Supposed there is a message set and a code  
such that for $({\hat{W}}_{\theta'},{\hat{V}}_{\theta'})$, ${\theta'}\in\hat{\Theta}$ the sender is able to convey a secure message
to the legal receiver without the wiretapper  knowing  anything. Then the wiretapper
can  correctly decode the same  encoded message
when it is transmitted with the same decoding operator in corresponding orthonormal bases
through $({\hat{W}}_{\theta''},{\hat{V}}_{\theta''})$, $\theta''\not= {\theta'}$, while
the legal receiver cannot decode anything.\vspace{0.2cm}

Let $\{W_{\theta}: {\theta}\in \Theta\}$ be defined as in
Section \ref{aewtdscoaavqwc}. We now consider 
 $\{\hat{W}_{\theta'}\otimes W_{\theta}: ({\theta'},\theta)\in \hat{\Theta}\times \Theta\}$.
For any $\sigma \in {H}^{\hat{\mathfrak{A}}\otimes \mathfrak{A}}$,
when the channel state is $({\theta'},\theta)$,
the quantum state which the environment obtains is
$\hat{V}_{\theta'}\otimes V_{\theta} (\sigma)$.
We have
\begin{align*}&C_{s}(\{(\hat{W}_{\theta'}\otimes W_{\theta}, \hat{V}_{\theta'}\otimes V_{\theta}): ({\theta'},\theta)\in \hat{\Theta}\times \Theta\},r)\\
&\geq C_{s}(\{(W_{\theta}, V_{\theta}): {\theta}\in \Theta\},r)\\
&=\frac{1}{2}\text{ .}\end{align*}

We define 
\[\rho_1 := \frac{1}{2}(|0\rangle+ |1\rangle)(\langle 0|+\langle 1|)^{\hat{\mathfrak{A}}}\otimes |0\rangle\langle 0|^{\mathfrak{A}}\]
and
\[\rho_2 := \frac{1}{2}(|2\rangle+ |3\rangle)(\langle 2|+\langle 3|)^{\hat{\mathfrak{A}}}\otimes |0\rangle\langle 0|^{\mathfrak{A}}\text{ .}\]

If  $\{\hat{W}_{\theta'}\otimes W_{\theta}: ({\theta'},\theta)\in \hat{\Theta}\times \Theta\}$
is symmetrizable, then there exists
 a
parametrized set of distributions $\{\tau(\cdot\mid \sigma):
 \sigma \in {H}^{\hat{\mathfrak{A}}\otimes \mathfrak{A}}\}$ on $ \hat{\Theta}\times \Theta$ such that
\begin{align}&
\sum_{({\theta'},\theta)\in \hat{\Theta}\times \Theta}\tau(({\theta'},\theta)\mid \rho_2)
\hat{W}_{\theta'}\otimes W_{\theta} (\rho_1)=
\sum_{({\theta'},\theta)\in \hat{\Theta}\times \Theta}\tau(({\theta'},\theta)\mid \rho_1)
\hat{W}_{\theta'}\otimes W_{\theta} (\rho_2) \notag\\
& \Rightarrow 
\frac{1}{2} (\tau((1,1^{(+)})\mid \rho_2) + \tau((1,1^{(-)})\mid \rho_2) ) 
 \left(|0\rangle+ |1\rangle\right)\left(\langle 0|+\langle 1|\right)^{\hat{\mathfrak{A}}}\otimes |0\rangle\langle 0|^{\mathfrak{A}}\notag\\
&~ +\frac{1}{2} (\tau((2,1^{(+)})\mid \rho_2) + \tau((2,1^{(-)})\mid \rho_2) ) 
 \left(|0\rangle\langle 0|^{\hat{\mathfrak{A}}} +|1\rangle\langle 1|^{\hat{\mathfrak{A}}}\right)\otimes |0\rangle\langle 0|^{\mathfrak{A}}\notag\\
&~ +\frac{1}{2} (\tau((1,2^{(+)})\mid \rho_2) + \tau((1,2^{(-)})\mid \rho_2) ) 
 \left(|0\rangle+ |1\rangle\right)\left(\langle 0|+\langle 1|\right)^{\hat{\mathfrak{A}}}\otimes |1 \rangle\langle 1|^{\mathfrak{A}}\notag\\
&~ +\frac{1}{2} (\tau((2,2^{(+)})\mid \rho_2) + \tau((2,2^{(-)})\mid \rho_2) ) 
 \left(|0\rangle\langle 0|^{\hat{\mathfrak{A}}} +|1\rangle\langle 1|^{\hat{\mathfrak{A}}}\right)\otimes |1\rangle\langle 1|^{\mathfrak{A}}\notag\\
&=\frac{1}{2} (\tau((1,1^{(+)})\mid \rho_1) + \tau((1,1^{(-)})\mid \rho_1) ) 
 \left(|2\rangle+ |3\rangle\right)\left(\langle 2|+\langle 3|\right)^{\hat{\mathfrak{A}}}\otimes |0\rangle\langle 0|^{\mathfrak{A}}\notag\\
&~ +\frac{1}{2} (\tau((2,1^{(+)})\mid \rho_1) + \tau((2,1^{(-)})\mid \rho_1) ) 
 \left(|2\rangle\langle 2|^{\hat{\mathfrak{A}}} +|3\rangle\langle 3|^{\hat{\mathfrak{A}}}\right)\otimes |0\rangle\langle 0|^{\mathfrak{A}}\notag\\
&~ +\frac{1}{2} (\tau((1,2^{(+)})\mid \rho_1) + \tau((1,2^{(-)})\mid \rho_1) ) 
 \left(|2\rangle+ |3\rangle\right)\left(\langle 2|+\langle 3|\right)^{\hat{\mathfrak{A}}}\otimes |1 \rangle\langle 1|^{\mathfrak{A}}\notag\\
&~ +\frac{1}{2} (\tau((2,2^{(+)})\mid \rho_1) + \tau((2,2^{(-)})\mid \rho_1) ) 
 \left(|2\rangle\langle 2|^{\hat{\mathfrak{A}}} +|3\rangle\langle 3|^{\hat{\mathfrak{A}}}\right)\otimes |1\rangle\langle 1|^{\mathfrak{A}}\notag\\
& \Rightarrow 
\tau((1,1^{(+)})\mid \rho_2) + \tau((1,1^{(-)})\mid \rho_2) \notag\\
&~ = \tau((2,1^{(+)})\mid \rho_2) + \tau((2,1^{(-)})\mid \rho_2)  \notag\\
&~ = \tau((1,2^{(+)})\mid \rho_2) + \tau((1,2^{(-)})\mid \rho_2)  \notag\\
&~ = \tau((2,2^{(+)})\mid \rho_2) + \tau((2,2^{(-)})\mid \rho_2)  \notag\\
&~ = 0 \notag\\
& \Rightarrow \lightning
\text{ .}\end{align}

 $\{\hat{W}_{\theta'}\otimes W_{\theta}: ({\theta'},\theta)\in \hat{\Theta}\times \Theta\}$
is thus not symmetrizable.
By Theorem \ref{lntvttitbaa}
\begin{align*}&C_{s}(\{(\hat{W}_{\theta'}\otimes W_{\theta}, \hat{V}_{\theta'}\otimes V_{\theta}): ({\theta'},\theta)\in \hat{\Theta}\times \Theta\})\\
&= C_{s}(\{(\hat{W}_{\theta'}\otimes W_{\theta}, \hat{V}_{\theta'}\otimes V_{\theta}): ({\theta'},\theta)\in \hat{\Theta}\times \Theta\},r)\\
&\geq \frac{1}{2}\text{ .}\end{align*}
\end{example}

\section{Secrecy Capacity of
Quantum Compound Channels}

In our earlier work
\cite{Bo/Ca/Ca/De} we determined the secrecy capacity of the
classical-quantum compound wiretap channel.
Our idea for the case with    CSI    
at the transmitter was  sending the information in two parts. Firstly, the sender sent
the state information with finite blocks of finite bits with a public code
 to the receiver, and then, depending on $\theta$, the sender
sent the message
with a secure code  in the second part.
The following Lemma has been proved in \cite{Bo/Ca/Ca/De}.\vspace{0.2cm}

\begin{lemma}\label{e1}

The secrecy capacity of the compound classical-quantum wiretap channel $\{(W_{\theta}, V_{\theta})\}: {\theta} \in \Theta\}$ in the case with CSI
 is given by
\begin{equation}\label{e1q}
C_{CSI} (\{W_{\theta}, V_{\theta}\}: {\theta} \in \Theta\})= \lim_{n \rightarrow \infty} \min_{{\theta} \in \Theta} \max_{P_{inp},
w_t}\frac{1}{n}( \chi(P_{inp};B_{\theta}^{ n})- \chi(P_{inp};Z_{\theta}^{
n}))\end{equation}
 where $B_{\theta}$ are the resulting random  quantum states
at the output of legal receiver channels  and $Z_{\theta}$ are the
resulting random quantum  states at the output of wiretap channels.
The maximum is taken over all probability distributions 
$P_{inp}$ on the input quantum  states $w_{\theta}$.\vspace{0.15cm}

Assume that the sender's encoding is restricted to transmitting an  indexed finite set of orthogonal quantum  states
$\{\rho_{x}: x\in A\}\subset \mathcal{S}({H'}^{\otimes n})$, then
 the secrecy capacity of the compound classical-quantum wiretap channel $\{W_{\theta}, V_{\theta}\}: {\theta} \in \Theta\}$ in the case with no CSI
 at the encoder is given by
\begin{align}&\label{qnocsie1q}
C_{S} (\{(W_{\theta}, V_{\theta})\}: {\theta} \in \Theta\})= \lim_{n\rightarrow \infty} \max_{\mathcal{U}\rightarrow A \rightarrow
(BZ)_{\theta}} \frac{1}{n}\biggl(\min_{{\theta}\in\Theta} \chi(\mathcal{U};B_{{\theta}}^{
n})\allowdisplaybreaks\notag\\
&- \max_{{\theta}\in\Theta} \chi(\mathcal{U};Z_{\theta}^{ n})\biggr)\text{ .}
\end{align} 

\end{lemma}\vspace{0.2cm}

Now we are going to prove Corollary \ref{bibqicbttcompg}.\vspace{0.2cm}

\begin{proof}
(\ref{bibqicbttcomps}) follows immediately from Lemma \ref{e1}.

(\ref{bibqicbttcomp}) can be proved in a similar way as  Lemma \ref{e1}:
The converse is 
trivial since
the secrecy capacity of $\{(N_{\theta}\circ F,{V}_{\theta}\circ F): {\theta}\in \Theta\}$  with CSI
 is
upper bounded by $\lim_{n\rightarrow \infty} \frac{1}{n} $ $\max_{U\rightarrow A \rightarrow \{(BZ)_{\theta}:t\}}$
	$\chi(p_U;B_{\theta}^n{\otimes n})$ $-\chi(p_U;Z_{\theta}^n{\otimes n})$.

 Now we are going to show the achievability.
When    \[\lim_{n\rightarrow \infty} \frac{1}{n}  \max_{U\rightarrow A \rightarrow \{(BZ)_{\theta}:t\}} 
	 \chi(p_U;B_{\theta}^n) -\chi(p_U;Z_{\theta}^n) \leq 0\]    holds,
	the secrecy capacity of $\{(W_{\theta},{V}_{\theta}): {\theta}\in \Theta\}$   with CSI
 is zero and there is nothing to prove.

Now we assume that    \[\lim_{n\rightarrow \infty} \frac{1}{n}  \max_{U\rightarrow A \rightarrow \{(BZ)_{\theta}:t\}} 
	 \chi(p_U;B_{\theta}^n) -\chi(p_U;Z_{\theta}^n) > 0\]    holds.
In this case we showed that
in \cite{Bo/Ca/Ca/De} the  secrecy capacity of $\{(N_{\theta}\circ F,{V}_{\theta}\circ F): {\theta}\in \Theta\}$  without CSI
is positive and  the sender can
build a  code $C_1$ $=  \bigl(E^{(1)}, \{D_{\theta}^n{(1)} : {\theta}\in \Theta\}\bigr)$
such that the CSI can be sent to the legal
receiver with a block with  finite length and the eavesdropper knowing nothing.
The first part
is of length O(1), which is negligible compared to the second
part.

If both the sender and the legal receiver
have the full knowledge of $\theta$, then we only have to look at the
single wiretap channel $(N_{\theta}\circ F,{V}_{\theta}\circ F)$.

In \cite{Ca/Wi/Ye} and \cite{De} it was shown that if $n$ is
sufficiently large, there exists an $(n, J_{n,\theta})$ code for the
quantum wiretap channel $(N_{\theta}\circ F,{V}_{\theta}\circ F)$ with
\[\log J_{n,t} = \max_{U\rightarrow A \rightarrow (BZ)_{\theta}}( \chi(P_{U};B_{\theta}^n{ n})- \chi(P_{U};Z_{\theta}^n{ n}))-\epsilon\text{ ,}
\] for any positive $\epsilon$ and $\zeta$ such that
\[  1- \frac{1}{J_n} \sum_{j=1}^{J_n}
\mathrm{tr}(N_{\theta}^{\otimes n}\circ F^n(E_{\theta}(~|j))D_j^{(\theta)}) < \epsilon\text{ ,}\] and
\[
\chi\left(R_{uni};Z_{\theta}^{\otimes n}\right) < \zeta\text{
.}\]

When the sender and the legal receiver both know $\theta$, they can build
an $(n, J_{n})$ secure code $C_2^{(\theta)}$ 
$=  \bigl(E^{(2)}, \{D_j^{(2)} : j = 1,\cdots J_n\}\bigr)$
to send the message,
where
\[\log J_{n} =  \min_{{\theta}\in \Theta} 
\max_{U\rightarrow A \rightarrow \{(BZ)_{\theta}:\theta\}}
	\chi(p_U;B_{\theta}^n)-\chi(p_U;Z_{\theta}^n) -\epsilon\text{ .}
	\]
	We define now a code
	$\mathcal{C}$ $:= \bigl(E^{(1)}E^{(2)}, \{D_{\theta}^n{(1)}\otimes D_j^{(2)}
	: {\theta}\in \Theta,  j = 1,\cdots J_n\}\bigr)$.
	Since the length of $C_1$ can be negligible compared to the second
part in asymptotic scenarios, we have
 \[C_{S,CSI} \geq \min_{{\theta}\in \Theta} 
\lim_{n\rightarrow \infty} \frac{1}{n}\max_{U\rightarrow A \rightarrow \{(BZ)_{\theta}:\theta\}}
	\chi(p_U;B_{\theta}^n)-\chi(p_U;Z_{\theta}^n)\text{ .}
	\]
\end{proof}

\section*{Acknowledgment}   
Support   by
Bundesministerium f\"ur Bildung und 
Forschung (BMBF) via Grant 16KIS0118K and 16KIS0117K,
the German Research Council (DFG) via Grant 
1129/1-1 and Bo 1734/20-1, and the ERC via Advanced 
Grant IRQUAT
 is   gratefully acknowledged.

\end{document}